\documentclass[pdflatex,sn-mathphys-num]{sn-jnl}

\usepackage{graphicx}%
\usepackage{multirow}%
\usepackage{amsmath,amssymb,amsfonts}%
\usepackage{amsthm}%
\usepackage{mathrsfs}%
\usepackage[title]{appendix}%
\usepackage{xcolor}%
\usepackage{textcomp}%
\usepackage{manyfoot}%
\usepackage{booktabs}%
\usepackage{algorithm}%
\usepackage{algorithmicx}%
\usepackage{algpseudocode}%
\usepackage{listings}%
\usepackage{hhline}
\usepackage{float}
\usepackage{placeins}
\usepackage{blindtext}
\usepackage{amsmath}
\usepackage{amssymb}
\usepackage{amsfonts}
\usepackage{physics}
\usepackage{comment}
\usepackage{natbib}
\usepackage{color}
\usepackage{tikz}
\usepackage{amsthm}
\usepackage{dsfont}
\usepackage{cancel}
\usepackage{hyperref}
\usepackage{mathtools}
\usepackage{multirow}
\usepackage{colortbl} 
\usepackage{caption}

\theoremstyle{thmstyleone}%
\theoremstyle{thmstyletwo}%

\theoremstyle{thmstylethree}%

\renewcommand{\eqref}[1]{Eq.~(\ref{#1})}
\newcommand\mysign{\stackrel{\mathclap{\normalfont\mbox{??}}}{=}}

\newcommand\checkmarkbis[1][]{
  \tikz[scale=0.4,#1]{\fill(0,.35) -- (.25,0) -- (1,.7) -- (.25,.15) -- cycle;}
}
\newcommand\crossmark[1][]{
  \tikz[scale=0.4,#1]{
    \fill(0,0)--(0.1,0) .. controls (0.5,0.4) .. (1,0.7)--(0.9,0.7) ..  controls (0.5,0.5) ..(0,0.1) --cycle;
    \fill(1,0.1)--(0.9,0.1) .. controls (0.5,0.3) .. (0,0.7)--(0.1,0.7) .. controls (0.5,0.4) ..(1,0.2) --cycle;
  }
}

\newcommand{\rt} {(\textbf{r},t)}

\newcommand{\unit}{\mathds{1}}

\newcommand{\BO} {\mathcal{O}}
\newcommand{\BE} {\mathcal{E}}
\newcommand{\BW} {\mathcal{W}}

\newcommand{\BN} {\mathcal{N}}

\newcommand{\V}{A}
\newcommand{\A}{\textbf{A}}
\newcommand{\p}{\psi}
\newcommand{\B}{\textbf{B}}
\newcommand{\E}{\textbf{E}}

\newcommand{\px}{\psi(\textbf{x},t)}

\newcommand{\vB}[1]{\textbf{v}_B^{#1}}
\newcommand{\vO}[1]{\textbf{v}_O^{#1}}

\newcommand{\opvn}[1]{V_{#1}}
\newcommand{\oppn}[1]{P_{#1}}
\newcommand{\An}[1]{A_{#1}}
\newcommand{\egxn}[1]{x_{#1}}
\newcommand{\Bn}[1]{B_{#1}}
\newcommand{\En}[1]{E_{#1}}
\newcommand{\vBn}[2]{{v}_{B,#1}^{#2}}
\newcommand{\vOn}[2]{{v}_{O,#1}^{#2}}

\newcommand{\g}{g}
\newcommand{\gx}{g(\textbf{x},t)}

\newcommand{\pxg}{\psi^g(\textbf{x},t)}

\newcommand{\Vg}{A^g}
\newcommand{\Ag}{\textbf{A}^g}
\newcommand{\pg}{\psi^g}

\newcommand{\opp}{\hat{\textbf{P}}}
\newcommand{\opv}{\hat {\textbf{V}}}
\newcommand{\opA}{\hat{\textbf{A}}}
\newcommand{\opV}{\hat{A}}
\newcommand{\opH}{\hat{{H}}}
\newcommand{\opx}{\hat{\textbf{X}}}
\newcommand{\opU}{\hat{U}}
\newcommand{\opUd}{\hat{U}^{\dagger}}

\newcommand{\opUt}[1]{\hat{U}_{{#1}}}
\newcommand{\opUtd}[1]{\hat{U}^{\dagger}_{{#1}}}
\newcommand{\opUtg}[1]{\hat{U}^g_{{#1}}}

\newcommand{\opG}{\hat{{G}}}
\newcommand{\opg}{\hat{{g}}}
\newcommand{\opF}{\hat{{F}}}
\newcommand{\opO}{\hat{{O}}}
\newcommand{\opC}{\hat{{C}}}
\newcommand{\opAg}{\hat{\textbf{A}^g}}
\newcommand{\opVg}{\hat{A^g}}
\newcommand{\opE}{\hat{\textbf{E}}}
\newcommand{\opB}{\hat{\textbf{B}}}

\newcommand{\opW}{\hat{{W}}}

\newcommand{\opphi}{\hat{\Lambda}}
\newcommand{\opomega}{\hat{\Omega}}
\newcommand{\opa}{\hat{{\Xi}}}

\newcommand{\evf}{\textsl{f\;}}

\newcommand{\egv}{\textbf{v}}
\newcommand{\ego}{o}

\newcommand{\egx}{\textbf{x}}
\newcommand{\egp}{\textbf{p}}
\newcommand{\egf}{{f}}

\newcommand{\egw}{w}

\newcommand{\egphi}{{\lambda}}

\begin{document}

\title[Time Derivatives of Weak Values]{Time Derivatives of Weak Values}

\author*[1]{\fnm{Xavier Oriols}}\email{xavier.oriols@uab.es}

\affil*[1]{\orgname{Universitat Aut\`onoma de Barcelona}, \postcode{08193}, \state{Bellaterra}, \country{Spain}}


\abstract{The time derivative of a physical property often gives rise to another meaningful property. Since weak values provide empirical insights that cannot be derived from expectation values, this paper explores what physical properties can be obtained from the time derivative of weak values. It demonstrates that, in general, the time derivative of a weak value that is invariant under an electromagnetic gauge transformation is neither a weak value nor a gauge-invariant quantity. Left- and right-hand time derivatives of a weak value are defined, and two necessary and sufficient conditions are presented to ensure that they are also gauge-invariant weak values.  Finite-difference approximations of the left- and right-hand time derivatives of weak values are also presented, yielding results that match the theoretical expressions when these are gauge invariant. With these definitions, a local Ehrenfest-like theorem can be derived, giving a natural interpretation for the time derivative of weak values. Notably, a single measured weak value of the system's position provides information about two additional unmeasured weak values: the system's local velocity and acceleration, through the first- and second-order time derivatives of the initial weak value, respectively. These findings also offer guidelines for experimentalists to translate the weak value theory into practical laboratory setups, paving the way for innovative quantum technologies. An example illustrates how the electromagnetic field can be determined at specific positions and times from the first- and second-order time derivatives of a weak value of position.}




\maketitle
\section{Introduction}
\label{intro}

The primary goal of science is to construct models that successfully predict empirical results. Weak values are an original way of predicting novel properties of quantum systems  \cite{aharonov1988,aharonov1990,vaidman1996,wiseman2007,brun2008,foo2022,  durr2009,review2012,destefani2023b,destefani2023,diosi2023,ahnert2004,botero2000,brout1995,campo2004,hofmann2010,devashish2021,oriols1996,aharonov1964,steinberg1995,ruseckas2002,leavens2005,kastner2004} which are empirically observable in the laboratory \cite{agarwal2007,sbramanian2023,pryde2005,jordan2010,marian2016,jordan2007,korotkov2006,romito2008,zilberberg2011,starling2009, kocsis2011observing,hariri2019experimental,ramos2020measurement,review2014,lundeen2011,zhu2021,ricthie1991,brunner2004,hosten2008,starling2010,williams2008,resch2004}. This new information offered by weak values cannot be accessed through expectation values. In physics, the time derivative of a physical property often leads to another meaningful property. It remains an open question whether this relationship holds for the time derivatives of weak values. This paper investigates such a relationship. 

Weak values are generating increasing interest in the scientific community. They have been employed to explore novel properties such as the position-dependent velocity of particles in either non-relativistic \cite{wiseman2007,durr2009} or relativistic \cite{foo2022} scenarios, thermalized kinetic energies \cite{destefani2023b,destefani2023} as well as tunneling and arrival times \cite{steinberg1995, oriols1996,ahnert2004,ruseckas2002}. Weak values have been invoked in cosmological contexts, such as inflation theory \cite{campo2004} and the back-reaction of the Hawking radiation from black holes \cite{brout1995}. In quantum information science, they have been applied to quantum computation \cite{brun2008}, quantum communication \cite{botero2000}, and quantum sensing  \cite{hofmann2010}. 

There are many experimental protocols for weak values in optical and solid-state platforms \cite{agarwal2007,jordan2010,jordan2007,korotkov2006,romito2008,marian2016,sbramanian2023,pryde2005,zilberberg2011}. Relevant experiments conducted with weak values include the three-box paradox \cite{resch2004}, the violation of the Leggett-Garg inequality \cite{williams2008}, the detection of the superluminal signals \cite{brunner2004} and the measurement of Bohmian trajectories \cite{kocsis2011observing}. Since weak values are expressed as quotients, a small denominator (i.e., quasi-orthogonal pre- and post-selected states) can be used to amplify the spin Hall effect of light \cite{hosten2008}, optimize the signal-to-noise ratio \cite{starling2009} and measure small changes in optical frequencies \cite{starling2010}.  

This paper aims to explore the additional physical insights that can be gained about the behavior of quantum systems by analyzing not only weak values, but also their time derivatives. Wiseman was the first to discuss the time derivative of weak values \cite{wiseman2007}, presenting a local velocity derived from the time derivative of a weak value of the position.\footnote{The word \textit{local} in this paper is defined in contrast to \textit{global}. The expectation value of the velocity represents a \textit{global} quantity, as it does not account for how velocity is distributed across different positions, while the weak value of the velocity, when post-selected at a given position, provides a \textit{local} velocity at that position.  In a two-particle system, the term local could be ambiguous because a \textit{local} velocity at a point in configuration space may correspond to two \textit{non-local} velocities in physical space \cite{braverman2013}. In any case, for the single-particle systems considered in this work, the term \textit{local} will be unproblematic.} This paper examines the time derivatives of weak values from a general perspective using arbitrary operators. In general, the time derivative of a weak value reveals a much richer phenomenology than the time derivatives of expectation values. In many cases, the time derivative of a weak value is neither a weak value nor invariant under an electromagnetic gauge transformation.

A quantity that is not invariant under an electromagnetic gauge transformation cannot be measured in a laboratory. The electromagnetic vector and scalar potentials are the typical examples of the utility and unmeasurability of gauge-dependent elements \cite{healey2007,kobe1982,jackson2001,stokes2021,cohen1986} (with the Aharonov-Bohm effect being its most iconic example \cite{aharonov1959}). Similarly, a weak value or its time derivatives are empirically observable in the laboratory only when they are gauge invariant. Several examples of observable and unobservable weak values and their time derivative will be discussed in the paper. For example, it is known that the wave function depends on the electromagnetic gauge \cite{optics, Ballentine2014,cohen1986}. Thus, strictly speaking, the wave function cannot be empirically measured through a weak value. 
 
The ontological meaning of weak values remains controversial, ranging from being mere mathematical transition amplitudes in most orthodox views \cite{sokolovski2016, matzkin2019}, to representing basic elements in alternative interpretations of quantum phenomena \cite{aharonov1964, reznik1995, hiley2012, kastner2017, cohen2017, sinclair2019}. For example, the meaning of weak values has been interpreted in the context of the two-state vector formalism of quantum mechanics \cite{aharonov1988,aharonov1990,vaidman1996,aharonov1964}, consistent histories \cite{kastner2004}, and Bohmian or modal interpretations \cite{leavens2005,devashish2021}. Controversies surrounding weak values also arise in debates about which empirical procedures can reliably measure a weak value in the laboratory.  It is argued that weak values are measured from conditional probabilities obtained from a large ensemble of identically prepared systems \cite{garretson2004}. In a recent experiment, however, a weak value was measured ``with just a single click,'' without requiring statistical averaging \cite{rebufello2021}. It has also been argued that weak values can, in fact, be measured using strong measurements \cite{cohen2018}. The widely accepted consensus around the concept of weak values is the mathematical expression that defines a weak value. This mathematical expression,  which is sufficient for all developments and conclusions done in this paper, can be derived from all common quantum theories, regardless of the ontological meaning attributed to weak values, and the resulting predictions can subsequently be tested in the laboratory, irrespective of the specific experimental protocol used. Thus, weak values (and their time derivatives) possess predictive power and offer novel ways to characterize quantum systems without the need to select a particular ontological interpretation or a fixed measurement protocol. Therefore, all results presented in this paper are ontologically neutral (valid under any interpretation of quantum mechanics). Similarly, we do not need to specify any ontological status for gauge-dependent elements to draw the conclusions of this paper.\footnote{The concept of a ``true physical quantity" is defined in Ref. \cite{cohen1986} as: ``a quantity whose value at any time does not depend (for a given motion of the system) on the gauge used to describe the electromagnetic field." This definition suggests ontological consequences, as other gauge-dependent properties are referred to as ``non-physical quantities." However, as mentioned in the introduction, such ontological implications are not necessary for discussing the results in this paper \cite{beche2016,tumulka2022}.  The analogy between electromagnetic potentials and angles, both exhibiting \emph{redundancies}, clarifies this point. The direction of a physical system specified by an angle $\theta$ or by another $\theta + 2\pi n$, with $n \in \mathbb{Z}$, is the same. Thus, the angle specifying the direction of a physical system cannot be directly measured as a unique number, because there are infinite equivalent angles corresponding to the same physical system. However, this \emph{redundancy} of the angle does not directly imply that $\theta$ is an ``unphysical'' parameter. Similarly, the discussion of how the \emph{redundancy} of electromagnetic potentials affects their direct measurability is indeed pertinent here. However, the question of how this \emph{redundancy} affects the ``physicality'' of the electromagnetic potentials is a different and much more subtle debate.}
 
The structure of the remainder of the paper is as follows. In the rest of this introduction, we revisit the role of electromagnetic gauge invariance in non-relativistic quantum mechanics, particularly regarding expectation values and their time derivatives. In Sec.~\ref{theory}, we present the time derivatives of weak values and the conditions under which they are gauge invariant.  In Sec.~\ref{theorylaboratory}, we present finite-difference approximations of the left- and right-hand time derivatives of weak values, and we show that they yield results identical to the theoretical ones when the latter are gauge invariant.  In Sec.~\ref{discussion1}, we explore how gauge invariance determines the empirical observability of typical weak values. Sec.~\ref{discussion2} introduces several examples of time derivatives of weak values like the local velocity, the local version of the work-energy theorem, and the local Lorentz force, leading to local quantum sensors of the electromagnetic field. In Sec.~\ref{which}, we discuss which pre- and post-selected states provide time derivatives of weak values consistent with the time derivative of expectation values.  Finally, Sec.~\ref{conclusion} explains the main conclusion. Several appendices detail the discussions presented in the paper.

\subsection{Gauge invariance in quantum mechanics}

Although they can also be formulated for many-particle scenarios \cite{destefani2023b,destefani2023}, weak values become especially relevant when discussing the properties of a single particle. 
All the results in the present paper are developed within the so-called semi-classical approach to light–matter interaction, in the sense that the electron (i.e., matter) is treated as a quantum particle, while the electromagnetic fields or potentials (i.e., light) are treated as classical fields, affecting the electron as external potentials, but without being affected themselves by the electron.\footnote{If the electromagnetic fields are treated quantum mechanically, allowing the self-consistent exchange of energy between light and matter, the problem of gauge invariance remains. A simple way to see this is through the so-called canonical quantization of the electromagnetic field—an approach sometimes criticized for being too heuristic but widely used in textbooks. The quantum Hamiltonian of light and matter is obtained from the semi-classical Hamiltonian in  \eqref{hami} by promoting the classical fields to field operators. However, other equally valid (gauge-dependent) quantum Hamiltonians can be obtained using the same canonical quantization procedure when starting from different (gauge-dependent) semi-classical Hamiltonians in  \eqref{hamig}. } 
 In particular, we will deal with a non-relativistic spinless particle of mass $m^*$ interacting with a classical electromagnetic field.

The Hamiltonian of the system, in the Coulomb gauge, is written as  \cite{optics, Ballentine2014,cohen1986}
\begin{eqnarray}
H=H(\A,\V)= \frac{1}{2m^*}\left( \opp -q\A\right)^2+ q\V,
 \label{hami}
\end{eqnarray}
where $q$ is the  charge of the particle,  $\opp=-i\hbar \grad $ is the canonical momentum operator, and $\V$ and $\A$ are the scalar and vector electromagnetic potentials, respectively. The evolution of the wave function $\px=\langle \egx|\p(t)\rangle$ is given by the Schrödinger equation
\begin{eqnarray}
i \hbar \frac{\partial \px}{\partial t}= H(\A, \V) \px.
 \label{scho}
\end{eqnarray}
It is well-known that a different set of potentials (indicated by the superscript $g$) given by \cite{optics,Ballentine2014,cohen1986} 
\begin{eqnarray}
\Ag = \A+{\nabla} \g \;\;\;\;\;\text{and}\;\;\;\;\;\Vg = \V-\frac{\partial \g}{ \partial t},
\label{vg}
\end{eqnarray}
keeps the overall theory invariant in the sense that the observable properties remain the same in any gauge. Here, $\g=\gx$ is any sufficiently regular real function \footnote{Apart from having space and time derivatives, $\gx$ must be single-valued.} over $3D$ space $\egx$ plus time $t$. For example, the electric and magnetic fields are gauge invariant and defined as $\E=-\grad \Vg-\frac{\partial \Ag}{\partial t}$ and $\B=\grad \times \Ag$, respectively. In this paper, according to the above notation \eqref{vg}, any symbol (for a state, eigenvalue, operator, etc.) without a superscript $g$ should be interpreted either as a gauge-invariant element or as an element defined within the Coulomb gauge. 

When using a general gauge, to keep the same structure of the Schrödinger equation in (\ref{scho}),  a gauge-dependent Hamiltonian $H^g$ and a gauge-dependent wave function $\pxg$ are needed \cite{optics,Ballentine2014,cohen1986}. See Appendix~\ref{ap1} for details and Ref. \cite{cohen1986,optics,Ballentine2014}. The new-gauge Hamiltonian $H^g$ is given by 
\begin{eqnarray}
H^g:=H (\Ag, \Vg)=e^{i\frac{q}{\hbar}\g}\left(H(\A, \V)-q\frac{\partial \g}{ \partial t}\right)e^{-i\frac{q}{\hbar}\g},
\label{hamig}
\end{eqnarray}
and the new-gauge wave function $\pxg$ is given by:
\begin{eqnarray}
\langle \egx|\p^g(t)\rangle=\langle \egx|\opG(t)|\p\rangle=\pxg = e^{i\frac{q}{\hbar}\gx} \px,
\label{wg}
\end{eqnarray}
where $\opG(t)$ is a local operator whose position representation is $\langle \egx|\opG(t)|\egx' \rangle=\langle \egx|e^{i\frac{q}{\hbar}\opg}|\egx' \rangle=e^{i\frac{q}{\hbar}\gx}\delta(\egx-\egx')$. This arbitrary gauge  allows an infinite set of possible Hamiltonians, electromagnetic potentials and wave functions \cite{healey2007,kobe1982,jackson2001,stokes2021}. 

Finally, depending on the gauge, the wave function evolution can be equivalently written in terms of the (Coulomb gauge) time-evolution operator as $|\p(t_R)\rangle = \opU(t) |\p(0) \rangle$ or in a general gauge as $|\pg(t)\rangle = \opU^g(t) |\pg(0) \rangle$. As shown in Appendix~\ref{ap2} (see also \cite{kobe1985}), the gauge dependence of the time evolution operator $\opU(t)$ is
\begin{eqnarray}
\opU(t)=\opG(t)\opU(t)\opG^{\dagger}(0)=\hat{ \mathds{T}} e^{-\frac{i}{\hbar}\int_0^{t} \opH (\Ag, \Vg) dt'},
\label{utg}
\end{eqnarray}
in agreement with the gauge-dependent Hamiltonian in \eqref{hamig} with $\hat {\mathds{T}}$ being the time-ordering integral operator and $\opG^{\dagger}(t)$ defined such that $\opG(t)\opG^{\dagger}(t)=\unit$.


\subsection{Time Derivative of Expectation values } 

The time derivative of weak values shows similarities and differences with the time derivative of expectation values. In this subsection, a brief explanation of the expectation values and their time derivatives is presented.  

We consider an ensemble of identical quantum states $|\p\rangle=|\p(0)\rangle$ prepared at the initial time $t=0$. Such an initial state evolves until time $t$ when a measurement of an output linked to the operator $\opO$ is made. The relationship between the measured output and the hermitian operator $\opO$ is determined through a positive-operator-valued measure (POVM) or a projection-valued measure (PVM) \cite{svensson2013pedagogical,wiseman2009}. The expectation value $\BE$ is given by
\begin{equation}
\BE\big(\opO,t\big||\p\rangle\big):=\langle \p|\opUtd{}\opO \opUt{} |\p\rangle=\langle \p(t)|\opO  |\p(t)\rangle,
\label{ev}
\end{equation}
The notation of the left-hand side of \eqref{ev} indicates that the expectation value is evaluated for the operator $\opO$ and conditioned on $|\p\rangle$ (i.e., the same property gives a different expectation value if the quantum state changes).   

As reported in the literature \cite{cohen1986,kobe1982}, the condition for  $\BE\big(\opO,t_R|\p\big)$ to be empirically observable (i.e., gauge invariant) is that the same value has to be obtained in the laboratory for the Coulomb or any other gauge, $\langle \p(t)|\opO | \p(t)\rangle=\langle \p^g(t)|\opO^g | \p^g(t)\rangle$. Using \eqref{wg} and \eqref{utg}, the necessary and sufficient   condition on the operator $\opO$ for the empirical observability of $\BE\big(\opO,t|\p\big)$ for any state $|\p\rangle$ is then,
\begin{equation}
\textbf{C1:}\;\;\;  \opO^g:=\opO(\opAg,\opVg)=\opG \opO(\opA, \opV) \opG^{\dagger}\nonumber.
\end{equation}
When this condition is satisfied, one gets the same expectation value in any gauge.

The preparation (pre selection) of the initial state $|\p\rangle$ used in \eqref{ev} to evaluate an expectation value is not free of gauge ambiguities. The Hilbert space representation of the quantum state  is a gauge-dependent object, as shown in \eqref{wg}. Thus, the preparation of the initial state $|\pg\rangle$ is achieved by detecting a measurable property $\egphi^g$ of the system (not the initial quantum state itself). Let us define $\opphi^g$ as the operator used to detect such a property as follows $\opphi^g |\pg\rangle=\egphi^g |\pg\rangle$. We discuss below the condition on $\opphi^g$ to ensure that $\egphi^g=\egphi$ is gauge invariant (i.e., empirically measurable). Using $|\pg \rangle=\opG|\p \rangle$ as in \eqref{wg}, the previous eigenvalue equation in an arbitrary gauge $\opphi^g |\pg\rangle=\egphi^g |\pg\rangle$ can be written as $\opphi^g \opG |\p\rangle=\egphi^g \opG |\p\rangle$. Thus, to have $\egphi^g=\egphi$, the  necessary and sufficient  condition on the operator $\opphi$ is 
\begin{equation}
\textbf{C2:}\;\;\;\opphi^g:=\opphi(\opA^g,\opV^g)=\opG \opphi(\opA, \opV) \opG^{\dagger}\nonumber,
\end{equation}
In agreement with \eqref{wg}, when \textbf{C2} is satisfied, the direct measurement of the initial state $|\pg\rangle$ is not accessible, but one can infer that the quantum system is $|\pg\rangle$ (in any gauge) by the measurement of the (gauge invariant) eigenvalue $\egphi$.  In summary, condition \textbf{C2} shows that, if \eqref{ev} wants to be measured in the laboratory, the initial state $|\p\rangle$ has to be empirically identified in the laboratory through the measurement of the gauge invariant value $\egphi$.  

A list of common operators and their accomplishment or not of \textbf{C1} and \textbf{C2} is presented in Table~\ref{table}.\footnote{The paper is not dealing with the gauge invariance of a weak value due to an arbitrary (global) phase in quantum states, but with the (electromagnetic) gauge invariance when not only the quantum states are changed as $|\egf\rangle \to|\egf^g\rangle$ and $|\p\rangle \to |\p^g\rangle$ as indicated in \eqref{wg} with an arbitrary and local phase determined by $\gx$, but also the operator is changed as $\opO\to\opO^g$ because the electromagnetic potentials have changed in \eqref{vg} with the same $\gx$.}

\begin {table}
\begin{tabular}{c|c|c}
\hline
Operator name  & Symbol $\opomega$ &  $\opomega^g=\opG \opomega \opG^{\dagger}$ \\
\hline
\hline
\color{teal} Position  &  $\opx$ &  \color{teal} \checkmarkbis \color{black} \\
\hline
\color{red} Canonical momentum   &  $\opp$  &  \color{red} \crossmark  \color{black}\\
\hline
\color{teal} Velocity  &  $\opv$  &   \color{teal} \checkmarkbis \color{black}  \\
\hline
\color{teal} Position projector &  $|\egx \rangle \langle \egx|$ & \color{teal} \checkmarkbis \color{black}   \\
\hline
\color{red} Canonical momentum  projector  &  $|\egp \rangle \langle \egp|$  &  \color{red} \crossmark  \color{black}\\
\hline
\color{teal} Velocity projector  & $|\egv \rangle \langle \egv|$  &   \color{teal} \checkmarkbis \color{black}  \\
\hline
\color{red} Vector Potential  &  $\opA$ & \color{red} \crossmark  \color{black}    \\
\hline
\color{red} Scalar Potential  &  $\opV$  &  \color{red} \crossmark  \color{black}  \\
\hline
\color{teal} Electric field  &  $\opE$ & \color{teal} \checkmarkbis \color{black}   \\
\hline
\color{teal} Magnetic field  &  $\opB$ & \color{teal} \checkmarkbis \color{black}   \\
\hline
\color{red} Hamiltonian  &  $\opH$  &  \color{red} \crossmark \color{black}  \\
\hline
\color{teal} Kinetic energy  &  $\opW$ & \color{teal} \checkmarkbis \color{black}   \\
\hline
\hline
\end{tabular}
\caption{\label{table} Typical operators satisfying (\color{teal}\checkmarkbis\color{black}) or not (\color{red}\crossmark\color{black}) the gauge conditions $\opomega^g=\opG \opomega \opG^{\dagger}$. The previous condition has to be satisfied by $\opomega=\opO,\opphi$ (i.e., \textbf{C1}, \textbf{C2}) to ensure a gauge-invariant expectation value. Satisfying the previous condition for $\opomega=\opO, \opphi, \opF$ (i.e., \textbf{C1}, \textbf{C2} and \textbf{C3}) are necessary and sufficient \color{black}  to define a gauge-invariant weak value.}
\end{table}

Writing explicitly its $t$-dependence in \eqref{ev} allows us to describe the time derivative of expectation values. From $\BE \big(\opO,t\big||\p\rangle\big)$, using
\begin{eqnarray}
\frac{\partial \opU(t)}{\partial t}=-\frac{i}{\hbar}\opH(t)\opU(t)
\label {dtu1}
\end{eqnarray}
and its complex conjugate $\frac{\partial \opUd(t)}{\partial t}=\frac{i}{\hbar}\opUd(t) \opH(t)$, it can be shown that:
\begin{eqnarray}
&&\frac{\partial \BE \big(\opO,t \big||\p\rangle \big)}{\partial t} =\langle \Psi(t)|\opC | \Psi(t)\rangle,
\label{eren}
\end{eqnarray}
where we have defined:  
\begin{equation}
\opC:=\frac{d \opO(t)}{dt}=\frac{i}{\hbar}[\opH,\opO]+\frac{\partial \opO}{\partial t}.
\label{eren2}
\end{equation}
As far as the initial state is properly identified by the eigenvalue $\egphi$, and the operator $\opO$ satisfies, $\opO^g:=\opG\opO\opG^{\dagger}$, it can be shown that the time derivative of the expectation value is also gauge invariant $\langle \Psi^g(t)|\opC^g | \Psi^g(t)\rangle=\langle \Psi(t)|\opC | \Psi(t)\rangle$. See Appendix~\ref{ap3} and also Refs.  \cite{Ballentine2014,cohen1986}. In summary, no additional condition (apart from \textbf{C1} and \textbf{C2}) is needed to ensure the gauge invariance of the time derivative of an expectation value. 

All the results presented thus far are well-known in the literature. Eq. \eqref{eren2} is just the Heisenberg equation of motion of the operator $\opO$. When selecting a particular wave function $|\p\rangle$, the expectation value applied to \eqref{eren2}, using position and momentum as operators, leads to the well-known Ehrenfest theorem \cite{ehrenfest1927}.


\section{Time Derivative of Weak values} 
\label{theory}

Before discussing their time derivatives, we briefly introduce the weak value.  As indicated first by Aharonov et al. \cite{aharonov1988},  additional empirical insights, which are not directly accessible from the expectation value  $\BE\big(\opO,t\big||\p\rangle\big)$, can be obtained in the laboratory when the protocol to measure the expectation values is modified. 

The prepared system $|\p\rangle$ is allowed to interact with another system, called the meter (or ancilla). During this interaction, the meter and the system become coupled, and the system undergoes a ``weak'' perturbation that can be modeled by a POVM linked to $\opO$. The subsequent measurement of the meter is expected to yield information about the property of the system associated with $\opO$ due to the previous system-meter coupling.  

Up to this point, the procedure follows the general approach used to measure expectation values (using a POVM instead of a PVM) and introduces no novelty. The distinctive aspect of the weak value is that the averaging over the different meter outputs is performed only on a sub-ensemble of the initial states. This sub-ensemble includes only those initial states that yield the specific eigenvalue $\egf$ when a \textit{strong} measurement is performed via the PVM of $\opF$ (i.e., $\opF| \evf \rangle = \egf |\evf\rangle$).  

\begin{figure}
  \centering
\includegraphics[height=16.5cm]{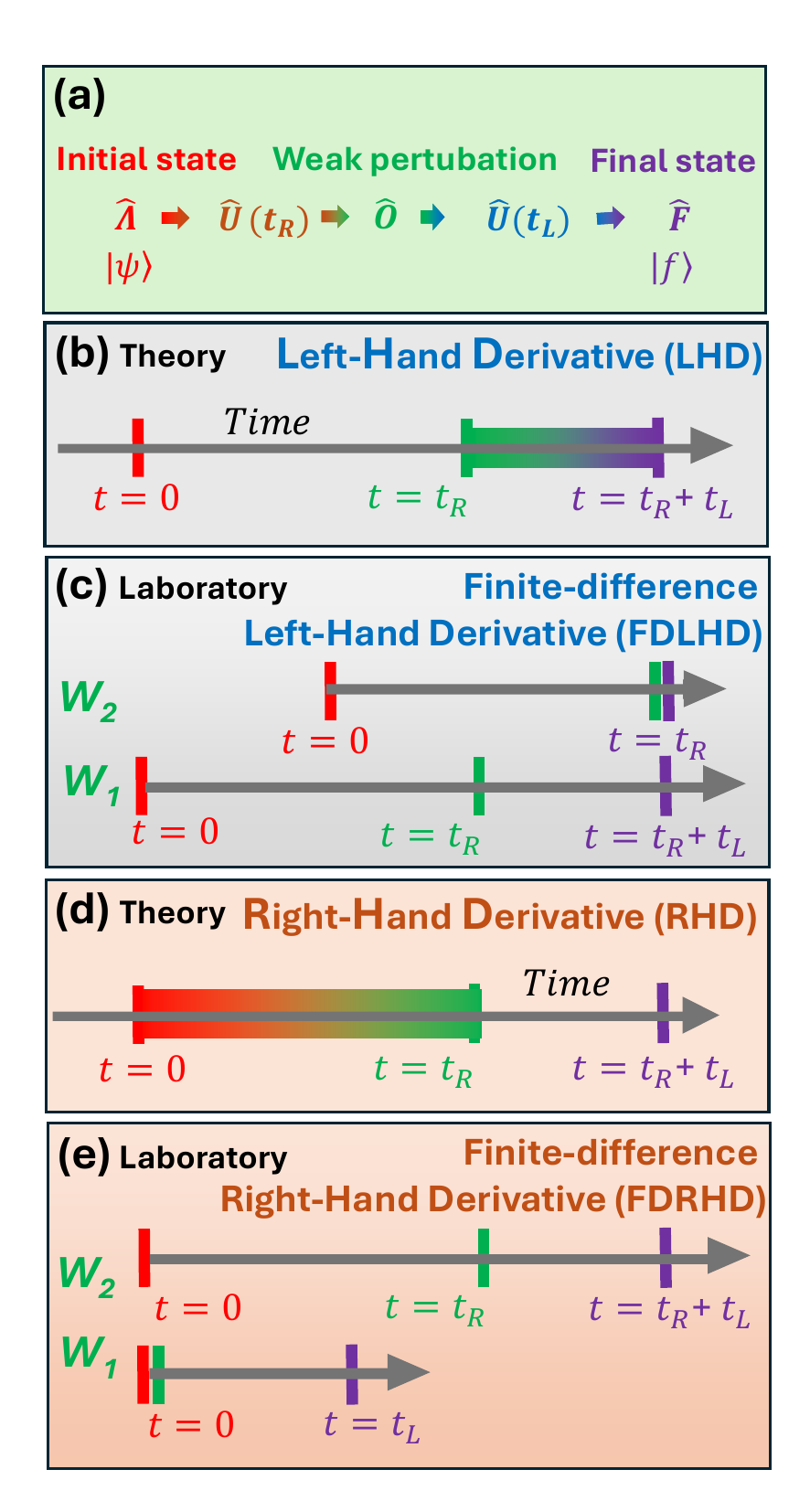}
\caption{ \textbf{(a):} The weak value in \eqref{ewv} involves five steps: preparation of $|\p\rangle$ (red) at the initial time $t=0$, unitary evolution $\opUt{t_R}$, weak perturbation linked to $\opO$ (green) at time $t=t_R$, unitary evolution $\opUt{t_L}$ and strong measurement linked to $\opF$ (violet) at the final time $t=t_R+t_L$. The weak value can be roughly understood as the property of the system during the intermediate green times. \textbf{(b):} The LHD computed when $t_L\to0$ while $t_R$ is keep constant.  \textbf{(c):} The FDLHD is evaluated at the final time and computed from the difference of the weak value $W_1$ and $W_2$, which are computed using two different values of $t_L$ (i.e., $t_L=0$ and $t_L>0$).  \textbf{(d):} The RHD computed when $t_R\to0$ while $t_L$ is keep constant.  \textbf{(e):} The FDRHD is evaluated at the initial time and computed from the difference of the weak value $W_1$ and $W_2$, which are computed using two different values of $t_R$ (i.e., $t_R>0$ and $t_R=0$). }
\label{f0}
\end{figure}

In Fig.~\ref{f0}(a), we have represented the five steps involved in the definition of a weak value. Considering that the initial state $|\p\rangle$ defined at $t=0$ evolves during a time interval $t_R$ between the preparation (pre-selection) and the (\textit{weak}) perturbation, modeled by $\opUt{t_R}:=\opU(t_R)$ , and a time interval $t_L$ between the (\textit{weak}) perturbation  and the \textit{strong} measurement (post-selection), modeled by $\opUt{t_L}:=\opU(t_L)$, the \textit{expectation} value of $\opO$ conditioned to the fixed eigenstate $\langle f|$ is given by: 
\begin{equation}
\BW\big(\opO,t_L,t_R\big||\evf\rangle,|\p\rangle\big):=\Re{ \frac{\langle \evf | \opUt{t_L} \opO  \opUt{t_R}|\p\rangle}{\langle \evf |\opUt{t_L} \opUt{t_R}|\p\rangle}}.
\label{ewv}
\end{equation}
The notation in the left-hand side of \eqref{ewv} indicates that the weak value $\BW$ is evaluated over the operator $\opO$ conditioned to the initial state $|\p\rangle$ and the final state $|\evf\rangle$ (i.e., a property measured through a weak value changes depending on the initial state $|\p\rangle$ and/or the final state $|\evf\rangle$). Notice that $\opUt{t_L}$ in \eqref{ewv}  is not complex-conjugated because it is a time evolution operator acting on the initial state $|\p\rangle$ (not on $|\evf\rangle$). In other words, $|\evf\rangle$ is an eigenstate of $\opF$ at the final time, not at the initial time.

To justify that the weak value provides \textit{simultaneous} information on the properties linked to the non-commuting operators $\opO$ and $\opF$ (or $\opphi$ and $\opO$), it is required $t_L\to 0$ ($t_R \to 0$). In the literature, the so-called weak value is typically defined as $\BW\big(\opO,t_L=0,t_R=0\big||\evf\rangle,|\p\rangle \big)$, and given by:
\begin{equation}
\BW\big(\opO\big||\evf\rangle,|\p\rangle \big):=\Re{\frac{\langle \evf | \opO |\p\rangle}{\langle \evf | \p\rangle}}.
\label{ewvnot}
\end{equation}
When $[\opF,\opO]=0$ or $[\opphi,\opO]$=0, we get $\BW\big(\opO\big||\evf\rangle,|\p\rangle \big)=\ego$ in \eqref{ewvnot},  where $\ego$ is the eigenstate of $\opO$. In general, when neither $[\opF,\opO]=0$ nor $[\opphi,\opO]$=0 is satisfied, the weak value can dramatically differ from an eigenvalue $\ego$ giving additional physical insight about the quantum system.  In the rest of the paper, we will use \eqref{ewv} or \eqref{ewvnot} depending on the need to make explicit the time dependence or not of the weak values.
   
The discussion of the gauge invariance of \eqref{ewv} requires first that the post-selected eigenvalue $\egf$ is gauge invariant (i.e., empirically observable). Using $|\evf^g \rangle=\opG|\evf \rangle$ as in \eqref{wg}, the eigenvalue equation in an arbitrary gauge $\opF^g |\evf^g\rangle=\egf^g |\evf^g \rangle$ can be written as $\opF^g \opG |\evf \rangle= \opG  \egf^g |\evf \rangle$. Thus, to have $\egf^g=\egf$  certifying that the eigenvalue is accessible in the laboratory, the new  necessary and sufficient  condition on the operator $\opF$ is
\begin{equation}
\textbf{C3:}\;\;\;\opF^g:=\opF(\opA^g,\opV^g)=\opG \opF(\opA, \opV) \opG^{\dagger}\nonumber,
\end{equation}
In agreement with \eqref{wg}, when \textbf{C3} is satisfied, the direct measurement of $|\evf^g\rangle$ is not accessible, but one can infer that the quantum system is $|\evf^g\rangle$ (in any gauge) by the measurement of the (gauge invariant) eigenvalue $\egf$. The role played by \textbf{C3} in identifying $|\egf\rangle$ is identical to the role played by \textbf{C2} in identifying $|\p\rangle$. 

Finally, the gauge invariance of \eqref{ewv} requires  ${\langle \evf^g | \opUtg{t_L} \opO^g  \opUtg{t_R}|\p^g\rangle}/{\langle \evf^g |\opUtg{t_L} \opUtg{t_R}|\p^g\rangle}={\langle \evf | \opUt{t_L} \opO  \opUt{t_R}|\p\rangle}/{\langle \evf|\opUt{t_L} \opUt{t_R}|\p\rangle}$. By construction, according to \eqref{wg} and \eqref{utg}, the denominator satisfies ${\langle \evf |\opUt{t_L} \opUt{t_R}|\p\rangle}={\langle \evf^g |\opUtg{t_L} \opUtg{t_R}|\p^g\rangle}$. Then, only the gauge invariance of the numerator, i.e. $\langle \evf | \opUt{t_L} \opO  \opUt{t_R}|\p\rangle=\langle \evf^g | \opUtg{t_L} \opO^g  \opUtg{t_R}|\p^g\rangle$, needs to be checked. By the same arguments done in discussing the gauge invariance of $\BE\big(\opO,t|\p\big)$, we conclude that the necessary and sufficient condition to ensure the gauge invariance of the weak value for any pre-selected $\langle \egf|$ and post-selected $|\p\rangle$ states is that \textbf{C1} is satisfied (together with \textbf{C2} and \textbf{C3}). Thus Table~\ref{table} can also be used to discern which operators produce gauge-invariant weak values. Practical examples of gauge-invariant (and gauge-dependent) weak values will be presented in Sec.~\ref{discussion1}.

To simplify the discussion, unless indicated, the weak value will be referred only to the real (not complex) value written in \eqref{ewv}. If needed, the weak value can be defined as a real plus an imaginary part, $\BW\big(\opO,t_L,t_R\big||\evf\rangle,|\p\rangle\big)+i \BW_i\big(\opO,t_L,t_R\big||\evf\rangle,|\p\rangle\big)$. The imaginary part is identified by a subscript $i$, indicating that the experimental measuring protocol to get the (imaginary) weak value $\BW_i$ is different from the experimental measuring protocol to get the (real) weak value $\BW$. See a detailed discussion in Ref. \cite{dressel2012,jozsa2007}. 

As mentioned in the introduction, the time derivative of a physical property frequently gives rise to another meaningful property. We are interested here in the time derivative of a weak value.  We also argue in the introduction that there is no consensus on either the ontological meaning of a weak value or the proper protocol to measure them. This lack of consensus will not affect the conclusions of this paper since we are only assuming the mathematical expression of the weak value given by \eqref{ewv} in all further developments.  The weak value in \eqref{ewv} involves three times $\{0,t_R,t_R+t_L\}$ separated by two time intervals $t_R$ and $t_L$. Thus, two different time derivatives can be envisioned.

\subsection{Left-hand derivative (LHD)} 
\label{left}

The left-hand derivative (LHD) is evaluated at the final post-selected time ($t=t_R+t_L$), and it captures time-variations of $\opUt{t_L}$ and $\opO$ (while $t_R$ is constant) as shown in Fig.~\ref{f0}(b). One can evaluate LHD as:
\begin{eqnarray}
\label{LHD}
&&\hspace*{-0.5cm}LHD:=\left.\frac{ \partial \BW\big(\opO,t_L,t_R\big||\evf\rangle,|\p\rangle \big)}{\partial t_L}\right|_{t_L=0} =\BW\big(\opC,0,t_R\big||\evf\rangle,|\p\rangle \big)\\
&&+\Re{\frac{i}{\hbar}\frac{\langle \evf  |\opO\opH \opUt{t_R}  |\p\rangle}{\langle \evf |\opUt{t_R}|\p\rangle}-\frac{i}{\hbar}\frac{\langle \evf  |\opO \opUt{t_R}|\p\rangle}{\langle \evf |\opUt{t_R} \p\rangle}\frac{\langle \evf |\opH \opUt{t_R} |\p\rangle}{\langle \evf |\opUt{t_R}|  \p\rangle}},\nonumber
\end{eqnarray}
with $\opC=\frac{i}{\hbar}[\opH,\opO]+\frac{d\opO}{dt}$ the operator mentioned in \eqref{eren2}. See the detailed demonstration in Appendix~\ref{ap4}.

Certainly, the shape of \eqref{LHD} is very different from the shape of \eqref{ewv}. Thus, \eqref{LHD} cannot be understood as a weak value of a property linked to $\opC$.  In addition, this expression presents a relevant impediment to be defined as a weak value because it is gauge dependent (i.e., empirically unobservable). The term $\BW\big(\opC,0,t_R\big||\evf\rangle,|\p\rangle \big)$ in \eqref{LHD} is gauge invariant, but the terms $\langle \evf  | \opO\opH \opUt{t_R} |\p\rangle$ and $\langle \evf |\opH \opUt{t_R} |\p\rangle$ are gauge dependent because they are an inner product of the Hamiltonian which is a gauge-dependent operator as shown in \eqref{hamig} and in Table~\ref{table}. 

In Appendix~\ref{ap4} we have shown that, in general, \eqref{LHD} becomes gauge dependent when arbitrary states $\langle \egf |$ are post-selected. However, the gauge dependence of the LHD of a weak value disappears, and it becomes empirically measurable, when we fix the post-selected state as an eigenstate of an operator satisfying 
\begin{equation}
\textbf{C4:}\;\;\;\;\;\;\;[\opO,\opF]=0\nonumber.
\end{equation}   
which means that $\opO |\egf \rangle =\ego |\egf \rangle$ . Thus, only some very particular states $\langle \egf |$ can be considered as post-selected states to compute LHD. Notice that  Appendix~\ref{ap4} shows that \textbf{C4} is a necessary and sufficient condition. Then, only the term $\BW\big(\opC,0,t_R\big||\evf\rangle,|\p\rangle \big)$ remains in \eqref{LHD} and the LHD of the weak value is also a (gauge invariant) weak value. In particular, under this condition \textbf{C4}, a (gauge invariant) Ehrenfest theorem \cite{ehrenfest1927} for weak values can be written as:
\begin{eqnarray}
\left.\frac{ \partial \BW\big(\opO,t_L,t_R\big||\evf\rangle,|\p\rangle \big)}{\partial t_L}\right|_{t_L =0}=\BW\big(\opC,0,t_R\big||\evf\rangle,|\p\rangle \big).
\label{dwvextra1}
\end{eqnarray}
 If $t_R=0$, the weak value provides simultaneous information on two operators.  The fact that $[\opO,\opF]=0$ in \textbf{C4} is compatible with $[\opC,\opF]\ne 0$ so that the right-hand side of \eqref{dwvextra1} gives simultaneous information on two non-commuting operators $\opC$ and $\opF$. Wiseman's result on the local velocity \cite{wiseman2007} can be interpreted as a result of \eqref{dwvextra1}, as seen in detail later in the development of  \eqref{weakvvelo} and in subsection~\ref{localvelocity}.

\subsection{Right-hand derivative (RHD)} 
\label{right}

The right-hand derivative (RHD) is evaluated at the initial pre-selection time ($t=0$), and it captures time-variations of $\opUt{t_L}$ and $\opO$ (while $t_L$ is constant) as shown in Fig.~\ref{f0}(d). One can evaluate RHD as:
\begin{eqnarray}
\label{RHD}
&&\hspace*{-0.5cm}RHD:=\left.\frac{ \partial \BW\big(\opO,t_L,t_R\big||\evf\rangle,|\p\rangle \big)}{\partial t_R}\right|_{ t_R =0}=\BW\big(\opC,t_L,0\big||\evf\rangle,|\p\rangle \big)\\
&&\hspace*{-0.5cm}+\Re{-\frac{i}{\hbar}\frac{\langle \evf  | \opUt{t_L} \opH \opO | \p\rangle}{\langle \evf | \opUt{t_L}|\p\rangle}+\frac{i}{\hbar}\frac{\langle \evf | \opUt{t_L} \opO|\p\rangle}{\langle \evf | \opUt{t_L}|\p\rangle}\frac{\langle \evf |\opUt{t_L}\opH |\p\rangle}{\langle \evf | \opUt{t_L}|\p\rangle}}.\nonumber
\end{eqnarray}
Again, in general, the right-hand side of \eqref{RHD} does not have the shape of a weak value,  and it is also gauge dependent. See the detailed demonstration in Appendix~\ref{ap5}.  The condition to achieve the gauge invariance of the RHD is 
\begin{equation}
\textbf{C5:}\;\;\;\;\;\;\;[\opO,\opphi]=0 \nonumber. 
\end{equation}
which means that $|\p\rangle$ is prepared at $t=0$ as an eigenstate of the operator $\opO$, i.e., $\opO |\p \rangle =\ego |\p \rangle$. Since the initial state is prepared by detecting an eigenstate $\egphi$ of the operator $\opphi$ (as discussed in \textbf{C2}), the operator $\opphi$ commutes with $\opO$ as indicated in \textbf{C5}. Notice that it is a necessary and sufficient condition for the gauge invariance of RHD, as shown in Appendix~\ref{ap5}. 

Finally, when \textbf{C5} is satisfied, similarly to what happens to LHD, \eqref{RHD} for evaluating the RHD can be rewritten as: 
\begin{eqnarray}
\left.\frac{ \partial \BW\big(\opO,t_L,t_R\big||\evf\rangle,|\p\rangle \big)}{\partial t_R}\right|_{ t_R =0}=\BW\big(\opC,t_L,0\big||\evf\rangle,|\p\rangle \big).
\label{dwvextra2}
\end{eqnarray}
Again, the fact that $[\opO,\opphi]=0$ in \textbf{C5} is compatible with $[\opC,\opphi]\ne 0$ so that the time-derivative of a weak value in \eqref{dwvextra2} provides simultaneous information on two non-commuting operators.When conditions \textbf{C4, C5} are not satisfied, LHD and RHD are different and gauge dependent.  

The sum of RHD in \eqref{RHD} and LHD in \eqref{LHD}, at $t_R=0$ and  $t_L=0$ respectively, gives:
\begin{eqnarray}
&&\hspace*{-0.5cm}\left.\frac{\partial \BW \big(\opO,0,t_R\big||\egf\rangle,|\p\rangle  \big)}{\partial t_R}\right|_{ t_R =0}+\left.\frac{\partial \BW \big(\opO,t_L,0\big||\egf\rangle,|\p\rangle  \big)}{\partial t_L}\right|_{t_L =0}\nonumber\\
&&\hspace*{-0.5cm}=\BW\big(\opC,0,0\big||\evf\rangle,|\p\rangle \big)+\frac{\langle \evf | \frac{\partial  \opO}{\partial t}| \p\rangle }{\langle \evf | \p\rangle }.
\label{twovalues}
\end{eqnarray}
Notice that \eqref{twovalues} introduces a strong relationship between LHD and RHD. The term $\BW\big(\opC,0,0\big||\evf\rangle,|\p\rangle \big)$ is gauge invariant. On the contrary, the term ${\langle \evf | \frac{\partial  \opO}{\partial t}| \p\rangle }/{\langle \evf | \p\rangle }$ can be gauge dependent for a time-dependent (Schrödinger picture) operator $\opO$. In this paper, we will only consider time-independent operators $\opO$. Then,  \eqref{twovalues} specifies that if LHD is gauge invariant, then RHD is also gauge invariant, and vice versa.

\section{Finite-difference approximations} 
\label{theorylaboratory}

In the laboratory, the time derivative of a weak value can be numerically computed after empirically evaluating two different weak values, $\BW_1$ and $\BW_2$, which differ in a time interval $\tau$. Then, the time derivative can be numerically approximated as the finite difference $(\BW_2 - \BW_1)/\tau$.  Again, we can distinguish between a finite-difference left-hand derivative (FDLHD) when $\tau=t_L$, as seen in Fig.~\ref{f0}(c), and a finite-difference right-hand derivative (FDRHD) when  $\tau=t_R$, as seen in Fig.~\ref{f0}(e).  
 
However, an apparent contradiction arises here because the finite-difference approximation of the time derivative of a weak value discussed in this section is, by construction, always gauge invariant (as long as the weak values $\BW_1$ and $\BW_2$ are gauge invariant by satisfying conditions \textbf{C1}, \textbf{C2}, and \textbf{C3}), while we have shown that LHD and RHD can be gauge dependent (even if conditions \textbf{C1}, \textbf{C2}, and \textbf{C3} are fully satisfied) when conditions \textbf{C4} and \textbf{C5} are not accomplished, as discussed in the previous section. 

A proper way to begin addressing the apparent contradiction between theoretical (LHD and RHD) and empirical (FDLHD and FDRHD) time derivatives of weak values is to recall that they are not the same. In quantum mechanics, theoretical properties and empirical properties do not always have a direct connection\footnote{For example, the theoretical projective measurement of the eigenvalue $x$ of the position operators implies that the theoretical wave function collapses into a Dirac delta function. But, this cannot be an exact explanation of what happens in a laboratory when the empirical position $x$ is measured because a delta function would contain an infinite amount of energy (which cannot be a realistic scenario in a laboratory)}. The theoretical time derivative is an operation on the Hilbert space, whereas the empirical time derivative is an operation on the empirical weak values of the laboratory. Therefore, if the theoretical LHD and RHD are not conceptually identical to the empirical FDLHD and FDRHD, the gauge behavior of the theoretical and empirical time-derivative of weak values can be different. This point is further elaborated in the following discussion.

\subsection{Finite-difference left-hand derivative (FDLHD)} 
\label{fd-left}

The finite-difference left-hand derivative (FDLHD) is evaluated at the final post-selected time $t=t_R+t_L$ as happens to LDH. The FDLHD requires the empirical evaluation of the weak value $\BW\big(\opO,t_L,t_R\big||\evf\rangle,|\p\rangle \big)$  as well as the weak value $ \BW\big(\opO,0,t_R\big||\evf\rangle,|\p\rangle \big)$, as depicted in Fig.~\ref{f0}(c). Then, the FDLHD for a small $t_L$ (while $t_R$ arbitrarily fixed) is:  
\begin{eqnarray}
FDLHD:=\frac{\BW\big(\opO,0,t_R\big||\evf\rangle,|\p\rangle \big)-\BW\big(\opO,t_L,t_R\big||\evf\rangle,|\p\rangle \big)}{t_L}.
  \label{FDLHD} 
\end{eqnarray}  
In both weak values, the pre-selected state $|\p(0)\rangle$ is prepared at $t=0$. However, the weak perturbation is produced at time $t=t_R$ in the first weak value and time $t=t_R+t_L$ in the second. Identically, the (same) post-selected state $\langle \egf|$ is considered at $t=t_R$ in the first weak value and at $t=t_R+t_L$ in the second. As indicated in Fig.~\ref{f0}(c), $\BW\big(\opO,0,t_R\big||\evf\rangle,|\p\rangle \big)$ with $t_L=0$ provides information on what \textit{is} the property of the quantum system (linked to $\opO$) at the time when the system \textit{is} post-selected in $\langle \egf|$. On the contrary, $\BW\big(\opO,t_L,t_R\big||\evf\rangle,|\p\rangle \big)$ provides information on what \textit{was} such property, not at the time when the post-selection  $\langle \egf|$ takes place, but at a time $t_L$ earlier. For example, we can expect $\BW\big(\opx,0,t_R\big||\evf\rangle,|\p\rangle \big) >\BW\big(\opx,t_L,t_R\big||\evf\rangle,|\p\rangle \big)$ for a quantum system with positive velocity. 

Let us notice that, apart from \eqref{FDLHD}, other expressions can also be invoked as finite-difference approximations to a first-order time derivative of \eqref{LHD}. The same conclusion deduced for FDLHD would apply to them. In any case, the FDLHD is obtained by a simple numerical manipulation of the empirical weak values. Thus, the fact that the two empirical weak values involved in \eqref{FDLHD} are gauge invariant (as long as conditions \textbf{C1}, \textbf{C2}, and \textbf{C3} are satisfied), implies that  FDLHD itself is gauge invariant too. In Appendix~\ref{ap4b}, it is shown that \eqref{FDLHD} becomes equal to \eqref{LHD} when condition \textbf{C4} is satisfied.

\begin{figure}
  \centering
\includegraphics[width=0.45\textwidth]{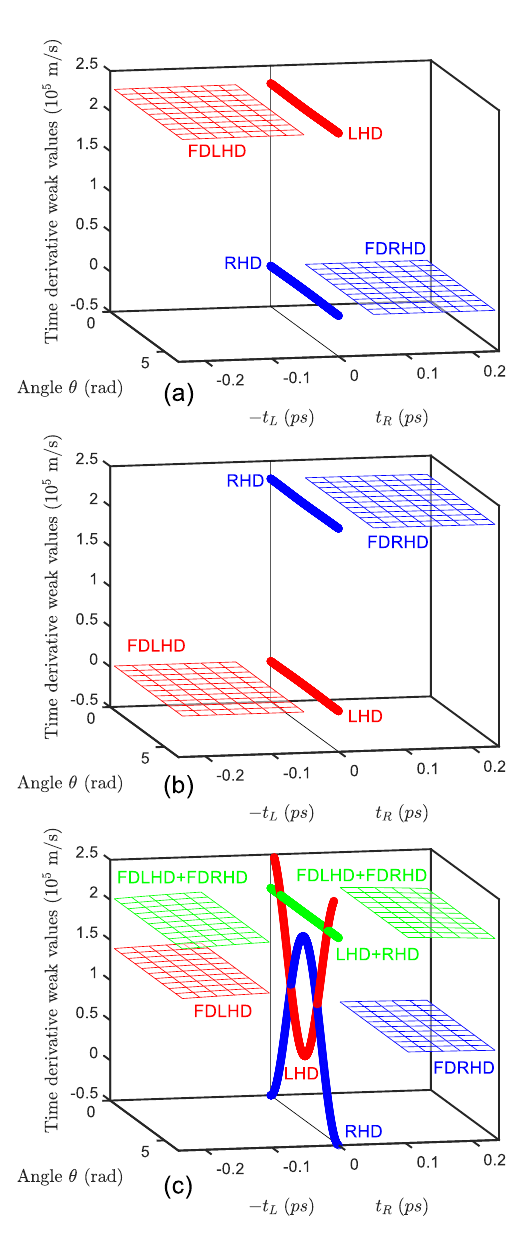}
\caption{ Time derivative of the weak value of the position for different pre- and post-selected states defined in Table~\ref{table1} at different angles \( \theta \) that parametrize different gauge functions $\gx$. At \( t_L=t_R=0 \), LHD (red) and RHD (blue); at negative \( t_L \), FDLHD (red); at positive \( t_R \), FDRHD (blue) are plotted. \textbf{(a):} $|\Phi_{1}\rangle$ for the pre-selected state and $\langle \Phi_{2}|$ for the post-selected state satisfying condition \textbf{C4}. \textbf{(b):} $|\Phi_{2}\rangle$ for the pre-selected state and $\langle \Phi_{1}|$ for the post-selected state satisfying condition \textbf{C5}. \textbf{(c):} $|\Phi_{3}\rangle$ for the pre-selected state and $\langle \Phi_{4}|$ for the post-selected state  without satisfying neither condition \textbf{C4} nor \textbf{C5}. The sum LHD+RHD (green) at $t_L=t_R=0$ and FDLHD+FDRHD (green) at positive $t_R$ and negative $t_L$ are also plotted.  }  
\label{f0bis}
\end{figure}

\subsection{Finite-difference right-hand derivative (FDRHD)} 
\label{fd-right}

The finite-difference right-hand derivative (FDRHD), as happens to RHD,  is evaluated at the initial  (pre-selection) time, i.e., $t=0$. The FDRHD requires evaluating the following two weak values as seen in Fig.~\ref{f0}(e):
\begin{eqnarray}
FDRHD:=\frac{ \BW\big(\opO,t_L,t_R\big||\evf\rangle,|\p\rangle \big)-\BW\big(\opO,t_L,0\big||\evf\rangle,|\p\rangle \big)}{t_R}.
\label{FDRHD} 
\end{eqnarray}  
Again, the pre-selected state $|\p(0)\rangle$ is prepared at $t=0$. But, the same post-selected state $\langle \egf|$ is evaluated at $t=t_L+t_R$ in the first weak value and at $t=t_L$ in the second.  Now, as indicated in Fig.~\ref{f0}(e), the weak value $\BW\big(\opO,t_L,0\big||\evf\rangle,|\p\rangle \big)$ with $t_R=0$ provides information on what \textit{is} the property of the quantum system (linked to $\opO$) at the time the system \textit{is} pre-selected in $|\p \rangle$, while $\BW\big(\opO,t_L,t_R\big||\evf\rangle,|\p\rangle \big)$ provides such information at a time interval $t_R$ after the system \textit{was} pre-selected in $|\p \rangle$. For example, we can expect $\BW\big(\opx,t_L,t_R\big||\evf\rangle,|\p\rangle \big) >\BW\big(\opx,t_L,0\big||\evf\rangle,|\p\rangle \big)$ for a quantum system with positive velocity. Again, both weak values are gauge invariant and observable in the laboratory (as long as conditions \textbf{C1}, \textbf{C2}, and \textbf{C3} are satisfied). In Appendix~\ref{ap5b}, it is shown that FDRHD becomes equal to RHD when condition \textbf{C5} is satisfied.  

As mentioned in the introduction to this section, when conditions \textbf{C4} and \textbf{C5} are not satisfied, the LHD and RHD are gauge dependent, whereas the FDLHD and FDRHD remain gauge independent. The physical reason, as explained earlier, is that the former involve operations in the Hilbert space, while the latter are defined in the laboratory frame. The mathematical distinction can be identified in the elements involved in their respective inner products.  

By construction, the inner products required for evaluating the FDLHD involve the same inner products needed to evaluate a weak value in \eqref{ewv}. Thus, the relevant elements are the pre- and post-selected states, the $\opO$ operator, and the time-evolution operators $\opUt{t_L}$ and $\opUt{t_R}$. In contrast, the evaluation of the inner products in the LHD in \eqref{LHD} involves the same states and $\opO$ operator but, apart from the time-evolution operators, it also involves explicitly the Hamiltonian operator $\opH$. This Hamiltonian operator depends on $\partial \gx/\partial t$, as shown in \eqref{hamig}. It is this dependence on $\partial \gx/\partial t$ that leads to the gauge-dependence of the inner products involving the Hamiltonian \footnote{The reader can argue that the time-evolution operators $\opUt{t_L}$ and $\opUt{t_R}$ in FDLHD also depend on $\opH$—which itself includes $\partial \gx/\partial t$— as seen in \eqref{utg}. However, the time integral in \eqref{utg} transforms $\partial \gx/\partial t$ into a gauge function $\gx$ (without its time derivative). The final dependence of $\opUt{t_L}$ and $\opUt{t_R}$ on $g(\egx,t_L)$ or $g(\egx,t_R)$ in the inner products of FDLHD is unproblematic, as demonstrated in Appendix~F. The same happens to FDRHD  as shown in  Appendix~G.} (for the same reason, the  Hamiltonian has gauge-dependent expectation values as seen in Table~\ref{table}).  In simple terms, the FDLHD in \eqref{FDLHD} can be obtained in the laboratory when $t_L$ is small (but not in the limit $t_L\to0$). Thus, it always involves $\opUt{t_L}$ and $\opUt{t_R}$ with finite $t_L$ and $t_R$, whose inner products remain gauge invariant. However, defining the LHD in \eqref{LHD} requires taking the limit $t_L\to0$, which implies dealing with the inner products that explicitly depend on $\opH$, making them gauge dependent when condition \textbf{C4} is not satisfied. The same reasoning applies to the FDRHD and RHD.   

We illustrate these important results with the following numerical simulations.

\subsection{Numerical comparison}
\label{numericalcomputation}

We consider a particle in free space and we evaluate the weak value of the position and its time derivative (which in most of the case, not all, will meant the velocity).  The expressions LHD, RHD, FDLHD, and FDRHD are evaluated using the different Gaussian wave packets listed in Table~\ref{table1} as pre- and post-selected states. The time derivative of the weak values are computed numerically as explained below.

\begin {table}
\begin{tabular}{c|c|c|c|c|c|c|}
\hline
 Wave packet  & $E_c$  & $k_c$ & $v_c$ & $x_c$ & $\sigma_x$ & Observation \\
     &  meV  & nm$^{-1}$ & $10^5$ m/s & nm & nm &    \\
\hline
\hline
$\langle x|\Phi_{1}\rangle$  &  10 &  0.132 &  2.28 & 400 & 84  &    \\
\hline
$\langle x|\Phi_{2}\rangle$  &  5 &    0.097 &  1.61 & 400 & 5 & $\approx \delta(x-x_c)$ \\
\hline
$\langle x|\Phi_{3}\rangle$ &  10 &  0.132 &  2.28 & 400 & 60  & \\
\hline
$\langle x|\Phi_{4}\rangle$  &  5 &     0.097 &  1.61 & 400 & 40  & \\
\hline
$\langle x|\Phi_{5}\rangle$ &  50 &  0.29 &  5.12 & 200 & 127 & $\approx e^{i k_c x}$\\
\hline
\hline
\end{tabular}
\caption{\label{table1} Different Gaussian wave packets $\langle x|\Phi_{i}\rangle=\psi_{E_c,x_c,\sigma_x}(x,t_I)$ used in the numerical simulations in this work and defined in \eqref{apnum5} in terms of the central energy $E_c$, central wave vector $k_c=\sqrt{2m^*E/\hbar^2}$, central velocity $v_c=\frac{\hbar k_c}{m^*}, $ central position $x_c$, and spatial dispersion $\sigma_x$ at the wave packet preparation time.  } 

\end{table}

\subsubsection{When \textbf{C4} is satisfied}

In Fig.~\ref{f0bis}(a), the pre-selected state is a Gaussian wave packet with a normal (not too short and not too large) spatial dispersion $|\p \rangle=|\Phi_{1}\rangle$, while the post-selected state is the Gaussian wave packet with a very short spatial dispersion $\langle \Phi_{2}|$, as seen in Table~\ref{table1}. The last Gaussian wave packet is so narrow that it mimics a position eigenstate $\langle \egf|=\langle \Phi_{2}|\approx \langle x_c|$. The theoretical LHD in \eqref{LHD} for the position operator $\opO=\opx$, when $t_R=0$,  can be written as:
\begin{eqnarray}
\label{numeric1}
&&\left.\frac{ \partial \BW\big(\opx,t_L,0\big||x_c \rangle,|\Phi_{1}\rangle \big)}{\partial t_L}\right|_{ t_L =0}=\BW\big(\opv,0,0\big||x_c\rangle,|\Phi_{1}\rangle \big)\nonumber\\
&&+\Re{\frac{i}{\hbar}\frac{\langle x_c  |\opx \opH | \Phi_{1}\rangle}{\langle x_c |\Phi_{1}\rangle}-\frac{i}{\hbar}\frac{\langle x_c | \opx|\Phi_{1} \rangle}{\langle x_c | \Phi_{1}\rangle}\frac{\langle x_c |\opH |\Phi_{1}\rangle}{\langle x_c |\Phi_{1}\rangle}}.
\end{eqnarray}
The numerical evaluation of inner products in \eqref{numeric1} just requires multiplying $|\Phi_{1}\rangle$ by the operators $\opH$ and/or $\opx$ and doing the appropriate inner  products with $\langle \Phi_{2}|\approx \langle x_c|$. The results are plotted in a solid red line in Fig.~\ref{f0bis}(a) at $t_R=t_L=0$. Since \textbf{C4} is satisfied, LHD is given by $\BW\big(\opv,0,0\big||v_o\rangle,|\p\rangle \big)=\Re{\frac{\langle x_c |\opv |\Phi_{1}\rangle}{\langle x_c |\Phi_{1}\rangle}}=\vB{|\Phi_{1}\rangle}(x_c,0)\approx {\hbar k_{c,1}}/{m^*}=2.28\cdot 10^5$ m/s which corresponds to the Bohmian velocity of $\langle x_c|\Phi_{1}(0)\rangle$. The formal development leading to the definition of the Bohmian velocity $\vB{|\Phi_{1}\rangle}$ will be elaborated in detail later in the development of \eqref{weakvvelo} (see also \eqref{extraeq}).

We now evaluate FDLHD following \eqref{FDLHD} for different $t_L$ (and $t_R=0$) as:
\begin{eqnarray}
&&\frac{ \BW\big(\opx,0,0\big||x_c\rangle,|\Phi_{1}\rangle \big)-\BW\big(\opx,t_L,0\big||x_c\rangle,|\Phi_{1}\rangle \big)}{t_L} \nonumber\\
&&\approx {  (x_c - x_L)}/{t_L},
\label{numeric2}
\end{eqnarray}
where we have defined $x_c=\BW\big(\opx,0,0\big||x_c\rangle,|\Phi_{1}\rangle \big)=\Re{\frac{\langle x_c |\opx |\Phi_{1}\rangle}{\langle x_c |\Phi_{1}\rangle}}$ and $x_L := \BW\big(\opx,t_L,0\big||x_c\rangle,|\Phi_{1}\rangle \big)=\Re{\frac{\langle x_c |\opUt{t_L}\opx |\Phi_{1}\rangle}{\langle x_c |\opUt{t_L}|\Phi_{1}\rangle}}$. The numerical evaluation of the last weak values requires preparing $|\Phi_{1}(0)\rangle$  at the preparation time $t_{1}=0$ and letting it evolve unitarily during a time interval $t_L$ following the algorithm described in  \eqref{apnum4}. The inner product of the evolved $|\Phi_{1}(t_L)\rangle$ with the not-evolved $\langle \Phi_{2}(t_L)|\approx \langle x_c|$, defined at its preparation time $t_{2}=t_L$, provides the required denominator $\langle x_c |\opUt{t_L}|\Phi_{1}\rangle$. The numerator i.e., $ \langle x_c| \opUt{t_L} \opx |\Phi_{1}(0)\rangle$ requires the time evolution following the algorithm described in \eqref{apnum4}, during the time interval  $t_L$,  of a state defined by the product of $|\Phi_{1}(0)\rangle$  by $\opx$. The final results in \eqref{numeric2} are plotted in red in Fig.~\ref{f0bis}(a) for different values of $t_L$.  The numerical values of results LHD and FDLHD are roughly identical, as justified in Appendix~\ref{ap4b} when \textbf{C4} is satisfied.  

The values of RHD and FDRHD are also computed numerically with similar procedures, giving both values close to zero as plotted in blue in Fig.~\ref{f0bis}(a). Such results can be easily justified through condition \eqref{twovalues} rewritten here as LHD+RHD$=\Re{ \frac{\langle x_c | \frac{i}{\hbar}[\opH,\opx] |\Phi_{1}\rangle }{\langle x_c |\Phi_{1}\rangle }}=\vB{|\Phi_{1}\rangle}(x_c,0)$,  which fixes RHD to zero because we have already shown that LHD gives $\vB{|\Phi_{1}\rangle}(x_c,0)$. The value FDRHD is also zero because it satisfies a similar relationship with FDLHD as shown in \eqref{aptwovalues} in Appendix~\ref{ap5b}. 

Notice that the results in Fig.~\ref{f0bis}(a) have been computed for different angles $\theta$. These angles $\theta$ parametrize different gauge functions $g(x,t)$ as descibed in \eqref{apnum7gauge}. The initial pre- and post-selected states are the ones mentioned above multiplied by the gauge function at their respective initial time, $e^{i\frac{q}{\hbar}g(x,0)}\langle x|\Phi_{1}(0)\rangle$ and $e^{-i\frac{q}{\hbar}g(x,t_L)} \langle \Phi_{2}(t_L) |x\rangle$, as described in \eqref{apnum8} and their time evolution is given now by the algorithm described in \eqref{apnum15}. Here, all the final outputs become gauge independent (i.e, they do not depend on the selected gauge function parametrized by the angle $\theta$). The FDLHD and FDRHD are gauge independent by construction, while LHD and RHD are gauge independent here because condition \textbf{C4} is satisfied.  

\subsubsection{When \textbf{C5} is satisfied}

In Fig.~\ref{f0bis}(b), we plot similar results for a scenario where $|\p\rangle$ and $\langle \egf|$ are interchanged. The pre-selected state is $|\p \rangle =|\Phi_{2}\rangle \approx | x_c \rangle$, while the post-selected states is $\langle \egf |=\langle \Phi_{1}|$.  The theoretical RHD in \eqref{RHD}, when $t_L=0$,  can be written as:
\begin{eqnarray}
\label{numeric3}
&&\hspace*{-0.5cm}\left.\frac{ \partial \BW\big(\opx,0,t_R\big||\Phi_{1} \rangle,|x_c\rangle \big)}{\partial t_R}\right|_{ t_R =0}=\BW\big(\opv,0,0\big|| \Phi_{1}\rangle,|x_c\rangle \big)\nonumber\\
&&\hspace*{-0.5cm}+\Re{-\frac{i}{\hbar}\frac{\langle \Phi_{1}  |\opH \opx | x_c \rangle}{\langle \Phi_{1} | x_c\rangle}+\frac{i}{\hbar}\frac{\langle \Phi_{1} | \opx|x_c\rangle}{\langle \Phi_{1} | x_c\rangle}\frac{\langle \Phi_{1} |\opH |x_c\rangle}{\langle \egf |x_c\rangle}}.
\end{eqnarray}
Since \textbf{C5} is satisfied RHD is given by $\BW\big(\opv,0,0\big||\Phi_{1}\rangle,|x_c\rangle \big)=\Re{\frac{\langle \Phi_{1} |\opv |x_c\rangle}{\langle \Phi_{1} |x_c\rangle}}=\vB{|\Phi_{1}\rangle}(x_c,0)\approx \frac{\hbar k_{c,1}}{m^*} \approx 2.28 \cdot 10^5$ m/s which corresponds again to the Bohmian velocity of the $\langle x_c|\Phi_{1}(0)\rangle$ computed in Fig.~\ref{f0bis}(a). This last result is a consequence of the hermiticity of the $\opv$, which implies that $\Re{\frac{\langle \Phi_{1} |\opv |x_c\rangle}{\langle \Phi_{1} |x_c\rangle}}=\Re{\frac{\langle x_c |\opv |\Phi_{1} \rangle}{\langle x_c |\Phi_{1}\rangle}}$.

We also evaluate FDRHD in the laboratory for different $t_R$ (and $t_L=0$) following \eqref{FDRHD}:
\begin{eqnarray}
\frac{ \BW\big(\opx,0,t_R\big||\Phi_{1} \rangle,|x_c\rangle \big)-\BW\big(\opx,0,0\big|| \Phi_{1}\rangle,|x_c\rangle \big)}{t_R} \nonumber\\
\approx{  (x_R- x_c) }/{t_R},
\label{numeric4}
\end{eqnarray}
where we have defined $x_c=\BW\big(\opx,0,0\big||x_c\rangle,|\Phi_{1}\rangle \big)=\Re{\frac{\langle x_c |\opx |\Phi_{1}\rangle}{\langle x_c |\Phi_{1}\rangle}}$ and $x_R =: \BW\big(\opx,0,t_R\big||x_c\rangle,|\Phi_{1}\rangle \big)=\Re{\frac{\langle x_c |\opx \opUt{t_R}|\Phi_{1}\rangle}{\langle x_c |\opUt{t_R}|\Phi_{1} \rangle}}$. The numerical evaluation of the second weak value implies preparing $|\Phi_{2}(0)\rangle \approx | x_c \rangle$ at the preparation time $t_{2}=0$ and allowing the system to evolve unitarily during a time interval $t_R$ following the evolution described in  \eqref{apnum4}. The inner products are done in a similar way as described for Fig.~\ref{f0bis}(a), involving now $\langle \Phi_{1}(t_R)|$ at the preparation time $t_{1}=t_R$. In Fig.~\ref{f0bis}(b), in blue, we show that the results RHD and FDRHD are roughly identical. The condition \eqref{twovalues} and \eqref{aptwovalues} also fix the values of LHD and FDLHD to roughly zero. The reasons for the gauge invariance (independence of $\theta$) are equivalent to the ones explained for Fig.~\ref{f0bis}(a). Now, the accomplishment of condition \textbf{C5} is involved.

\subsubsection{When \textbf{C4} and  \textbf{C5} are not satisfied}

The next analysis provided in Fig.~\ref{f0bis}(c) explores a different scenario where neither the pre-selected state $|\Phi_{3}\rangle$ nor the post-selected state $\langle \Phi_{4}|$ can be considered position eigenstates. Thus, since neither condition \textbf{C4} nor \textbf{C5} is  satisfied, the LHD computed from \eqref{numeric1} and RHD computed from \eqref{numeric3} are gauge dependent as seen for the red and blue curves at $t_L=t_R=0$ (i.e., their values depend on the angle $\theta$ which parametrizes different gauge functions $g(x,t)$ as described in \eqref{apnum7gauge}). On the contrary, by construction, the FDLHD computed from \eqref{numeric2} and FDRHD computed from \eqref{numeric4} remain gauge invariant as described in Appendixes~\ref{ap4b} and \ref{ap5b} but their values coincide neither with the Bohmian velocity of $|\Phi_{3}\rangle$ nor  $\langle \Phi_{4}|$ written in Table~\ref{table1}. 

In green, we have plotted LHD+RHD at $t_R=t_L=0$ and FDLHD+FDRHD for different values of $t_L$ and $t_R$ and $\theta$. Both sums become equal and gauge invariant as indicated in \eqref{twovalues} and \eqref{aptwovalues} (for the time-independent operator $\opx$), but the value of the sum does not coincide with the sum of the Bohmian velocities. In summary, when neither condition \textbf{C4} nor \textbf{C5} are satisfied, LHD and RHD cannot be obtained in the laboratory. On the contrary, FDLHD and FDRHD can be empirically evaluated in the laboratory, but it is not clear which physical insight can be attributed to them.

\begin{center}
\captionsetup{type=table}
\includegraphics[width=0.7\textwidth]{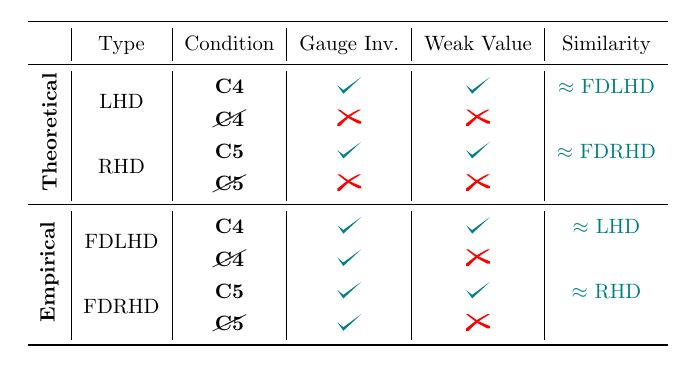} 
\caption{Theoretical (LHD and RHD) and empirical (FDLHD and FDRHD) time-derivative of weak values satisfying (\color{teal}\checkmarkbis\color{black}) or not (\color{red}\crossmark\color{black}) being invariant under gauge transformations (column \textbf{Gauge. Inv}) and having the mathematical shape of a wave value (column \textbf{Weak Value}) as a function of the accomplishment or not of conditions \textbf{C4} or \textbf{C5}. The scenarios when LHD $\approx$ FDLHD and RHD $\approx$ FDRHD are indicated (column \textbf{Similarity}).}
\label{summary}
\end{center}

Finally, it is interesting to revisit expressions \eqref{numeric2} to acquire an intuitive motivation for \textbf{C4}. It seems \textit{natural} to define the time derivative of a weak values as the difference between two weak values: one weak value $x_c$ evaluated without time evolution, i.e., $\opUt{t_L}=\unit$, and the other weak value $x_L$ with a time evolution during the time interval $t_L$. It seems obvious to define the post-selected position $x_c$, when $\opUt{t_L}=\unit$ in terms of weak values as $x_c={\langle x_c  |\opx | \p \rangle}/{\langle x_c |\p \rangle}$ with $\opx |x_c\rangle=x_c|x_c\rangle$. Thus, the weak value $x_L$ with arbitrary $\opUt{t_L}$ has to be defined as $x_L={\langle x_c  |\opUt{t_L}\opx | \p \rangle}/{\langle x_c |\opUt{t_L}|\p \rangle}$, and we conclude that \textbf{C4} becomes a quite \textit{natural} requirement for the time derivative of a weak value because the condition $\opx |x_c\rangle=x_c|x_c\rangle$, i.e. $[\opO,\opF]=0$ for $\opO=\opx$ and $\opF=\opx$, is required at the origin of this argumentation.  This is exactly what happens for the LHD in \eqref{LHD} when \textbf{C4} is satisfied. An identical conclusion occurs for \textbf{C5} from \eqref{numeric4}.  All these results regarding the gauge condition and the shape of the theorethical, LHD and RHD, and empirical,  FDLHD and FDRHD, weak values are summarized in Table \ref{summary}.


\section{Examples on Gauge invariance of weak values}
\label{discussion1}

After the theoretical development and finite-difference approximations done in Secs.~\ref{theory} and \ref{theorylaboratory}, here, we show several weak values often mentioned in the literature. In particular, we discuss whether or not they satisfy the gauge invariance implicit in properties \textbf{C1, C3}. To simplify the discussion, we will assume that the condition \textbf{C2} in the preparation of the initial state $|\pg \rangle$ is always satisfied. Since this section does not deal with time-derivative yet, we will use expression \eqref{ewvnot} for the evaluation of the weak values, rather than \eqref{ewv}. 

To properly understand the results in this section, first, we need to clarify what we mean when we say, for example, that the canonical momentum on Table~\ref{table} is not empirically observable in the laboratory because it is gauge dependent. It is well-known that the canonical momentum $\opp^g := \opp(\opAg,\opVg) = \opp \neq \opG \opp \opG^{\dagger}$ does not satisfy condition \textbf{C1}. Thus, strictly speaking, it cannot be empirically measured in the laboratory either from an expectation value or from a weak value in a direct way. However, in many experimental works, the measurement of the ``momentum'' is inferred from the measurement of another property that is gauge invariant. For example, by measuring the position of the quantum system on the screen \cite{freericks2023}  and knowing the initial and final time of the experiment. As we will discuss below, this measuring procedure is a measurement of the velocity (or the velocity multiplied by the mass) rather than a direct measurement of the canonical momentum.\footnote{We emphasize that the above discussions and all the results in this paper about non-measurable gauge-dependent results do not imply ontological consequences for the canonical momentum or other gauge-dependent properties. }

In other words, the fact that the electromagnetic potentials are not gauge invariant does not imply that they are not physically relevant~\cite{aharonov1959}, but only that they cannot be directly measured in the laboratory.   

\subsection{Weak value of the canonical momentum post-selected in position}
 
As discussed above, the canonical momentum operator $\opp^g:=\opp(\opAg,\opVg)=\opp\neq\opG \opp \opG^{\dagger}$ does not satisfy condition \textbf{C1} and cannot be measured through weak values. Let us confirm this point by writing the theoretical expression corresponding to the weak measurement of the canonical momentum plus a subsequent strong measurement of the position operator $\opx^g:=\opx(\opAg,\opVg)=\opx=\opG \opx \opG^{\dagger}$. Such weak value would be given by $\BW\big(\opp\big||\egx\rangle,|\p\rangle \big)$ which, when written in any gauge, becomes
\begin{equation}
\Re{\frac{\langle \egx^g |\opp |\pg \rangle}{\langle \egx^g|\pg \rangle}}=\Re{-i\hbar \frac{\grad \pg }{\pg }}=\grad S-q \grad \g.
\label{pvelo}
\end{equation}
where $[\opG,\opx]=0$ and $|\pg\rangle=\opG|\p\rangle$. We use $\px=R(\textbf{x},t)e^{iS(\textbf{x},t)/\hbar}$ in polar form to better elaborate ${\grad \pg }/{\pg }$.  The explicit dependence of \eqref{pvelo} on $\grad \g$ confirms that this weak value cannot be observed in the laboratory.

In the Coulomb gauge representation ($\g=0$ and $\grad \cdot \A=0$), in the absence of a magnetic field $\A=0$, the velocity and canonical momentum operators are identical $m^*\opv=\opp$. This coincidence in one particular gauge does not mean that the velocity and momentum operators are the same in general, i.e.,  $m^*\opv^g\neq\opp^g$. In the same way, the knowledge of $\A=0$ in the Coulomb gauge does not mean that one can assume $\Ag=0$ in another gauge.

\subsection{Weak value of the velocity post-selected in position}
 
Contrary to the canonical momentum, the velocity operator $\opv^g$ defined as 
\begin{eqnarray}
\opv^g&:=\opv(\opAg,\opVg)=(\opp^g-q\opA^g)/m^*=\opv=\opG \opv \opG^{\dagger},\nonumber
\label{velo}
\end{eqnarray}
is gauge invariant and satisfies \textbf{C1} for $\opO=\opv$. Hence, the weak measurement of the velocity post-selected in position gives the empirically observable weak value $\BW\big(\opv\big||\egx\rangle,|\p\rangle\big)$ defined as
\begin{eqnarray}
\Re{\frac{\langle \egx | \opv^g |\pg \rangle}{\langle \egx|\pg \rangle}}&&=\frac{1}{m^*}\Re{-i\hbar \frac{\grad \pg }{\pg } -q \textbf{A}^g}\nonumber\\
&&=\frac{\grad S-\cancel{q \grad G}-q\textbf{A} +\cancel{q\grad G}}{m^*}=:\vB{|\p\rangle}.
\label{weakvvelo}
\end{eqnarray}
The weak value gives the so-called Bohmian velocity \cite{wiseman2007,durr2009,destefani2023},  $\vB{|\p\rangle}=(\grad S-q\textbf{A})/m^*$, which is gauge invariant because it satisfies \textbf{C1} for $\opO=\opv$ and \textbf{C3} for $\opF=\opx$. See an alternative demonstration in Appendix~\ref{ap6} and also \cite{deotto1998}.  The superscript ${|\p\rangle}$ in $\vB{|\p\rangle}$ just indicates that the Bohmian velocity is a functional of the wave function $\p$.

As discussed previously (see  \cite{dressel2012,jozsa2007}), another weak value can be designed to give the imaginary part of \eqref{weakvvelo}, $\Im{{\langle \egx | \opv^g |\pg \rangle}/{\langle \egx|\pg \rangle}}$, which is called the ``osmotic'' velocity $\vO{|\p\rangle}$  \cite{destefani2023,hiley2012,nelson1966,bohm1989}  and given by $\vO{|\p\rangle}:=-\Im{{\langle \egx | \opv^g |\pg \rangle}/{\langle \egx|\pg \rangle}}=\frac{\hbar}{m^*}\frac{\grad R^g}{R^g}$, which is also gauge invariant because $R^g=R$.  Notice that there is some ambiguity regarding the sign in the definition of osmotic velocity in the literature \cite{destefani2023,hiley2012,nelson1966,bohm1989}.

The consistency in interpreting $\BW\big(\opv\big||\egx\rangle,|\p\rangle\big)$ as a velocity is provided by the LHD in \eqref{LHD} when $\opO=\opx$ and $|\evf\rangle=|\egx\rangle$, as will be seen in  \eqref{extraeq} in next section.  

\subsection{Weak value of the position projector post-selected in the canonical momentum }

Let us now analyze the weak value corresponding to the position projector $|\egx^g \rangle \langle \egx^g|=\opG |\egx\rangle \langle \egx|\opG^{\dagger}=|\egx \rangle \langle \egx|$ post-selected by the  canonical momentum operator $\opp$ defined as:
\begin{equation}
 \BW\big(|\egx\rangle \langle \egx|\big||\egp\rangle,|\p\rangle\big)=\frac{{\langle \egp^g|\egx \rangle \langle \egx |\pg \rangle}}{{\langle \egp^g|\pg \rangle}}.\nonumber
 \label{pxx}
\end{equation}
It is argued in the literature \cite{lundeen2011,zhu2021} that such a weak value, together with the corresponding conditional ensemble for the imaginary part \cite{dressel2012,jozsa2007}, is proportional to the wave function $\langle \egx |\pg \rangle$ when post-selected at zero momentum. Despite the fact that   this weak value satisfies \textbf{C1} for $\opO=|\egx\rangle\langle \egx|$,  the selection $\opF=\opp^g$, to define $\langle \egp^g|$, does not satisfy \textbf{C3} because the canonical momentum is gauge dependent. The gauge-dependent eigenstate $\langle \egx|\egp^g\rangle=e^{i(q\gx+\egp \egx)/\hbar}$ has a gauge-dependent eigenvalue $-i\hbar \grad \langle \egx|\egp^g\rangle=(\gx+\egp)\langle \egx|\egp^g\rangle$. Thus, this weak value of the canonical momentum cannot be obtained in the laboratory.

\subsection{Weak value of the position projector post-selected in the velocity }

It is true, however, that the non-measurability of the canonical momentum described above can be avoided by substituting $\opp$ by $\opv$.  Then, $\BW\big(\opx\big||\egv\rangle,|\p\rangle\big)$, is given by
\begin{eqnarray}
\frac{\langle \egv^g|\egx \rangle \langle \egx |\pg \rangle}{\langle \egv^g|\pg \rangle} =\frac{R_{\egv} R_{\p} e^{i\frac{S_{\egv}-S_{\p}}{\hbar}}}{\langle \egv|\p\rangle},
\label{wfunction2}
\end{eqnarray}
where $\langle \egv^g|\egx \rangle =R_{\egv} e^{i (S_{\egv}+qG)/\hbar}$ and $ \langle \egx |\pg \rangle = R_{\p} e^{i (S_{\p}+qG)/\hbar}$. Clearly, this weak value satisfies \textbf{C1} for $\opO=|\egx\rangle\langle \egx|$ and \textbf{C3} for $\opF=\opv$. Hence, the (phase of the) weak value $\BW\big(\opx\big||\egv\rangle,|\p\rangle\big)$ is now gauge invariant, as seen here:  $S_{\egv}^g-S_{\p}^g=S_{\egv}-\cancel{qG}-S_{\p}+\cancel{qG}=S_{\egv}-S_{\p}$. 

However, arguing that \eqref{wfunction2} measures the (phase of the)  Hilbert space representation of the  wave function $\langle \egx |\pg \rangle$ is not rigorous, because all wave functions are gauge dependent and unmeasurable, as seen in \eqref{wg}. The argument in the literature is based on assuming that for a magnetic field equal to zero, in the Coulomb gauge, $m\opv=\opp$  with $\langle \egv|\egx \rangle \propto e^{i(\egp \egx)/\hbar}$ and $ \langle \egx |\p\rangle = R_{\p} e^{i S_{\p}/\hbar}$. Then,  \eqref{wfunction2} gives $\BW\big(|\egx\rangle \langle \egx |\big||\egv\rangle,|\p\rangle\big)_{\egv=0} \propto \langle \egx |\p\rangle$ \cite{lundeen2011}.  While this protocol, which effectively gives the wave function defined in the Coulomb gauge, can indeed be very useful for developing new quantum technologies, there is no reason  to affirm that the Coulomb-gauge representation of the wave function is the \textit{true} wave function of the system, in the same way as the Coulomb gauge representation of the electromagnetic potentials are not the \textit{true} potentials. At the fundamental level, there should be no preferred gauge, and hence, the wave function and the potentials are unmeasurable, as indicated in \eqref{wg} and \eqref{vg}, respectively.


\section{Examples on the time derivative of weak values}
\label{discussion2}

Here, we discuss three examples of the time derivative of weak values, dealing with local velocity, the local energy theorem, and the local Lorentz force. These numerical examples explicitly demonstrate the usefulness and abilities of time-dependent weak values.

We compute the time derivative of the weak value through numerical simulations. As discussed in Appendix~\ref{apnumerical}, the numerical simulations require spatial and temporal grids. Thus, strictly speaking, the time derivatives computed in this section correspond to the finite-difference expressions FDLHD and FDRHD outlined in Sec.~\ref{theorylaboratory}.  In all simulations, we consider scenarios where conditions \textbf{C4}, \textbf{C5}, or both are satisfied so that the FDLHD and FDRHD can be interpreted as the LHD and RHD defined in Sec.~\ref{theory}. In addition, in these numerical simulations, the time step is chosen to be sufficiently small so that the FDLHD and FDRHD provide excellent approximations to the LHD and RHD.\footnote{The small numerical time step used in the numerical computations in  this paper does not need to coincide with the experimental time step required to evaluate the time derivatives in the laboratory, which must be determined by the time interval over which relevant time-dependent variations occur in the quantum system.}  

To gain additional physical insight into the behavior of such weak values and their time derivatives, after the numerical evaluation of the weak values, we re-interpret the results as Bohmian properties of a quantum system.

\subsection{Local velocity of particle}
\label{localvelocity}

Wiseman formulated the first attempt to discuss time derivatives of weak values \cite{wiseman2007} when presenting a local velocity from the time-derivative of a weak value of the position (post-selected in position eigenstates). 

In \eqref{weakvvelo} we have found how the expression $\BW\big(\opv\big||\egx\rangle,|\p\rangle\big)$ can be interpreted as the local velocity of a quantum particle.  The consistency in interpreting $\BW\big(\opv\big||\egx\rangle,|\p\rangle\big)$ as a velocity is provided by the LHD in \eqref{LHD} when $\opO=\opx$ and $|\evf\rangle=|\egx\rangle$.   
\begin{eqnarray}
\left.\frac{ \partial \BW\big(\opx,t_L,0\big||\egx\rangle,|\p\rangle\big)}{\partial t_L}\right|_{t_L=0} =\BW\big(\opv,0,0\big||\egx\rangle,|\p\rangle\big),
\label{extraeq}
\end{eqnarray}
where $\opv=\frac{i}{\hbar}[\opH,\opx]$ as seen in Appendix~\ref{ap7}. Notice that \eqref{extraeq} satisfies  \textbf{C4} (gauge invariance) because $[\opO,\opF]=[\opx,\opx]=0$. Thus, the local velocity can be obtained for an arbitrary pre-selected quantum state $|\p\rangle$ because the condition \textbf{C4} just fixes the post-selected state $\langle \egf|=\langle \egx|$. We only require a Hamiltonian that satisfies   $\opv=\frac{i}{\hbar}[\opH,\opx]$. 

As suggested by Wiseman \cite{wiseman2007}, \eqref{extraeq} shows that the  Bohmian velocity is just the quantum version of the classical time-of-flight procedure  to evalute the velocity: two positions are consecutively measured and the velocity defined as the distance divided by the time interval between the two measurements.  In the discussion at the end of Sec.~\ref{theorylaboratory}, we have identified the first weak value as $x_c={\langle x_c  |\opx | \p \rangle}/{\langle x_c |\p \rangle}$ and the second as $x_L={\langle x_c  |\opUt{t_L}\opx | \p \rangle}/{\langle x_c |\opUt{t_L}|\p \rangle}$. Notice that ${\langle x_c  |\opUt{t_L}\opx | \p \rangle}/{\langle x_c |\opUt{t_L}|\p \rangle}$ can be understood as giving insight on which \textit{was} the (weak value of the) position of the system at a time $t_L$ earlier than the time when the (strong) position of the system \textit{is} $x_c$. This reasoning implies $x_c=x_L+v_B t_L$ giving $(x_c-x_L)/t_L=v_B$, as seen from \eqref{numeric2}.

\subsection{Local work-energy theorem} 
\label{localwork}
The weak value of the Hamiltonian operator in \eqref{hami} is not measurable, because $\opH$ does not satisfy \textbf{C1} as seen in \eqref{hamig}. On the contrary, the kinetic energy operator defined as $\opW:=\frac{1}{2}m\opv^2$ is gauge invariant. Then, we can use   \eqref{RHD} to compute its RHD  as: 
\begin{eqnarray}
\left.\frac{\partial \BW\big(\opW,0,t_R\big||\egx\rangle,|\egv\rangle\big)}{\partial t_R}\right|_{t_{1}=0}=q\;\vB{|\egv\rangle}(\egx) \cdot \textbf{E}(\egx),
\label{power}
\end{eqnarray}
where the operator $\opW$ is gauge invariant and satisfies \textbf{C5} because $|\egv\rangle$ is an eigenstate of $\opO=\opW$.  The details can be found in Appendixes~\ref{ap8} and \ref{ap9}.  The expression (\ref{power}) is just the local version (i.e., $\egx$-density) of the corresponding global work-energy theorem. If $q\;\vB{|\egv\rangle}(\egx) \cdot \textbf{E}(\egx)$ is positive at some particular location $\egx$, one can interpret that the kinetic energy of the particle at such position increases, while decreases otherwise. 

\subsubsection{Numerical example}

We show how  an example of \eqref{power} can be implemented in  a practical scenario. We consider a single electron with a uniform electric field in the $x$-direction, $\E = (E, 0, 0)$ with $E=-1\cdot10^{6}$ V/m. To proceed,
we need to find a gauge potential for $\A$ and $\V$. There is, of course, no
unique choice. We select the usual Coulomb gauge $\Ag=0$ and $\Vg=-Ex$. Since the system becomes separable, to predict the electrical field from the time derivative of a weak value \eqref{power}, we can focus only on the $x$ component of such a weak value. Then, the 1D version of the Hamiltonian in \eqref{hami} becomes just:
\begin{eqnarray}
H=H(\A,\V)= -\frac{\hbar^2}{2m^*}\frac{\partial^2}{\partial x^2}+ q E x,
 \label{hamiE}
\end{eqnarray}
To evaluate RHD (in fact FDRHD), the initial (pre-selected) state has to be an eigenstate of the velocity operator that  satisfies \textbf{C5}. We roughly approximate such eigenstate by the Gaussian $\langle x|\Phi_{5}(0) \rangle$ described in Table~\ref{table1}. The wave packet spatial dispersion ${{\sigma }_{x}}=127$ nm corresponds to a width in the reciprocal space ${\sigma }_{k} = 1/{\sigma }_{x}=7.8$ $\mu m^{-1}$.  Thus,  the wave packet with a large spatial dispersion ${\sigma }_{x}\to \infty$ (  ${\sigma }_{k} \to 0$) mimics a plane wave, which is an effective eigenstate of the velocity operator. 

\begin{figure}
  \centering
\includegraphics[width=0.75\textwidth]{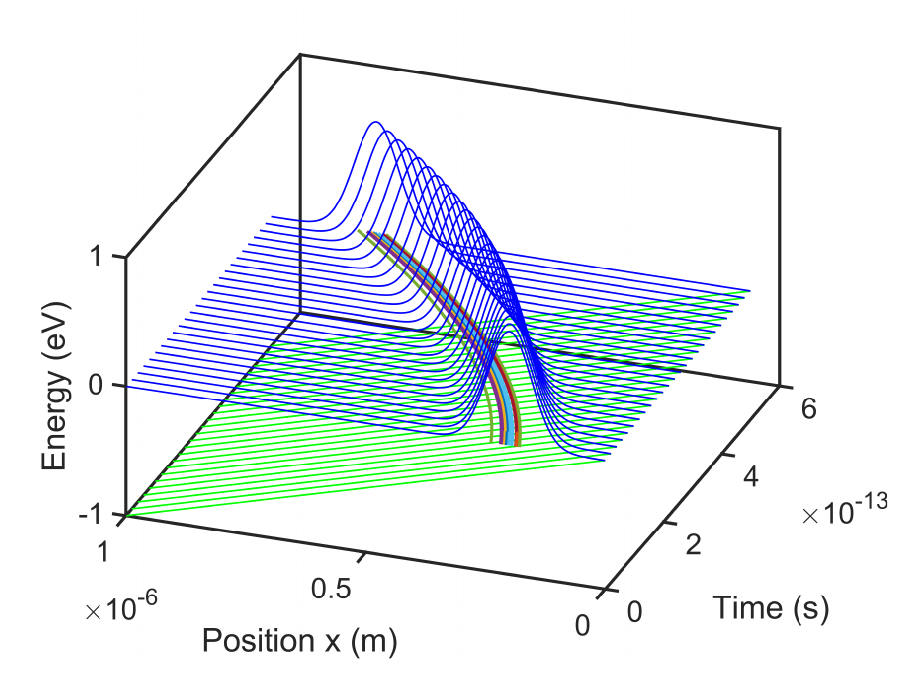}
\caption{Time evolution of the modulus squared of $\langle x|\Phi_{5}(t) \rangle$ (blue) in a time-independent potential energy profile (green). Different Bohmian trajectories $x_B^j(t)$ for $j=1,..,10$ are computed by integrating the $x$-component of the weak value of the velocity in  \eqref{weakvvelo} as indicated in \eqref{apnum7}. The different $x_B^j(t)$ will be used later to evaluate the weak values in different relevant positions and times.}
\label{f1}
\end{figure}

 The numerical time-evolution of $\langle x|\Phi_{5}(t) \rangle$ is computed from \eqref{apnum4} in Appendix~\ref{apnumerical}.   
In Fig.~\ref{f1}, we have plotted such time-evolution of the modulus squared of $\langle x|\Phi_{5}(t) \rangle$ (blue)  together with the time-independent potential energy profile (green) as $q E x$ in \eqref{hamiE}. For a posterior discussion, we also plot the Bohmian trajectories $x_B^j(t)$  for $j=1,..,N$  computed following the numerical algorithm in \eqref{apnum7}. The velocity of each trajectory can  be evaluated from the $x$-component of the weak value $\egv_B^{|\egv\rangle}=\{v_{B,x}^{|\egv\rangle},v_{B,y}^{|\egv\rangle},v_{B,z}^{|\egv\rangle}\}$ in \eqref{weakvvelo}. In this example, to simplify the notation, we refer to this $x$-component of the velocity as $v_{B}^{|\egv\rangle}$. 

\begin{figure}
  \centering
\includegraphics[width=0.75\textwidth]{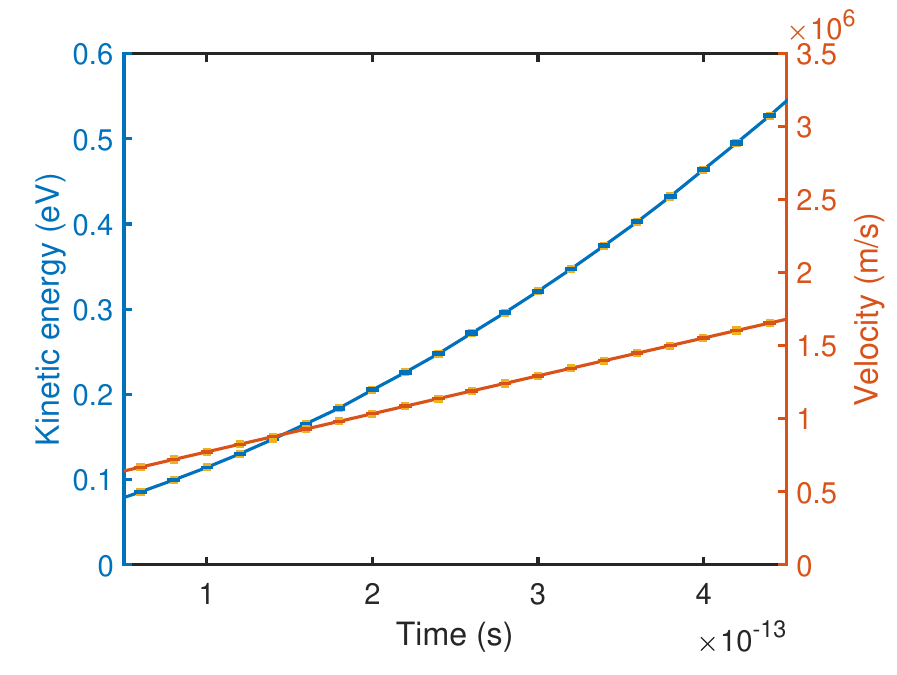}
\caption{In blue, mean value and standard deviation of the $x$-component of the weak value of the kinetic velocity of the 10 trajectories plotted in Fig.~\ref{f1} at 20 different times computed from  \eqref{power} (left axis). In orange, the mean value and standard deviation of the $x$-component of the weak value of the velocity in \eqref{weakvvelo} of the 10 trajectories in Fig.~\ref{f1} at 20 different times (right axis), which corresponds to the velocity $v_{B}^{|\Phi_{5}\rangle}(t)$. }
\label{f2}
\end{figure}

In Fig.~\ref{f2}, we have plotted in blue the $x$-component of the weak value of the kinetic energy at different times:
\begin{equation}
\BW\big(\opW_x,0,t_R\big||x\rangle,|\Phi_{5} \rangle \big)=\Re{ \frac{\langle x| \frac{1}{2}m^*\hat V_x ^2  \opUt{t_R} |\Phi_{5}(t) \rangle}{\langle x |\opUt{t_R}|\Phi_{5}(t) \rangle}}.
\label{wvk}
\end{equation}
with $\opW_x=\frac{1}{2}m^*\hat V_x ^2$ .  The inner products in \eqref{wvk}, i.e. ${\langle x| \frac{1}{2}m^*\hat V_x ^2 \opUt{t_R} |\Phi_{5}(t) \rangle}$ and ${\langle x|\opUt{t_R} |\Phi_{5}(t) \rangle}$, are computed in the same way as explained in subsection \ref{numericalcomputation}. In orange, the $x$-component of the Bohmian velocity, i.e., the weak value in \eqref{weakvvelo} or in \eqref{extraeq}, is also plotted as a function of time. 

\begin{figure}
  \centering
\includegraphics[width=0.75\textwidth]{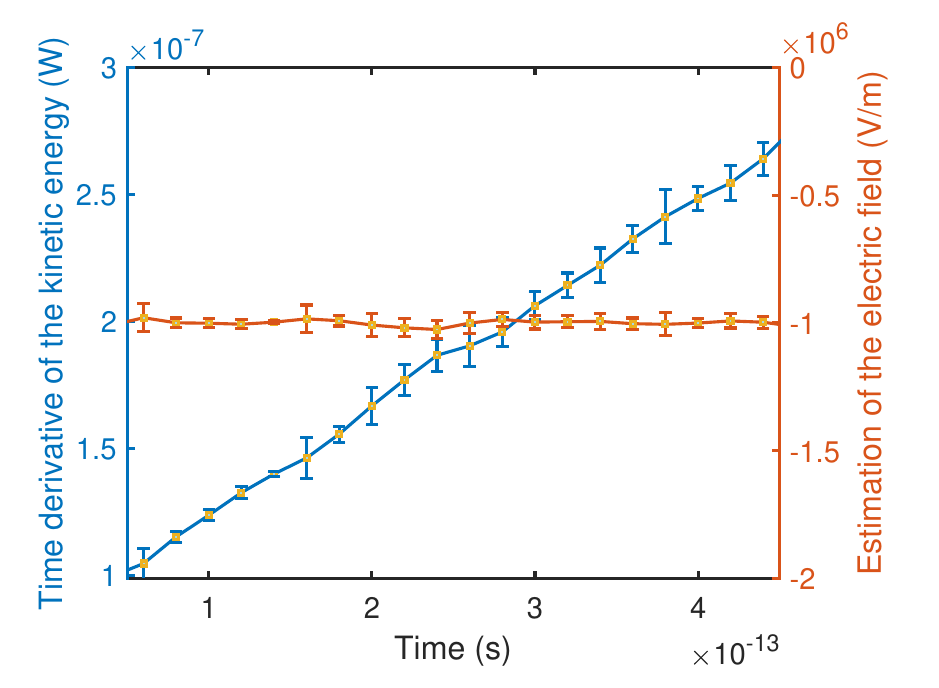}
\caption{In blue, mean value and standard deviation of the time derivative of  \eqref{wvk} of the 10 trajectories in Fig.~\ref{f1} at 20 different times (left axis). In orange, the mean value and standard deviation of the electrical field of the 10 trajectories in Fig.~\ref{f1} at 20 different times computed from \eqref{power} (right axis).}
\label{f3}
\end{figure}

Finally, in Fig.~\ref{f3}, the time derivative of \eqref{wvk} shown in Fig.~\ref{f2} in blue is also plotted.  In orange, we plot the estimation of the electrical field $E=E(x,t)$ from \eqref{power} as  
\begin{eqnarray}
E(x,t) \approx \frac{\frac{\partial}{\partial t_R} \Re{ \frac{\langle x | \frac{1}{2}m^*\hat V_x^2 \opUt{t_R}|\Phi_{5}(t) \rangle}{\langle x |\opUt{t_R}|\Phi_{5}(t) \rangle}}}{q\;v_{B}^{|\Phi_{5}\rangle}(x,t) },
\label{predictE}
\end{eqnarray}
There is an excellent agreement between the electric field used in this numerical simulation, i.e.,  $E=-1\cdot10^{6}$ V/m, and the value predicted through \eqref{predictE}. As mentioned,  and as would also happen in a laboratory, in this numerical simulation the time derivative $\frac{\partial}{\partial t_R} \Re{ \frac{\langle x | \frac{1}{2}m^*\hat V_x^2 \opUt{t_R}|\Phi_{5} \rangle}{\langle x |\opUt{t_R}|\Phi_{5} \rangle}}$ is done through FDRHD in \eqref{FDRHD}, instead of RHD, as they both give identical results because \textbf{C5} is satisfied.

 Despite one position $x$ and one time $t$ are enough to predict $E(x,t)$ through \eqref{predictE}, we have enriched the figures by considering several times $t$ and positions $x$ where the weak values are computed. At each time, $t$, the value $E(x,t)|_{x=x_B^j(t)}$ are numerically evaluated for a set of positions $x=x_B^j(t)$, where $x_B^j(t)$ are the different $j=1,...,10$  Bohmian trajectories plotted in Fig.~\ref{f1}. The initial positions of the trajectories are selected according to the initial wave function probability distribution to ensure that they are relevant positions in this scenario during all times. The mean value and the standard deviation of the different $E(x,t)|_{x=x_B^j(t)}$ are indicated in Fig.~\ref{f3}.  The same procedure is done in Fig.~\ref{f2} to plot the mean value and the standard deviation of the weak values of the velocity and of the kinetic energy. The information on the standard deviation shows how robust are the predictions when different times and positions are considered in this example with a uniform and constant electric field. 

We emphasize that the velocity at each position $x$ and time $t$ in the trajectories of Fig.~\ref{f1} requires a specific weak measurement protocol that is independent of previous positions and times. In the laboratory, the computation of the weak value at each time $t$ would assume that the initial (pre-selected) Gaussian wave packet is defined at such time $t$, i.e. the pre-selected state at $t$ is $\langle x|\Phi_{5}(t) \rangle$. Therefore, the connection of different velocities in Fig.~\ref{f2} to construct the continuous Bohmian trajectory shown in Fig.~\ref{f1} is natural within the Bohmian interpretation. However, such a connection is not necessary in the ontologically-free approach used to develop the time-derivative weak values presented in this paper. 

In any case, it will be illustrative to provide an additional justification of the above results from a Bohmian perspective. The weak value of the kinetic energy in \eqref{wvk} can be rewritten in a Bohmian language as:

\begin{eqnarray}
\Re{ \frac{\langle x | \frac{1}{2}m^*\hat V_x^2  |\Phi_{5} \rangle}{\langle x |\Phi_{5} \rangle}}=\frac{1}{2}m^* \left(v_{B}^{|\Phi_{5}\rangle}(x,t)\right)^2+Q^{|\Phi_{5}\rangle}(x,t)
\approx \frac{1}{2}m^* \left(v_{B}^{|\Phi_{5}\rangle}(x,t)\right)^2
\label{esder}
\end{eqnarray}

where $Q^{|\Phi_{5}\rangle}(x,t))$ is the so-called (Bohmian) quantum potential that depends on the curvature of the modulus of the wave function $\langle x|\Phi_{5}(t) \rangle$. For the spatially large wave packet considered here, such last contribution can be neglected in front of the first one.  Then, the last line in \eqref{esder} shows one possible procedure to evaluate the weak value of the kinetic energy in \eqref{wvk} for this particular scenario in the laboratory: once the velocity $v_{B}^{|\Phi_{5}\rangle}(x,t)$ is computed, the kinetic energy $\frac{1}{2}m^* \left(v_{B}^{|\Phi_{5}\rangle}(x,t)\right)^2$  in \eqref{esder} can be obtained by numerical post-processing the empirically obtained Bohmian velocity by squaring the value and multipling the result by $\frac{1}{2}m^*$ \cite{destefani2023}. Finally, the derivative in \eqref{predictE} can be written from \eqref{esder} as  $\frac{d }{dt}\left( \frac{1}{2}m^* \left(v_{B}^{|\Phi_{5}\rangle}(x,t)\right)^2\right) \approx m^*\; v_{B}^{|\Phi_{5}\rangle}(x,t)\; \frac{d v_{B}^{|\Phi_{5}\rangle}(x,t)}{dt}$. Invoking again the negligible role of the quantum potential in this scenario (the spatial derivative of $Q^{|\Phi_{5}\rangle}(x,t))$ is negligible in front of the spatial derivative of scalar potential $\Vg=-Ex$), one recovers a Newton's law for the Bohmian trajectory $m^*\; \frac{d v_B^{|\Phi_{5}\rangle}(t)}{dt} \approx qE$ to reach again expression \eqref{predictE} from a Bohmian perspective.

Finally, we present a summary of the protocol to be followed in a laboratory to obtain the electric field $E(x,t)$ at a specific time $t$ and a specific position $x$ (corresponding to one of the orange points in Fig.~\ref{f3}). For the quantum system described in \eqref{hamiE}, the steps are as follow:  

\begin{itemize}  
\item[] 1).- Take an ensemble of identically prepared (pre-selected at time $t$) states $|\Phi_{5}(t) \rangle$ and empirically evaluate the weak value $\BW\big(\hat V_x,0,0\big||x\rangle,|\Phi_{5}(t) \rangle\big)$.  

\item[] 2).- From the weak value in step 1), which corresponds to the Bohmian velocity, numerically evaluate the weak value of the kinetic energy as shown in \eqref{esder}.  

\item[] 3).- Repeat steps 1) and 2) to empirically evaluate $\BW\big(\hat V_x,0,t_R\big||x\rangle,|\Phi_{5}(t) \rangle \big)$ and to numerically compute the corresponding weak value of the kinetic energy at time $t+t_R$ where $t_R$ is small (as defined in subsection~\ref{fd-left}).  

\item[] 4).- Numerically evaluate the time derivative of the weak value of the kinetic energy  

${\frac{\partial}{\partial t_R} \Re{ {\langle x | \frac{1}{2}m^*\hat V_x^2 \opUt{t_R}|\Phi_{5}(t) \rangle}/{\langle x |\opUt{t_R}|\Phi_{5}(t) \rangle}}}$  

using FDRHD from \eqref{FDRHD} and the final numerical outputs obtained in steps 2) and 3).  

\item[] 5).- Numerically evaluate the electric field $E(x,t)$ from \eqref{predictE}, using the output in 4) as an input and the $v_{B}^{|\Phi_{5}\rangle}(x,t)$ empirically evaluated in step 1) as the other input.  
\end{itemize}  

Notice that the above protocol can be extended further if the empirical weak value of the velocity is replaced by a numerical time derivative of an empirical weak value of the position, as seen in \eqref{extraeq}. This possible refined protocol, which involves a second-order time derivative, is demonstrated in the next example.

\subsection{Local Lorentz force} 
\label{Local Lorentz force}

We consider a consecutive application of  FDLHD in \eqref{FDLHD} and FDRHD in \eqref{FDRHD}, ensuring the gauge invariance of all intermediate expressions, to get:
\begin{equation}
m^*\left.\frac{\partial^2 \BW\big(\opx,t_L,t_R\big||\egx\rangle,|\egv\rangle\big)}{\partial t_R \partial t_L}\right|{\begin{matrix}_{ t_R=0}\\ _{t_L=0} \end{matrix}}=q(\E(\egx)+\vB{|\egv\rangle} \times \B(\egx)),
\label{lorentz}
\end{equation}
where $\E(\egx)$ and $\B(\egx)$ are the (gauge invariant) electric and magnetic fields at position $\egx$ and $\vB{|\egv\rangle}$ is the (gauge invariant) Bohmian velocity of the state $|\egv\rangle$ computed from \eqref{weakvvelo}. See Appendixes~\ref{ap10} and \ref{ap11}. 

Notice that the second-order time derivative in \eqref{lorentz} is, in fact, a compact way of referring to a first-order time derivative of another first-order time derivative. The output of the FDLHD given by $\left.{\partial \BW\big(\opx,t_L,t_R\big||\egx\rangle,|\egv\rangle\big)}/{ \partial t_L}\right|_{t_L=0}=\BW\big(\opv,0,t_R\big||\egx\rangle,|\egv\rangle\big)$ that satisfies \textbf{C4} (i.e., $[\opO,\opF]=[\opx,\opx]=0$), is used as the input of the FDRHD given  by $\left.{\partial \BW\big(\opv,0,t_R\big||\egx\rangle,|\egv\rangle\big)}/{\partial t_R}\right|_{ t_R=0}=\BW\big(\opa,0,0\big||\egx\rangle,|\egv\rangle\big)$ that satisfies \textbf{C5} (i.e., $[\opO,\opphi]=[\opv,\opv]=0$), where $\opa=:\frac{i}{\hbar}[\opH,\opv]+\frac{\partial \opv}{\partial t}$ is defined in Appendix~\ref{ap10}.  
This sequence is precisely what occurs in the Ehrenfest theorem. Two consecutive first-order time derivatives, $\frac{d}{dt}\langle \p|\opx  |\p\rangle=\langle \p|\opv  |\p\rangle$ and $\frac{d}{dt}\langle \p|\opv  |\p\rangle=\langle \p|\opa  |\p\rangle$, can be compactly written together as $\frac{d^2}{dt^2}\langle \p|\opx  |\p\rangle=\langle \p|\opa  |\p\rangle$. There are several possible finite-difference approximations to evaluate a mathematical second-order time derivative in \eqref{lorentz}. The technical differences among them is not relevant in our discussion \footnote{\label{seconderivative}If one uses \eqref{FDLHD} and \eqref{FDRHD} for the first order time-derivatives, then a simple finite-difference approximation for the second-order time derivative in \eqref{lorentz} is $\left(\BW_1-\BW_2-\BW_3+\BW_4 \right)/(t_L t_R)$ where $\BW_1=\BW\big(\hat{y},0,t_R\big||\egx\rangle,|\egv\rangle\big)$, $\BW_2=\BW\big(\hat{y},t_L,t_R\big||\egx\rangle,|\egv\rangle\big)$, $\BW_3=\BW\big(\hat{y},0,0\big||\egx\rangle,|\egv\rangle\big)$ and $\BW_4=\BW\big(\hat{y},t_L,0\big||\egx\rangle,|\egv\rangle\big)$.  Of course, one can find other finite-difference approximations for the second-order time derivative.}.  The final result in \eqref{lorentz} is simply a quantum and local version of the (Newton) Lorentz force, which can be used, for example, as a quantum sensor of electromagnetic fields at specific position and time as shown in the numerical example below.

\subsubsection{Numerical example}

We consider a quantum system with a magnetic field in the $z$-direction, so that $\B = (0, 0, B)$ with $B=0.19$ Tesla, and no electric field. To proceed,
we need to find a gauge potential $\Ag$ which obeys $\B =\nabla \times \Ag$. There is, of course, no
unique choice. Here we pick $\A^g = (0, xB, 0)$ and $\Vg=0$. This is called the Landau gauge. In this scenario, the Lorentz force in \eqref{lorentz} is:
\begin{equation}
\E(\egx)+\textbf{v}_B(\egx) \times \B(\egx)=\{\; v_{B,y} B, \;- v_{B,x} B,\; 0\}.
\label{vectorial}
\end{equation}
The dynamics of the $z$ direction are not relevant in this scenario so we assume that an electron moves only in the $x-y$ plane. Then, the Hamiltonian in \eqref{hami} becomes:
\begin{eqnarray}
H=\frac{1}{2m^*}\left( -{\hbar^2} \frac{\partial^2}{\partial x^2} + \left( -i{\hbar} \frac{\partial}{\partial y} -qBx  \right)^2\right).
\label{hamiB}
\end{eqnarray}
Since this Hamiltonian \eqref{hamiB} has translational invariance in the $y$ directions, the component of the energy eigenstate in such direction will be a plane wave. On the other hand, it is well-known that the global energy eigenstates of \eqref{hamiB} correspond to those of a displaced harmonic oscillator as described in Appendix~\ref{apnumerical}. Since we are interested in initial states that are eigenvalues of the velocity operator, we consider a  superposition of $10$ energy eigenstates, all with the same $k_y=0.0118$ ${nm}^{-1}$: 
\begin{eqnarray}
	\label{gausB}
	\psi_H(x,y,t) = \sum_{n=0}^9  c_{n,k_y}(t) \psi_{n,k_y}(x,y),
\end{eqnarray}
where the state $\psi_{n,k_y}(x,y)$ is defined in \eqref{egvh} and $c_{n,k_y}(t)$ in \eqref{chb}.  
By construction, the state $\psi_H(x,y,t)$ in \eqref{gausB} is an eigenstate of the velocity in the $y$-direction  (notice that the dependence on $y$ of the term $\psi_{n,k_y}(x,y)$ is given by the global term $e^{ik_y y}$ in \eqref{egvh}) , but not in the $x$ direction. Thus, we focus on the $y$-component of expression \eqref{lorentz} that gives $-v_{B,x} B$ in \eqref{vectorial}, allowing to predict $B$.

\begin{figure}
  \centering
\includegraphics[width=0.75\textwidth]{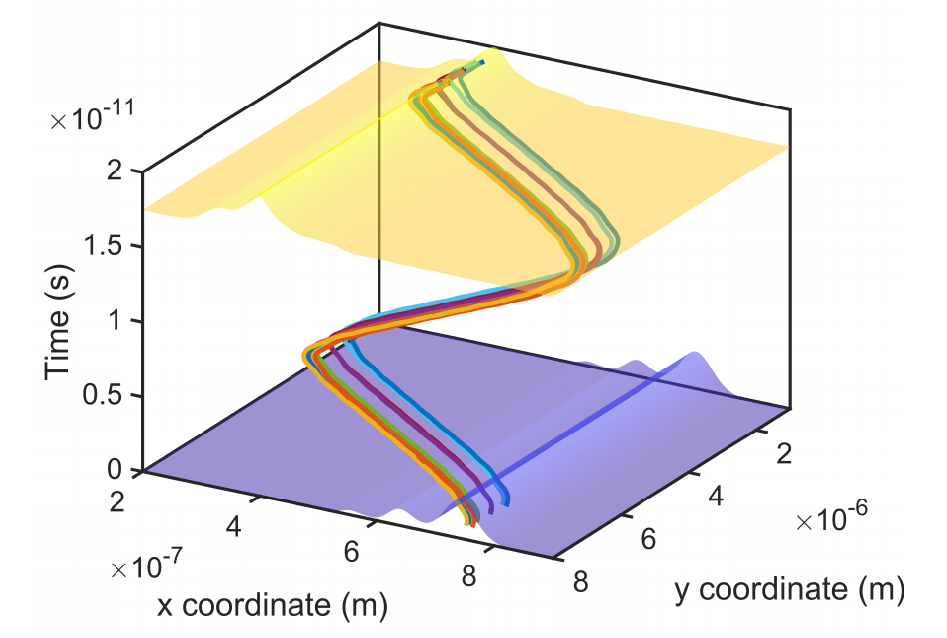}
\caption{Probability distribution in the $x-y$ space for the quantum state $|\psi_H(x,y,t)|^2$ in \eqref{gausB} at times $0$ and $0.2$ ps. A set of Bohmian trajectories $x_B^j(t)$ for $j=1,..,10$ are computed by time-integrating the ($x-$velocity) weak value in \eqref{velox} and the ($y$-velocity) weak value in \eqref{veloy}. }
\label{f4}
\end{figure}

\begin{figure}
  \centering
\includegraphics[width=0.75\textwidth]{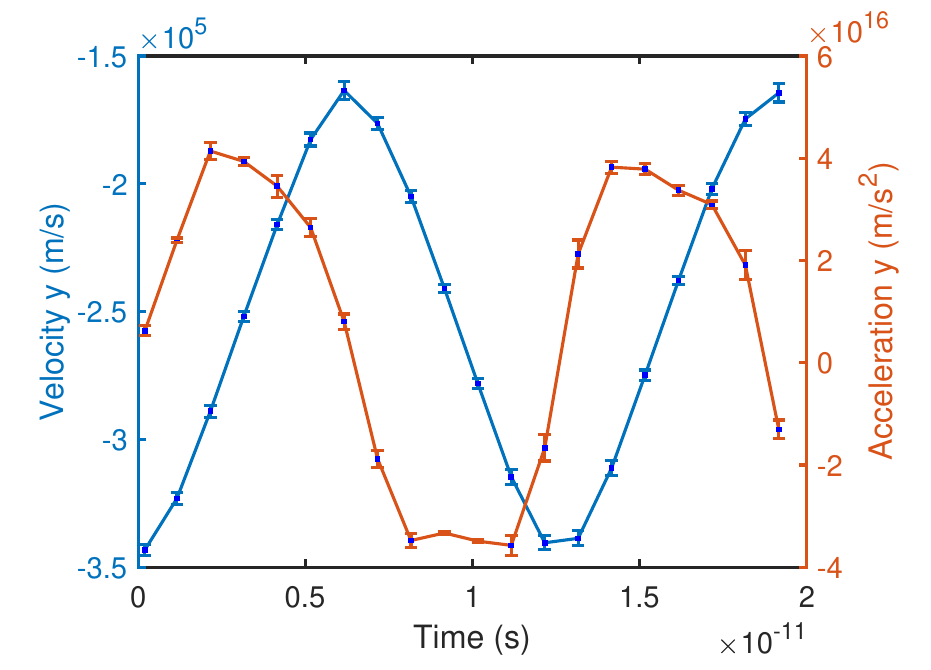}
\caption{In blue, mean value and standard deviation of the $y$-velocity of the 10 trajectories in Fig.~\ref{f4} at 20 different times (left axis) computed from the weak value in \eqref{veloy}, which also corresponds to the (first order) time derivative of the weak value of the $y-$position. In orange, the mean value and standard deviation of the $y-$acceleration  of the 10 trajectories in Fig.~\ref{f4} at 20 different times (right axis) computed by the time-derivative of the $y$-velocity, which also corresponds to the (second order) time derivative of the weak value of the $y-$position.}
\label{f5}
\end{figure}

\begin{figure}
  \centering
\includegraphics[width=0.72\textwidth]{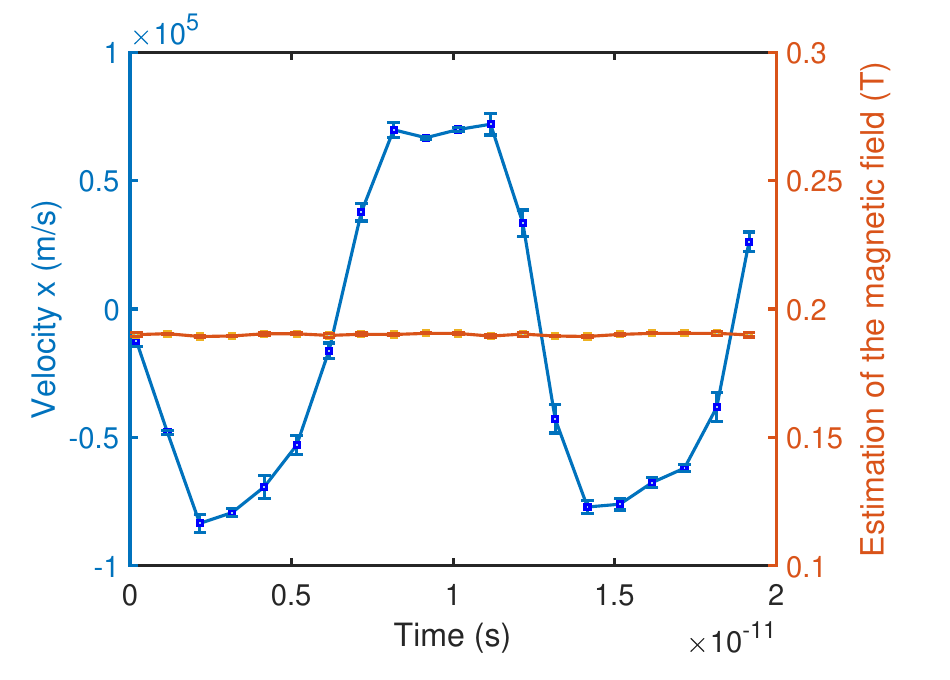}
\caption{In blue, mean value and standard deviation of the $x$-velocity of the 10 trajectories in Fig.~\ref{f4} at 20 different times (left axis) computed from the weak value in \eqref{velox}, which corresponds to the (first order) time derivative of the weak value of the post-selected in $x-$position. In orange, the mean value and standard deviation of the predicted value of the magnetic field computed from the expression \eqref{predictB} for the 10 trajectories in Fig.~\ref{f4}. Notice that all values are computed from the weak value of the position post-selected in position and their (first or second-order) time-derivative.}
\label{f6}
\end{figure}

In Fig.~\ref{f4}, we have plotted the probability distribution of $|\psi_H(x,y,t)|^2$ at the preparation time ($t=0$ ps) and final ($t=t_f=20$ ps) times.  The evaluation of  $\psi_H(x,y,t)$ at different times is explained in Appendix~\ref{apnumerical}. We also plot a set of 10 Bohmian trajectories $x_B(t)$ whose velocity in the $x$-component $v_{B,x}^{|\p_H\rangle}(x,y,t)$ is computed from the $x$-component of the weak value in \eqref{weakvvelo} as:
\begin{eqnarray}
v_{B,x}^{|\p_H\rangle}= \Re{ \frac{\langle x,y| \hat V_x  |\p_H(t)\rangle}{\langle x,y |\p_H(t)\rangle}}=\frac {\hbar}{m^*} \Im\left( \frac{\frac{\partial \Psi_H(x,y,t)}{\partial x}}{\Psi_H(x,y,t)} \right),
\label{velox}
\end{eqnarray}
and velocity in the $y$-component  $v_{B,y}^{|\p_H\rangle}(x,y,t)$ is computed from the $y$-component of the weak value in \eqref{weakvvelo} as:
\begin{eqnarray}
v_{B,y}^{|\p_H\rangle}= \Re{ \frac{\langle x,y| \hat V_y  |\p_H(t)\rangle}{\langle x,y |\p_H(t)\rangle}}\nonumber\\
=\frac {\hbar}{m^*} \Im\left( \frac{\frac{\partial \Psi_H(x,y,t)}{\partial y}}{\Psi_H(x,y,t)}\right) -\frac{q\; B\; x}{m^*}.
\label{veloy} 
\end{eqnarray}
 These weak values of the velocity can be computed in the laboratory from time derivatives of weak values of the positions as in \eqref{extraeq}. As discussed for the local work-energy theorem,  only one position in configuration space $\{x,y\}$ and one time $t$ would have been enough to compute the magnetic field $B=B(x,y,t)$. We consider several times and several positions (i.e., the Bohmian trajectories depicted in Fig.~\ref{f4}) to enrich the discussion and to test how robust these predictions are in front of variations of positions and times.  The development of \eqref{velox} and \eqref{veloy},  plus the computation of the trajectories are explained in Appendix~\ref{apnumerical}.  
Both weak values (velocities) in \eqref{velox} and \eqref{veloy} are plotted in blue in Figs.~\ref{f5} and \ref{f6}. Due to the magnetic field and the quantum state $\psi_H(x,y,t)$, the Bohmian trajectories show oscillations with a period of $12$ ps (corresponding to the mentioned $\omega=\frac{|q|B}{m^*}=0.49$ Trad/s define in Appendix~\ref{apnumerical}) in the $x-$direction (as seen in the positive and negative $x-$velocities in Fig.~\ref{f6}) and the same oscillation with a net translation in the $y-$direction (as seen in the negative $y-$velocities in Fig.~\ref{f5}).  

In Fig.~\ref{f5}, we have also plotted, in orange, the numerical evaluation of the time derivative of the velocity in \eqref{veloy}. Such a first-order derivative of the velocity corresponds to the second-order derivative of the position of the trajectories plotted in Fig.~\ref{f4}, as indicated in \eqref{extraeq}, i.e., $\left.{ \partial \BW\big(\hat y,t_L,t_R\big||x,y\rangle,|\p_H\rangle\big)}/{\partial t_L}\right|_{t_L=0}=\BW\big(\opv_y,0,t_R\big||x,y\rangle,|\p_H\rangle\big)$. Finally, in Fig.~\ref{f6}, we have plotted the predicted value of the magnetic field $B=B(x,y,t)$ through the y-component of the  \eqref{final} in the particular case of \eqref{vectorial}, giving:
\begin{eqnarray}
B(x,y,t)&\approx&-\frac{m^*}{q}\frac{\left.\frac{ \partial^2 \BW\big(\hat{y},t_L,t_R\big||x,y\rangle,|\p_H\rangle\big)}{\partial t_L \partial t_R}\right|{\begin{matrix}_{ t_R=0}\\ _{t_L=0} \end{matrix}}}{\left.\frac{ \partial \BW\big(\hat x,t_L,0\big||x,y\rangle,|\p_H\rangle\big)}{\partial t_L}\right|_{t_L=0}},
\label{predictB}
\end{eqnarray}
There is an excellent agreement between the magnetic field $B=0.19$ Tesla and the value predicted through the time derivative of weak values in  \eqref{predictE}.  Notice that the last identity in \eqref{predictB} shows that, in the laboratory, only the weak value of the x-position, i.e., $\BW\big(\hat{x},t_L,0\big||x,y\rangle,|\p_H\rangle$, and y-position, i.e., $\BW\big(\hat{y},t_L,t_R\big||x,y\rangle,|\p_H\rangle$ needs to be evaluated at different times $t_R$ and $t_L$. 

The result in \eqref{predictB} can be alternatively justified from a Bohmian perspective, as follows. The first term of the velocity in the $y$ direction in \eqref{veloy} can be considered roughly as a constant for a plane wave, i.e.  $ \frac{\hbar}{m^*} \Im\left( {\frac{\partial \Psi_H(x,y,t)}{\partial y}}/{\Psi_H(x,y,t)}\right)=\frac{\hbar k_y}{m^*}$, where the mentioned quantum potential becomes negligible. Thus, such velocity can be written as $v_{B,y}^{|\p_H\rangle}(t)={\hbar k_y}/{m^*}-q B x_B(t)/m^*$ where $x_B(t)$ is the x-component of the Bohmian trajectory. Finally, the acceleration (derivative) of this Bohmian velocity in \eqref{veloy} becomes just $\frac{d v_{B,y}^{|\p_H\rangle}(x,y,t)}{d t}=-\frac{q\; B\; }{m^*}\frac{d x_{B}(t)}{dt} =-\frac{q\; B\; }{m^*}v_{B,x}^{|\p_H\rangle}(t)$ which is another way of rewriting the expected result in \eqref{predictB}.

Finally, we present a summary of the protocol to be followed in a laboratory to obtain the magnetic field $B=B(x,y,t)$ at a specific time $t$ and at specific positions in the a point of the configuration space $\{x,y\}$ (corresponding to one of the orange points in Fig.~\ref{f6}). For the quantum system described in \eqref{hamiB}, the steps are as follow:  

\begin{itemize}  
\item[] 1).- Take an ensemble of identically prepared (pre-selected at $t$) states $|\p_H\rangle$ and empirically evaluate the weak value $\BW\big(\hat y,0,0\big||\egx\rangle,|\p_H\rangle\big)$.  

\item[] 2).- Repeat step 1) to empirically evaluate $\BW\big(\hat y,t_L,0\big||\egx\rangle,|\p_H\rangle\big)$, $\BW\big(\hat y,0,t_R\big||\egx\rangle,|\p_H\rangle\big)$, and $\BW\big(\hat y,t_L,t_R\big||\egx\rangle,|\p_H\rangle\big)$ for $t_L$ and $t_R$ small enough (as defined in subsection~\ref{fd-left}).  

\item[] 3).- Numerically evaluate the second-order time derivative ${ \partial^2 \BW\big(\hat{y},t_L,t_R\big||x,y\rangle,|\p_H\rangle\big)}/{\partial t_L \partial t_R}$ using the four empirical outputs from steps 1) and 2), applying a finite-difference approximation \footref{seconderivative}.  

\item[] 4).- Take an ensemble of identically prepared (pre-selected at $t$) states $|\p_H\rangle$ and empirically evaluate the weak value $\BW\big(\hat x,0,0\big||\egx\rangle,|\p_H\rangle\big)$.  

\item[] 5).- Repeat step 4) to empirically evaluate $\BW\big(\hat x,t_L,0\big||\egx\rangle,|\p_H\rangle\big)$ for small $t_L$.

\item[] 6).- Using FDLHD in \eqref{FDLHD}, numerically evaluate ${ \partial \BW\big(\hat x,t_L,0\big||x,y\rangle,|\p_H\rangle\big)}/{\partial t_L}$ from the empirical weak values obtained in steps 4) and 5).  

\item[] 7).- Numerically evaluate the magnetic field $B(x,y,t)$ from \eqref{predictB}, using as inputs the numerical outputs in step 3) and 6).  
\end{itemize}

\section{Which modal theories exhibit time-dependent consistency of weak values?}
\label{which}

In the results presented in the previous Sec.~\ref{discussion2}, we have shown that the weak values post-selected in position and its time derivative can be linked to the physical properties of the Bohmian theory. For example, the weak value of the velocity post-selected in position gives the Bohmian velocity of a particle \cite{wiseman2007}. Thus, it seems that the Bohmian theory is a good framework for predicting weak values post-selected in position and its time derivative. However, Ref. \cite{pang2020} explicitly shows that other Bohm-like theories can be used to predict weak values post-selected in eigenstates other than position eigenstates. The reason why there is an infinite number of Bohm-like theories (also known as modal theories \cite{devashish2021}) that can provide a \textit{consistent} picture of quantum phenomena in terms of weak values can be easily evidenced by unraveling the expectation value of the position operator in terms of weak values with an arbitrary post-selected state $|\egf\rangle$. From the identity $\langle \p|\opx  |\p\rangle= \langle  \p| \int d\egf  |\egf\rangle \langle \egf | \opx  |\p\rangle$, we get:  
\begin{eqnarray}
\langle \p|\opx  |\p\rangle&=&\int d\egf|\langle  \p|\egf\rangle|^2\Re{\frac{\langle \egf | \opx  |\p\rangle}{\langle \egf |\p\rangle}} \label{which1}.
\end{eqnarray}
Thus, interpreting the expectation value as a weighted sum of weak values post-selected in $\egf$ is \textit{consistent} with \eqref{which1}. The function $\Re\{..\}$ ensures that the expectation value has no imaginary part. Therefore, each set $\{|\egf\rangle\}$ that satisfies $\int d\egf |\egf\rangle \langle \egf |=\unit$ can be identified as the primary ontological element of a new Bohm-like theory (i.e., a new $\egf$-modal theory). In particular, the authors of Ref. \cite{pang2020} define the traditional Bohmian theory, where position is the primary ontology, as the x-Bohm theory. Alternatively, they define a p-Bohm theory as a Bohm-like theory where velocity (momentum) is the primary ontological element (i.e., the p-Bohm theory can be used to predict weak values post-selected in velocity eigenstates satisfying \eqref{which1}). By comparing the phase-space dynamics offered by the x-Bohm and p-Bohm theories for an arbitrary pre-selected state $\p$, the authors of Ref. \cite{pang2020} conclude that our previous argument that the x-Bohm theory provides a \textit{consistent} explanation of quantum phenomena is quite speculative. Their argument does not negate that the x-Bohm theory provides a \textit{consistent} explanation (i.e., weak values post-selected in position satisfy \eqref{which1}), but rather that the x-Bohm theory is merely one \textit{consistent} explanation among an infinite number of possible \textit{consistent}, yet distinct, Bohm-like explanations of quantum phenomena (i.e., weak values post-selected in any eigenstate $|\egf\rangle$ that also satisfy \eqref{which1}). The summary of the argumentation of Ref. \cite{pang2020} is the following: \textit{“When everyone is somebody, then no one's anybody”}.

The developments elaborated in this paper on time derivatives of weak values using LHD and RHD also establish an additional and more restrictive criterion on which post-selected states exhibit also a time-dependent \textit{consistency}. The Ehrenfest theorem deduced from \eqref{eren} shows the following time-dependent relationship $\frac{d}{dt}\langle \p|\opx  |\p\rangle=\langle \p|\opv  |\p\rangle$ for any quantum state $|\p\rangle$. Therefore, it seems reasonable to demand a similar \textit{time-dependent consistency} for the weak value that unravels the time derivative of the expectation values. For the position operator, this additional criterion can be written as:
\begin{eqnarray}
&&\frac{d}{dt}\langle \p|\opx  |\p\rangle=\langle \p|\opv  |\p\rangle=\int d\egf |\langle  \p|\egf\rangle|^2 \Re{\frac{\langle \egf | \opv  |\p\rangle}{\langle \egf |\p\rangle}}\nonumber\\
&&\mysign\int d\egf |\langle  \p|\egf\rangle|^2\left.\frac{\partial }{\partial t_L} \Re{\frac{\langle \egf | \opUt{t_L}\opx  |\p\rangle}{\langle \egf |\opUt{t_L}|\p\rangle}}\right|_{t_L=0}.\label{which4}
\end{eqnarray}
We have shown in Sec.~\ref{theory} that the accomplishment of the last identity in \eqref{which4}, which we named \textit{time-dependent consistent} criterion, is not true, in general, for arbitrary post-selected states $\langle \egf|$. The identity $\left.\frac{\partial }{\partial t_L} \Re{\frac{\langle \egf | \opUt{t_L}\opx  |\p\rangle}{\langle \egf |\opUt{t_L}|\p\rangle}}\right|_{t_L=0}=\Re{\frac{\langle \egf | \opv  |\p\rangle}{\langle \egf |\p\rangle}}$ holds only for very specific post-selected states $\langle \egf|$.  In the particular case of the operator $\opx$, from \textbf{C4}, we have shown that only the weak values post-selected in position, i.e., $|\egf\rangle=|x\rangle$, satisfy both \eqref{which1} and \eqref{which4} for an arbitrary pre-selected state $|\p\rangle$.  See appendix \ref{apenfinal} for an alternative demonstration of this result.   Thus, the Bohmian theory has a quite privileged status when dealing with weak value post-selected in position and their time derivatives. A similar \textit{time-dependent consistent} criterion can be obtained by replacing the left-hand derivative (LHD) with the right-hand derivative (RHD) in \eqref{which4}, and ensuring condition \textbf{C5}, at the price of fixing the pre-selected state as $|\p\rangle = |\egx\rangle$.

This previous criterion, on which pre-selected states have \textit{time-dependent consistency}, is an operator-dependent criterion. By replacing the operator $\opx$ in \eqref{which4} by another operator, for example $\opv$, other post-selected states are satisfying the  \textit{time-dependent consistent} criterion. Again,  we notice that there is an infinite set of arbitrarily post-selected states $|\egf\rangle$ satisfying: 
\begin{eqnarray}
\langle \p|\opv  |\p\rangle&=&\int d\egf|\langle  \p|\egf\rangle|^2\Re{\frac{\langle \egf | \opv  |\p\rangle}{\langle \egf |\p\rangle}}.
\label{which2}
\end{eqnarray}
The requirement imposed by the Ehrenfest theorem deduced from \eqref{eren} is now $\frac{d}{dt}\langle \p|\opv |\p\rangle=\langle \p|\opa |\p\rangle$, with $\opa := \frac{i}{\hbar}[\opH,\opv] + \frac{\partial \opv}{\partial t}$ being the operator defined in Appendix~\ref{ap10}.  The Ehrenfest theorem can be translated as an additional requirement for the \textit{time-dependent consistency} of the weak values as follows:
\begin{eqnarray}
&&\frac{d}{dt}\langle \p|\opv  |\p\rangle=\langle \p|\opa  |\p\rangle=\int d\egf |\langle  \p|\egf\rangle|^2 \Re{\frac{\langle \egf | \opa  |\p\rangle}{\langle \egf |\p\rangle}}\nonumber\\
&&\mysign\int d\egf |\langle  \p|\egf\rangle|^2\left.\frac{d}{d t_L} \Re{\frac{\langle \egf | \opUt{t_L}\opv  |\p\rangle}{\langle \egf |\opUt{t_L}|\p\rangle}}\right|_{t_L=0}.\label{which5}
\end{eqnarray}
In the case of $\opv$, from \textbf{C4}, we have shown that weak values post-selected in velocity (momentum), i.e., $\langle \egf|=\langle \egv|$, satisfy both \eqref{which2} and \eqref{which5} as they accomplish $\left.\frac{d}{d t_L} \Re{\frac{\langle \egf | \opUt{t_L}\opv  |\p\rangle}{\langle \egf |\opUt{t_L}|\p\rangle}}\right|_{t_L=0}=\Re{\frac{\langle \egf | \opa  |\p\rangle}{\langle \egf |\p\rangle}}$ for arbitrary post-selected states $|\p\rangle$. A similar \textit{time-dependent consistent} criterion can be obtained by changing LHD by RHD in \eqref{which5} and ensuring condition \textbf{C5}, at the price of fixing the pre-selected state as $|\p\rangle=|\egv\rangle$. 

In principle, it appears that only the first-order time derivative of weak values can be obtained because the post-selected states in \eqref{which4} differ from those in \eqref{which5}. However, by using the equality $\Re{{\langle \egv | \opv | x\rangle}/{\langle \egv |x \rangle}}=\Re{{\langle x | \opv | \egv\rangle}/{\langle x |\egv \rangle}}$, which follows from the hermiticity of $\opv$, both \eqref{which4} and \eqref{which5} can still be satisfied by considering $\langle x|$ as the post-selected state and $|\egv\rangle$ as the pre-selected state.  This is what happens in the numerical results in subsection~\ref{Local Lorentz force}.

\section{Conclusions}
\label{conclusion}

In physics, the time derivative of a property frequently gives rise to another meaningful property. For example, in classical mechanics, the knowledge of the position allows one to predict the value of the velocity without measuring it, just by taking a first-order time derivative of the position. Even the acceleration of a system (related to the forces acting on the system) can be anticipated from the second-order time derivatives of the position.  Since weak values offer empirical insights that cannot be derived from expectation values, this paper investigates what physical insight can be obtained from the time derivative of weak values.

A time derivative of a weak value is not always either a weak value or a gauge-invariant quantity. The necessary and sufficient conditions for the gauge invariance (i.e., empirical observability) of weak values are presented in this paper, i.e., conditions \textbf{C1, C2} and \textbf{ C3}. There are weak values in the literature that do not satisfy these conditions, as shown in Table~\ref{table} and discussed in Sec.~\ref{discussion2}. In addition, two more necessary and sufficient conditions (\textbf{C4, C5}) are presented to ensure that the time derivative of a gauge-invariant weak value also becomes a gauge-invariant weak value. In particular, condition \textbf{C4} applies to LHD and condition \textbf{C5} to RHD. 

 We have also discussed how the LHD and RHD can be empirically evaluated in the laboratory through finite-difference approximations of the left- and right-hand time derivatives of weak values, referred to as FDLHD and FDRHD, respectively. We show that FDLHD and FDRHD yield results that are roughly identical to the theoretical LHD and RHD when the latter are gauge invariant.

The time-derivative of weak values can be used, for example, to deduce a local Ehrenfest-like theorem as follows: one weak value (position) measured in the laboratory provides information on two other unmeasured weak values obtained through first-order (velocity) and second-order (acceleration) time derivatives of the first measured weak value. As an example, an electromagnetic field quantum sensor at a local position and time is proposed, utilizing the empirical evaluation of weak values of the position, and computing their time derivatives (through  FDLHD and FDRHD).

The findings in this paper reveal an important asymmetry between different post-selected states used to compute weak values. If we want to describe the \textit{local} position, \textit{local} velocity, and \textit{local} acceleration of quantum systems in terms of weak values, and we want these weak values also to satisfy the additional \textit{time-dependent consistency} criteria explained in \eqref{which4} and \eqref{which5}, we conclude that weak values post-selected in position, and somehow also post-selected in velocity, play a quite privileged role—one that cannot be attributed to other weak values post-selected in other quantum states.  This privilege, however, implies a limitation on the types of scenarios that are empirically accessible for obtaining simultaneous insight into the local dynamics of quantum systems through second-order time derivatives of weak values of the position. This second-order time derivative can be used in scenarios pre-selected in velocity eigenstates $ |\egv\rangle $ and post-selected in position eigenstates $ \langle \egx|$ (or vice versa). In any case, the second-order time derivative of the weak values involving these wave functions $\langle \egx|\egv(t)\rangle$ is enough to detect forces, in general, and electromagnetic fields, in particular.   

In this paper, the electric and magnetic fields are treated as external elements of the single-particle Schrödinger equation: the simulated particle used to compute the weak value (which later senses the electromagnetic fields) is affected by these fields, but the fields themselves are not affected by the particle. This approach is the so-called semi-classical treatment of light and matter. The possibility of measuring how electromagnetic fields are distributed in space and time demonstrates the richness of using local-in-position weak values, rather than ensemble values, to characterize physical systems. Future work aims to extend the ideas discussed here to weak values in which both the simulated particle (i.e., matter) and the electromagnetic fields (i.e., light) are quantized.\footnote{For example, in the framework of cavity quantum electrodynamics~\cite{optics}, where light is confined within an optical resonator, if we assume that energy restrictions ensure that only the fundamental mode of the resonator is available, the quantum treatment of light can be readily incorporated into the Schrödinger equation used in this work. This essentially requires adding the pair of non-commuting operators that arise from the canonical quantization of the classical canonical degrees of freedom, i.e., the Fourier amplitude of the single mode, to describe the quantum light inside the optical cavity~\cite{destefani2022}.}

We emphasize that no fundamental (ontological) conclusions can be drawn from the results of this paper, as all mathematical predictions presented here hold under any interpretation of quantum mechanics. The purpose of this work is not to address the ontological meaning of weak values but rather to investigate their time derivatives and practical applicability. Similarly, the privileged connection between weak values post-selected in position and the Bohmian properties of a quantum system is not intended to argue that Bohmian mechanics is true, but only to highlight its usefulness in envisioning empirical protocols for characterizing novel \textit{local} properties of quantum systems.


\section*{Acknowledgments}
The work has benefited from discussions with Xabier Oianguren-Asua. This research was funded by Spain's Ministerio de Ciencia, Innovaci\'on y Universidades under Grants  PID2021-127840NB-I00 (MICINN/AEI/FEDER, UE) and PDC2023-145807-I00 (MICINN/AEI/FEDER, UE), and European Union's Horizon 2020 research and innovation program under Grant 881603 GrapheneCore3.

\section*{List of abbreviations}

\begin{itemize}
\item POVM:Positive-operator-valued measure
\item PVM:Projection-valued measure
\item LHD:Left-hand derivative 
\item RHD: Right-hand derivative
\item FDLHD: Finite-difference left-hand derivative
\item FDRHD: Finite-difference right-hand derivative
\end{itemize}

\begin{appendix}

\section{Gauge invariance of the Schrödinger equation}
\label{ap1}

\renewcommand{\theequation}{\Alph{section}\arabic{equation}}
\setcounter{equation}{0}

One can show how the structure of the Schrödinger equation (in the Coulomb gauge) in \eqref{scho} in the paper, and rewritten in the right-hand side of \eqref{ap1.1}, is equivalent to the Schrödinger equation in another gauge (changing $\V\to\V^g$, $\A\to\A^g$ and $\p\to\p^g$) as written in the left-hand side of \eqref{ap1.1}: 
\begin{equation}
i\hbar \frac{\partial \pg}{\partial t}=\underbrace{\left(\frac{1}{2m^*}\left( \opp -q\Ag\right)^2+ q\Vg\right)}_{\opH^g}\pg\;\;\;\;\to\;\;\;\;\;i\hbar \frac{\partial \p}{\partial t}=\underbrace{\left(\frac{1}{2m^*}\left( \opp -q\A\right)^2+ q\V\right)}_{\opH}\p.
\label{ap1.1}
\end{equation}
First, using \eqref{vg} and \eqref{wg} in the paper, one can show 
\begin{equation}
( \opp -q\A^g)^2\pg=e^{i\frac{q}{\hbar}\g}( \opp -q\A)^2\p. 
\label{ap1.2}
\end{equation}
By using $-i\hbar\grad \pg=e^{i\frac{q}{\hbar}\g}(-i\hbar\grad \p+q(\grad \g )\p)$, we get:
\begin{equation}
( \opp -q\A^g)\pg=-i\hbar\grad \pg -q\A\pg -q(\grad \g)\pg=e^{i\frac{q}{\hbar}\g}(-i\hbar\grad -q\A)\p.
\label{ap1.2bis}
\end{equation}
By defining $\phi=(\opp -q\A)\p$, it is straightforward to show that \eqref{ap1.2} is satisfied, $( \opp -q\A^g)^2\pg=( \opp -q\A^g) e^{i\frac{q}{\hbar}\g}\phi$, by re-using the result \eqref{ap1.2bis} with $\p$ substituted by $\phi=(\opp -q\A)\p$. 

Second, using \eqref{vg} in the paper, we evaluate: 
\begin{equation}
q\Vg \pg=(q\V-q\frac{\partial \g}{ \partial t})\pg=e^{i\frac{q}{\hbar}\g} \left(q\V-q\frac{\partial \g}{ \partial t}\right)\p.
\label{ap1.4}
\end{equation}
Joining \eqref{ap1.2} and \eqref{ap1.4} we get:
\begin{equation}
H^g\pg=e^{i\frac{q}{\hbar}\g} \left(H-q\frac{\partial \g}{ \partial t}\right)\p,
\label{ap1.5}
\end{equation}
which is equivalent to \eqref{hamig} written in the paper. 
Third, we evaluate $\frac{\partial \pg}{\partial t}$ as:
\begin{equation} 
\frac{\partial \pg}{\partial t}=e^{i\frac{q}{\hbar}\g}\left(\frac{\partial \p}{\partial t} + i\frac{q}{\hbar}\frac{\partial \g}{\partial t} \p\right).
\label{ap1.6}
\end{equation}
Finally,  putting \eqref{ap1.5} and \eqref{ap1.6} together and eliminating $e^{i\frac{q}{\hbar}\g}$, \eqref{scho} in the paper (The Coulomb gauge Schrödinger equation). i.e.,  the right-hand side of \eqref{ap1.1}, is exactly recovered. 

Notice that, even when $\A=0$ and $\V=0$ in the right-hand side of \eqref{ap1.1}, we get $\Ag = {\nabla} \g$ and $\Vg = -{\partial \g}/{ \partial t}$ in the left-hand side of  \eqref{ap1.1}. Thus, the electromagnetic potentials are present in the left-hand side of \eqref{ap1.1}, even when we do not write them in the (Coulomb gauge) right-hand side of \eqref{ap1.1}.  A similar development about the gauge invariance of the Schrödinger equation and the gauge dependence of the Hamiltonian can be found in many textbooks. For example, in Refs. \cite{optics,Ballentine2014,cohen1986}.

\section{Gauge dependence of the time evolution operator}
\label{ap2}

From the left-hand side  of \eqref{utg} in the paper, one gets the result:
\begin{equation}
\pg(\egx,t_R)=\langle \egx| \opG(t_R)\opUt{t_R}\opG(0)^{\dagger} \opG(0) |\p(0)\rangle=\langle \egx| \opG(t_R)\opUt{t_R} |\p(0)\rangle=e^{i\frac{q}{\hbar}\g}\p(\egx,t_R).
\label{ap2.0bis}
\end{equation}
Notice that different operators are evaluated at different times. This result is already discussed in Ref. \cite{kobe1985}. Now, we want to show that the same result \eqref{ap2.0bis} is obtained from the right-hand side of \eqref{utg} in the paper. That is:
\begin{equation}
\pg(\egx,t_R)=\langle \egx|\hat{ \mathds{T}} e^{-\frac{i}{\hbar}\int_0^{t_R} H(\Ag, \Vg) dt}|\egx \rangle \p^g(\egx,0).
\label{ap2.0}
\end{equation}
Notice that the Hamiltonian in \eqref{hamig} is explicitly time-dependent because of $\gx$. The  time ordering integral operator in \eqref{utg} can be defined as
\begin{equation}
\hat{ \mathds{T}} e^{-\frac{i}{\hbar}\int_0^{t_R} \opH (\Ag, \Vg) dt}:=\lim_{N\to\infty}\Pi_{n=0}^{N-1}e^{-i\frac{\tau\opH^g(n \tau)}{\hbar}}
\label{ap2.1}
\end{equation} 
when the time interval $[0,t_R]$ is divided into $N$ infinitesimal time steps $\tau=t_R/N$ with $\opH^g(n \tau)=\opH(\Ag(n \tau), \Vg(n \tau))$. 

To show \eqref{ap2.0}, let us start by showing its validity when $t_R=\tau$, with $\tau$ a very small time step ($N\to 0$). From \eqref{utg} in the paper and $|\pg(t_R)\rangle = \opUtg{1} |\pg(0) \rangle$, we get $\pg(\tau)=  e^{-i\frac{\tau H^g(0)}{\hbar}} \pg(0)= e^{-i\frac{\tau H^g(0)}{\hbar}} e^{i\frac{q\g(0) }{\hbar}} \p(0)$. Neglecting terms on the order $\BO( \tau^2)$,  we have $e^{-i\frac{\tau H^g(0)}{\hbar}}=\unit-i\frac{\tau}{\hbar} H^g (0)$. Then, we get $\pg(\tau)=e^{i\frac{q\g(0) }{\hbar}}\p(0) -i\frac{\tau}{\hbar}H^g (0) e^{i\frac{q\g(0) }{\hbar}} \p(0)$. From the Hamiltonian in \eqref{hamig}, and demonstrated in \eqref{ap1.5}, we have $\opH^g (0) e^{i\frac{q\g(0) }{\hbar}} \p(0)=e^{i\frac{q\g(0) }{\hbar}} \left( H(0) -q\frac{\partial \g}{\partial t}\right)  \p(0)$, where $\frac{\partial \g}{\partial t}=\left.\frac{\partial \gx}{\partial t}\right|_{t=0}$. Then, we get:
\begin{equation}
\pg(\tau)= e^{i\frac{q\g(0) }{\hbar}}\left(\unit-i\frac{\tau}{\hbar}H(0)+i\frac{q\tau}{\hbar}\frac{\partial \g}{\partial t}\right)  \p(0) . 
\label{ap2.4}
\end{equation}
Let us now show that \eqref{ap2.4} can be identically recovered from \eqref{ap2.0bis}. By using a first-order  Taylor expansion, we get  $e^{i\frac{q}{\hbar}\g(\tau)}=e^{i\frac{q}{\hbar}\g(0)}\left(\unit+i\frac{q\tau}{\hbar}\frac{\partial \g}{\partial t}\right)$, and $\opU_{\tau}= \left( \unit-i\frac{\tau}{\hbar}H(0)\right) $. Their product acting on the wave function gives:
\begin{eqnarray}
e^{i\frac{q}{\hbar}\g(\tau)}\opU_{\tau}\p(0)&=&e^{i\frac{q}{\hbar}\g(0)}\left(\unit+i\frac{q\tau}{\hbar}\frac{\partial \g}{\partial t}\right) \left( \unit-i\frac{\tau}{\hbar}H(0)\right) \p(0)\nonumber\\
&=&e^{i\frac{q}{\hbar}\g(0)} \left(\unit+i\frac{q\tau}{\hbar}\frac{\partial \g}{\partial t}-i\frac{\tau}{\hbar}H(0)+\frac{q\tau^2}{\hbar^2}\frac{\partial \g}{\partial t}H(0)\right)  \p(0) . 
\label{ap2.5}
\end{eqnarray}
Then, neglecting terms on the order $\BO( \tau^2)$, which in this case is just the last summand in the second line of \eqref{ap2.5}, we reproduce \eqref{ap2.4}. Thus, we have shown that:
\begin{equation}
\p^g(\tau)=e^{i\frac{q}{\hbar}\g(\tau)}e^{-i\frac{\tau H(0)}{\hbar}}\p(0)=e^{i\frac{q}{\hbar}\g(\tau)}\p(\tau).\nonumber 
\end{equation}
By repeating the same arguments for each infinitesimal time interval $\tau$ in \eqref{ap2.1}, we get \eqref{ap2.0bis}.
 
\section{Gauge invariance of time derivative of expectation values}
\label{ap3}

From $\BE \big(\opO,t\big||\p\rangle\big)$ defined in \eqref{ev} in the paper, its time derivative gives:
\begin{eqnarray}
\frac{\partial \BE \big(\opO,t\big||\p\rangle )}{\partial t}&=&\langle \Psi(0)|\frac{\partial \opUd(t)}{\partial t}\opO \opU(t) |\Psi(0)\rangle\nonumber\\
&+&\langle \Psi(0)| \opUd(t) \frac{\partial \opO}{\partial t}\opU(t) | \Psi(0)\rangle+\langle \Psi(0)| \opUd(t)|\opO \frac{\partial \opU(t)}{\partial t} |\Psi(0)\rangle.
\label{ap3.1}
\end{eqnarray}
Putting \eqref{dtu1}  of the paper, and its complex conjugate, in \eqref{ap3.1}, we get:
\begin{eqnarray}
&&\frac{\partial \BE \big(\opO,t \big||\p\rangle \big)}{\partial t} =\langle \Psi(t)| \left( \frac{i}{\hbar}[\opH,\opO]+\frac{\partial \opO}{\partial t} \right) | \Psi(t)\rangle =\langle \Psi(t)|\opC | \Psi(t)\rangle,\nonumber
\label{ap3.2}
\end{eqnarray}
where we have defined:  
\begin{equation}
\opC:=\frac{i}{\hbar}[\opH,\opO]+\frac{\partial \opO}{\partial t}.
\label{ap3.33}
\end{equation}
Next, we evaluate the gauge invariance of $\opC$ assuming the accomplishment of \textbf{C1} in the paper for the operator $\opO^g:=\opG\opO\opG^{\dagger}$. Using $\frac{\partial \opG(t)}{\partial t}=e^{i\frac{q}{\hbar}\opg}i\frac{q}{\hbar} \frac{\partial \opg}{\partial t}$ from the definition in the text, we get:

\begin{eqnarray}
&&\frac{i}{\hbar}[\opH^g,\opO^g]+\frac{\partial \opO^g}{\partial t}=\frac{i}{\hbar}\opH^g\opG\opO\opG^{\dagger}-\frac{i}{\hbar}\opG\opO\opG^{\dagger}\opH^g+\opG\frac{\partial \opO}{\partial t}\opG^{\dagger} +\frac{iq}{\hbar} \left( \opG\frac{\partial\opg}{\partial t}\opO\opG^{\dagger} -\opG \opO\frac{\partial \opg}{\partial t}\opG^{\dagger}  \right)\nonumber\\
&&=\frac{i}{\hbar}\opG \opH\opO \opG^{\dagger}-\frac{iq}{\hbar}\opG \frac{\partial \opg}{\partial t}\opO\opG^{\dagger}-\frac{i}{\hbar}\opG \opO \opH\opG^{\dagger}+\frac{iq}{\hbar}\opG \opO \frac{\partial \opg}{\partial t}\opG^{\dagger}+\opG\frac{\partial \opO}{\partial t}\opG^{\dagger} \\
&&+\frac{iq}{\hbar} \left( \opG\frac{\partial\opg}{\partial t}\opO\opG^{\dagger} -\opG \opO\frac{\partial \opg}{\partial t}\opG^{\dagger}  \right)
=\frac{i}{\hbar}\opG [\opH,\opO] \opG^{\dagger}+\opG\frac{\partial \opO}{\partial t}\opG^{\dagger}=\opG\left(\frac{i}{\hbar} [\opH,\opO] +\frac{\partial \opO}{\partial t}\right) \opG^{\dagger},\nonumber
\label{ap6.0}
\end{eqnarray}
Eq. (\ref{ap1.5}) has been used and written here as: $H^g \pg=H^g \opG \p=\opG \left(H-q\frac{\partial \g}{ \partial t}\right)\p=\opG \left(H-q\frac{\partial \g}{ \partial t}\right)\opG^{\dagger}\pg$.  
Thus, we showed the gauge invariance of the time derivative of the expectation value $\langle \Psi^g(t)|\opC^g | \Psi^g(t)\rangle=\langle \Psi(t)|\opC | \Psi(t)\rangle\nonumber$.  A similar development can be found in Ref. \cite{Ballentine2014,cohen1986}.

\section{Left-hand derivative (LHD) of weak values in \eqref{LHD}}
\label{ap4}

We are interested in the time derivative at the final time of the post-selection. See Fig.~\ref{f0}(c).  To simplify the notation,  in this Appendix~\ref{ap4}, we define this final time of the post-selection as $t=0$ so that weak perturbation occurs at $t=-t_L$ and the preparation at $t=-t_L-t_R$.  

\subsection{What is the general expression for the LHD? }

We have to evaluate the time derivative of $\opUt{t_L}$ and $\opO(-t_L)$ as:
\begin{eqnarray}
&&\left.\frac{\partial \BW \big(\opO,t_L,t_R\big||\egf\rangle,|\p\rangle  \big)}{\partial t_L}\right|_{t_L =0}=\Re{ \frac{\partial}{\partial t_L} \frac{\langle \evf | \opUt{t_L} \opO  \opUt{t_R}|\p\rangle }{\langle \evf |\opUt{t_L} \opUt{t_R}|\p\rangle }}\nonumber\\
&&= \Re{\frac{\langle \evf | \frac{\partial \opUt{t_L}}{\partial t_L} \opO  \opUt{t_R}\p\rangle }{\langle \evf | \opUt{t_R}|\p\rangle }
+\frac{\langle \evf  |\frac{\partial \opO}{\partial t_L} \opUt{t_R} | \p\rangle }{\langle \evf | \opUt{t_R}|\p\rangle }
-\frac{\langle \evf  |\opO \opUt{t_R}|\p\rangle }{\langle \evf | \opUt{t_R}|\p\rangle }
\frac{\langle \evf | \frac{\partial   \opUt{t_L}}{\partial t_L} \opUt{t_R}|\p\rangle }{\langle \evf | \opUt{t_R}|\p\rangle }}\nonumber.
\label{ap4.1}
\end{eqnarray}
Notice that $|\evf\rangle$ is an eigenstate of $\opF$ at time $t=0$ so that, using \eqref{utg}, we have $\opUt{t_L}=e^{-\frac{i}{\hbar}\int_{-t_L}^0 \opH dt}$. Using the Leibniz integral rule, we get $\frac{d \opUt{t_L}}{d t_L}=\frac{d }{d t_L}\left( e^{-\frac{i}{\hbar}\int_{-t_L}^0 \opH dt}\right)=e^{-\frac{i}{\hbar}\int_{-t_L}^0 \opH dt}\left(-\frac{i}{\hbar}\right) \frac{d \int_{-t_L}^0 \opH dt}{d t_L}=\opUt{t_L}\left(-\frac{i}{\hbar}\right) \opH \frac{d (-t_L)}{d t_L}=\frac{i}{\hbar}\opUt{t_L}\opH$. Notice the change of sign with respect to the \eqref{dtu1}. Then:
\begin{eqnarray}
&&\left.\frac{\partial \BW \big(\opO,t_L,t_R\big||\egf\rangle,|\p\rangle  \big)}{\partial t_L}\right|_{t_L =0}=\nonumber\\ 
&&\Re{\frac{i}{\hbar}\frac{\langle \evf | \opH \opO \opUt{t_R}|\p\rangle }{\langle \evf | \opUt{t_R}|\p\rangle }
+\frac{\langle \evf  |\frac{\partial  \opO}{\partial t_L}  \opUt{t_R}|\p\rangle }{\langle \evf | \opUt{t_R}|\p\rangle }
-\frac{i}{\hbar}\frac{\langle \evf | \opO \opUt{t_R}|\p\rangle }{\langle \evf | \opUt{t_R}|\p\rangle }
\frac{\langle \evf |\opH  \opUt{t_R}|\p\rangle }{\langle \evf | \opUt{t_R}|\p\rangle }}.
\label{ap4.2}
\end{eqnarray}
The time derivative of the weak value has some additional complexities not present in the time derivative of an expectation value: (i) the expectation value in \eqref{ev} in the paper can be defined as $\langle \psi|\hat{U}(t_R)^\dagger \hat{O} \hat{U}(t_R) |\psi\rangle/\langle \psi|\hat{U}(t_R)^\dagger \hat{U}(t_R) |\psi\rangle$, where the denominator of the expectation value is just the norm of the state, which is time-independent ( while the denominator of the weak value in \eqref{ewv} in the paper is explicitly time-dependent), (ii) Instead of $\langle \psi|\hat{U}(t_R)^\dagger$ in \eqref{ev}, we use $\langle \egf|\opUt{t_L}$ in \eqref{ap4.2}. The former $\hat{U}(t_R)^\dagger$ implies a time evolution in the initial bra state $\langle \psi(0)|$, while the latter $\opUt{t_L}$ implies a time evolution of the corresponding ket state.

By using $\opH\opO=[\opH,\opO]+\opO\opH$, we can rewrite  \eqref{ap4.2} as:
\begin{eqnarray}
&&\left.\frac{\partial \BW \big(\opO,t_L,t_R\big||\egf\rangle,|\p\rangle  \big)}{\partial t_L}\right|_{t_L =0}=\nonumber\\ 
&&\Re{ \frac{\langle \evf | \opC \opUt{t_R}|\p\rangle }{\langle \evf | \opUt{t_R}|\p\rangle }}
+\Re{\frac{i}{\hbar}\frac{\langle \evf | \opO  \opH \opUt{t_R}|\p\rangle }{\langle \evf | \opUt{t_R}|\p\rangle }-\frac{i}{\hbar}\frac{\langle \evf |\opO  \opUt{t_R}|\p\rangle }{\langle \evf | \opUt{t_R}|\p\rangle }
\frac{\langle \evf |\opH \opUt{t_R}|\p\rangle }{\langle \evf | \opUt{t_R}|\p\rangle }},
\label{ap4.3}
\end{eqnarray}
where $\opC:=\frac{i}{\hbar}[\opH,\opO]+\frac{\partial \opO}{\partial t}$. Thus, in general, the time derivative of a weak value given by \eqref{ap4.3} does not have the shape of a weak value given by \eqref{ewv}. 

\subsection{What is the gauge condition of the LHD? }

The discussion about the gauge invariance of \eqref{ap4.3} requires, first, to acknowledge that $\langle \evf | \opC |\p\rangle $ is gauge invariant when $\opF$ and $\opO$ satisfy \textbf{C1} and \textbf{C3}. The same is true for $\langle \evf | \opC  \opUt{t_R}|\p\rangle $ by just interpreting $\opUt{t_R}|\p\rangle $ as a new wave function. Thus, we have to check the gauge dependence of the other two terms in the right hand side of \eqref{ap4.3}. We define $\Delta G$ as the difference between these two terms written in the Coulomb gauge (no superindex) and in a different gauge (superindex $g$) as: 
\begin{eqnarray}
\Delta G&=&\frac{\langle \evf | \opO  \opH \opUt{t_R}|\p\rangle }{\langle \evf | \opUt{t_R}|\p\rangle }-\frac{\langle \evf |\opO  \opUt{t_R}|\p\rangle }{\langle \evf | \opUt{t_R}|\p\rangle }
\frac{\langle \evf |\opH  \opUt{t_R}|\p\rangle }{\langle \evf | \opUt{t_R}|\p\rangle }\nonumber\\
&-&\left(\frac{\langle \evf^g | \opO^g  \opH^g  \opUtg{1}|\p^g\rangle }{\langle \evf^g | \opUtg{1}|\p^g\rangle }-\frac{\langle \evf^g |\opO^g \opUtg{1}|\p^g\rangle }{\langle \evf^g | \opUtg{1}|\p^g\rangle }\frac{\langle \evf^g |\opH^g \opUtg{1}|\p^g\rangle }{\langle \evf^g | \opUtg{1}|\p^g\rangle }\right).
\label{ap4.4}
\end{eqnarray}
We get $\langle \evf | \opUt{t_R}|\p\rangle =\langle \evf^g | \opUtg{1}|\p^g\rangle $ and  $\langle \evf |\opO  \opUt{t_R}|\p\rangle =\langle \evf^g |\opO^g \opUtg{1}|\p^g\rangle $. Now we evaluate $\langle \evf^g | \opO^g  \opH^g  \opUtg{1}|\p^g\rangle $ using \eqref{hamig} in the paper, $\langle \evf^g | \opO^g  \opH^g \opUt{t_R}|\p^g\rangle =\langle \evf | \opO  \opH \opUt{t_R}|\p\rangle -q\int d\egx \langle \evf | \opO |\egx\rangle\frac{\partial \gx}{\partial t} \langle \egx |\opUt{t_R}|\p \rangle$. Identically, we have: $\langle \evf^g | \opH^g  \opUtg{1}|\p^g\rangle =\langle \evf | \opH \opUt{t_R}|\p\rangle -q\int d\egx \langle \evf  |\egx\rangle\frac{\partial  \gx}{\partial t} \langle \egx |\opUt{t_R}|\p \rangle$. Then \eqref{ap4.4} can  be rewritten as:
\begin{eqnarray}
\Delta G&=&q\frac{\int d\egx  \left( \langle \evf | \opO |\egx\rangle-\frac{\langle \evf |\opO \opUt{t_R}|\p\rangle }{\langle \evf | \opUt{t_R}|\p\rangle } \langle \evf  |\egx\rangle	 \right)\frac{\partial  \gx}{\partial t} \langle \egx |\opUt{t_R}|\p \rangle}{\langle \evf | \opUt{t_R}|\p\rangle }
\label{ap4.5}
\end{eqnarray}
Thus, in general, the left hand time derivative of a weak value is gauge dependent. 

\subsection{Necessary and sufficient condition for the LHD being a gauge invariant weak value}
 
A sufficient condition to ensure $\Delta G=0$ in \eqref{ap4.5} is to ensure that $\langle \evf | \opO = o\langle \evf |$ which means that $| \evf \rangle$ is an eigenstate of $\opO$, i.e. $[\opF,\opO]=0$. If    $[\opF,\opO]=0$ then $\Delta G=0$. We are now showing that it is also a necessary condition: if   $\Delta G=0$ then $[\opF,\opO]=0$. Since $\gx$ in \eqref{ap4.5} can be any function, we select $\frac{\partial  g_1(\egx,t)}{\partial t} = \delta_\epsilon(\egx-\egx_1)\frac {\langle \evf | \opUt{t_R}|\p\rangle} { q\langle \egx |\opUt{t_R}|\p \rangle}$ where $\delta_\epsilon(\egx-\egx_1)$ is a function that is as close as possible to a delta (functional) function. Then,  \eqref{ap4.5} becomes $\Delta G_1=\langle \evf | \opO |\egx_1\rangle-\frac{\langle \evf |\opO \opUt{t_R}|\p\rangle }{\langle \evf | \opUt{t_R}|\p\rangle }\langle \evf  |\egx_1\rangle$. Then, the assumption $\Delta G_1=0$ implies $\frac{\langle \evf | \opO |\egx_1\rangle}{\langle \evf  |\egx_1\rangle}=\frac{\langle \evf |\opO \opUt{t_R}|\p\rangle }{\langle \evf | \opUt{t_R}|\p\rangle }$. If we select another gauge function $\frac{\partial  g_2(\egx,t)}{\partial t} = \delta_\epsilon(\egx-\egx_2)\frac {\langle \evf | \opUt{t_R}|\p\rangle} { q\langle \egx |\opUt{t_R}|\p \rangle}$, then  $\Delta G_2=0$ implies $\frac{\langle \evf | \opO |\egx_2\rangle}{\langle \evf  |\egx_2\rangle}=\frac{\langle \evf |\opO \opUt{t_R}|\p\rangle }{\langle \evf | \opUt{t_R}|\p\rangle }$. But, the results $\Delta G_1=\Delta G_2=0$ for any pairs of points $\egx_1$ and $\egx_2$ implies $\frac{\langle \evf | \opO |\egx_1\rangle}{\langle \evf  |\egx_1\rangle}=\frac{\langle \evf | \opO |\egx_2\rangle}{\langle \evf  |\egx_2\rangle}$. This last identity can only be true if the dependence on position $\egx$ disappears. This can only happen if  $\langle \evf | \opO =\egf \langle \evf |$, i.e  $[\opO,\opF]=0$. In conclusion, \textbf{C4} mentioned in the text is a necessary and sufficient condition to ensure that LHD is gauge invariant (we have disregarded uninteresting results as $\langle \egf|$ being the zero state or $\opO=\unit$). 

In summary, when condition \textbf{C4} is satisfied, \eqref{ap4.2} is gauge invariant  (can be measured in the laboratory) and it can be compactly written as a weak value:
\begin{eqnarray}
\left.\frac{\partial \BW \big(\opO,t_L,t_R\big||\egf\rangle,|\p\rangle  \big)}{\partial t_L}\right|_{t_L =0}=\Re{ \frac{\langle \evf | \opC \opUt{t_R}|\p\rangle }{\langle \evf | \opUt{t_R}|\p\rangle }}\nonumber.
\label{ap4.7}
\end{eqnarray}
We have used $\langle \evf |\opO \opUt{t_R}|\p\rangle =\ego \langle \evf | \opUt{t_R}|\p\rangle $ and $\ego\langle \evf |\opH  \opUt{t_R}\p\rangle =\langle \evf |\opO \opH  \opUt{t_R}|\p\rangle $ and the definitions in \eqref{ap3.33}.

\section{Right-hand derivative (RHD) of weak values in \eqref{RHD}}
\label{ap5}

From the definition of $\BW \big(\opO,t_L,t_R\big||\egf\rangle,|\p\rangle  \big)$ in \eqref{ewv}, we compute the right time derivative by using the time axis as seen  Fig.~\ref{f0}(d). 

\subsection{What is the general expression for the RHD? }

We get:
\begin{eqnarray}
&&\left.\frac{\partial \BW \big(\opO,t_L,t_R\big||\egf\rangle,|\p\rangle  \big)}{\partial t_R}\right|_{ t_R =0}=\Re{\frac{\partial}{\partial t_R} \frac{\langle \evf | \opUt{t_L} \opO  \opUt{t_R}|\p\rangle }{\langle \evf |\opUt{t_L} \opUt{t_R}|\p\rangle }}\nonumber\\
&&= \Re{\frac{\langle \evf |\opUt{t_L} \opO  \frac{\partial  \opUt{t_R}}{\partial t_R} \p\rangle }{\langle \evf | \opUt{t_L}|\p\rangle }
+\frac{\langle \evf | \opUt{t_L} \frac{\partial  \opO}{\partial t_R} \p\rangle }{\langle \evf | \opUt{t_L}|\p\rangle }
-\frac{\langle \evf | \opUt{t_L} \opO | \p\rangle }{\langle \evf | \opUt{t_L}|\p\rangle }
\frac{\langle \evf | \opUt{t_L} \frac{\partial  \opUt{t_R}}{\partial t_R}|\p\rangle }{\langle \evf | \opUt{t_L}|\p\rangle }}\nonumber.
\label{ap5.1}
\end{eqnarray}
We use $ t_R =0$ to indicate that we are approaching $t=0$ by the right as seen in Fig.~\ref{f0}(d). Notice that $|\evf\rangle$ is an eigenstate of $\opF$ at time $t=t_R+t_L$. Using \eqref{dtu1} in the text, where all time derivatives are evaluated at a time equal to zero, we get:
\begin{eqnarray}
&&\left.\frac{\partial \BW \big(\opO,t_L,t_R\big||\egf\rangle,|\p\rangle  \big)}{\partial t_R}\right|_{ t_R =0}\nonumber\\
&&=\Re{ -\frac{i}{\hbar}\frac{\langle \evf |\opUt{t_L} \opO  \opH |\p\rangle }{\langle \evf | \opUt{t_L}|\p\rangle }
+\frac{\langle \evf | \opUt{t_L} \frac{\partial  \opO}{\partial t_R} | \p\rangle }{\langle \evf | \opUt{t_L}|\p\rangle }
+\frac{i}{\hbar}\frac{\langle \evf |\opUt{t_L} \opO | \p\rangle }{\langle \evf | \opUt{t_L}|\p\rangle }
\frac{\langle \evf |\opUt{t_L}  \opH |\p\rangle }{\langle \evf | \opUt{t_L} |\p\rangle }}.
\label{ap5.2}
\end{eqnarray}
By using $-\opO\opH=[\opH,\opO]-\opH\opO$, we can rewrite  \eqref{ap5.2} as:
\begin{eqnarray}
&&\left.\frac{\partial \BW \big(\opO,t_L,t_R\big||\egf\rangle,|\p\rangle  \big)}{\partial t_R}\right|_{ t_R =0}\nonumber\\
&&=\Re{ \frac{\langle  \egf | \opUt{t_L} \opC| \p\rangle }{\langle  \egf | \opUt{t_L}| \p\rangle }
-\frac{i}{\hbar}\frac{\langle  \egf | \opUt{t_L} \opH \opO | \p\rangle }{\langle  \egf | \opUt{t_L}| \p\rangle }+\frac{i}{\hbar}\frac{\langle  \egf | \opUt{t_L}\opO | \p\rangle }{\langle  \egf | \opUt{t_L}| \p\rangle }
\frac{\langle  \egf | \opUt{t_L}\opH |\p\rangle }{\langle  \egf | \opUt{t_L}| \p\rangle }},
\label{ap5.3}
\end{eqnarray}
where $\opC:=\frac{i}{\hbar}[\opH,\opO]+\frac{\partial \opO}{\partial t}$.  In general, the final \eqref{ap5.3} of the time derivative of a weak value is not a weak value. 

\subsection{What is the gauge condition of the RHD? }

We study now if \eqref{ap5.3} is a gauge invariant expression or not. One can show that $\langle  \egf | \opUt{t_L} \opC| \p\rangle $ is gauge invariant when $\opF$ and $\opO$ satisfy \textbf{C1} and \textbf{C3}. Similarly to the previous appendix, in general, the other terms in \eqref{ap5.3} are gauge dependent. We define the difference of these terms in different gauges as $\Delta G$, which is given by: 
\begin{eqnarray}
\Delta G&=&\frac{\langle \evf |  \opUt{t_L} \opH \opO |\p\rangle }{\langle \evf | \opUt{t_L}|\p\rangle }-\frac{\langle \evf |\opUt{t_L} \opO  |\p\rangle }{\langle \evf | \opUt{t_L}|\p\rangle }
\frac{\langle \evf |\opUt{t_L} \opH |\p\rangle }{\langle \evf | \opUt{t_L}|\p\rangle }\nonumber\\
&-&\left(\frac{\langle \evf^g | \opUtg{t_L}  \opH^g  \opO^g  |\p^g\rangle }{\langle \evf^g | \opUtg{t_L}|\p^g\rangle }-\frac{\langle \evf^g |\opUtg{t_L} \opO^g|\p^g\rangle }{\langle \evf^g | \opUtg{t_L}|\p^g\rangle }\frac{\langle \evf^g |\opUtg{t_L} \opH^g |\p^g\rangle }{\langle \evf^g | \opUtg{t_L}|\p^g\rangle }\right).
\label{ap5.4}
\end{eqnarray}
We get $\langle \evf | \opUt{t_L}|\p\rangle =\langle \evf^g | \opUtg{t_L}|\p^g\rangle $ and  $\langle \evf |\opUt{t_L} \opO |\p\rangle =\langle \evf^g |\opUtg{t_L} \opO^g |\p^g\rangle $. Now we evaluate $\langle \evf^g | \opUtg{t_L} \opH^g \opO^g    |\p^g\rangle $ using \eqref{hamig} and \eqref{utg} in the paper, to get $\langle \evf^g | \opUtg{t_L} \opH^g \opO^g   |\p^g\rangle =\langle \evf | \opUt{t_L} \opH \opO  |\p\rangle -q\int d\egx \langle \evf |\opUt{t_L} |\egx\rangle\frac{\partial \gx}{\partial t} \langle \egx | \opO |\p \rangle$. Identically, we have $\langle \evf^g | \opUtg{t_L} \opH^g |\p^g\rangle =\langle \evf | \opUt{t_L} \opH |\p\rangle -q\int d\egx \langle \evf | \opUt{t_L}|\egx\rangle\frac{\partial  \gx}{\partial t} \langle \egx |\p \rangle$. Then \eqref{ap5.4} can  be rewritten as:
\begin{eqnarray}
\Delta G&=&q\frac{\int d\egx  \left( \langle \egx | \opO|\p\rangle-\frac{\langle \evf |\opUt{t_L} \opO |\p\rangle }{\langle \evf | \opUt{t_L}|\p\rangle } \langle \egx |\p\rangle	 \right)\frac{\partial  \gx}{\partial t} \langle \egf | \opUt{t_L}|\egx \rangle}{\langle \evf | \opUt{t_L}|\p\rangle }
\label{ap5.5}
\end{eqnarray} 
Thus, in general, the right hand time derivative of a weak value is gauge dependent. 

\subsection{Necessary and sufficient condition for RHD being a gauge invariant weak value}

 A sufficient condition to ensure $\Delta G=0$ in \eqref{ap5.5} is to ensure that $| \opO|\p\rangle = o|\p\rangle$ which means that $|\p\rangle$ is an eigenstate of $\opO$, i.e. $[\opphi,\opO]=0$. If    $[\opphi,\opO]=0$ then $\Delta G=0$. We are now showing that it is also a necessary condition: if   $\Delta G=0$ then $[\opphi,\opO]=0$. Since $\gx$ in \eqref{ap5.5} can be any function, we select $\frac{\partial  g_1(\egx,t)}{\partial t} = \delta_\epsilon(\egx-\egx_1)\frac {\langle \evf | \opUt{t_L}|\p\rangle} { q\langle \evf | \opUt{t_L}|\egx\rangle}$ where $\delta_\epsilon(\egx-\egx_1)$ is a function that is as close as possible to a delta (functional) function. Then,  \eqref{ap5.5} becomes $\Delta G_1=\langle \egx_1 | \opO |\p\rangle-\frac{\langle \evf |\opUt{t_L}\opO |\p\rangle }{\langle \evf | \opUt{t_L}|\p\rangle }\langle \egx_1  |\p \rangle$ and $\Delta G_1=0$ implies $\frac{\langle \egx_1 | \opO |\p\rangle}{\langle \egx_1  |\p \rangle}=\frac{\langle \evf |\opUt{t_L}\opO |\p\rangle }{\langle \evf | \opUt{t_L}|\p\rangle }$. If we select another gauge function $\frac{\partial  g_2(\egx,t)}{\partial t} = \delta_\epsilon(\egx-\egx_2)\frac {\langle \evf | \opUt{t_L}|\p\rangle} { q\langle \evf | \opUt{t_L}|\egx\rangle}$, then  $\Delta G_2=0$ implies $\frac{\langle \egx_2 | \opO |\p\rangle}{\langle \egx_2  |\p \rangle}=\frac{\langle \evf |\opUt{t_L}\opO |\p\rangle }{\langle \evf | \opUt{t_L}|\p\rangle }$. But, the results $\Delta G_1=\Delta G_2=0$ for any pairs of points $\egx_1$ and $\egx_2$ implies $\frac{\langle \egx_1 | \opO |\p\rangle}{\langle \egx_1  |\p \rangle}=\frac{\langle \egx_2 | \opO |\p\rangle}{\langle \egx_2  |\p \rangle}$. This last identity can only be true if the dependence on position $\egx$ disappears. This can only happen if  $\opO |\p\rangle =\egphi  |\p\rangle$, i.e  $[\opO,\opphi]=0$. In conclusion, \textbf{C5} mentioned in the text is a necessary and sufficient condition to ensure that RHD is gauge invariant (we have disregarded uninteresting results as $|\p\rangle$ being the zero state or $\opO=\unit$).  
 
In summary, when condition \textbf{C5} is satisfied, \eqref{ap5.2} is gauge invariant  (can be measured in the laboratory) and it can be compactly written as a weak value:
\begin{eqnarray}
&&\left.\frac{\partial \BW \big(\opO,t_L,t_R\big||\egf\rangle,|\p\rangle  \big)}{\partial t_R}\right|_{ t_R =0}= \Re{\frac{\langle  \egf | \opUt{t_L} \opC| \p\rangle }{\langle  \egf | \opUt{t_L}|\p\rangle }}\nonumber,
\end{eqnarray}

\section{Finite-difference left-hand derivative (FDLHD) of weak values in \eqref{FDLHD}}
\label{ap4b}

As discussed in Appendix~\ref{ap4}, to simplify the notation in the computation of FDLHD, it is more convenient to define $t=0$ as the final time when the post-selection takes place, which is the time where the time-derivative is evaluated.  We consider two weak values whose difference is the time interval between the weak perturbation and the post-selection. The time interval $t_R$ is fixed in both. In the first weak value, such a time interval is $0$ and in the second is $t_L$. Notice that we define $t=0$ as the final time (post-selected), as seen in Fig.~\ref{f0}(c), we define this final time of the post-selection as $t=0$ so that weak perturbation occurs at $t=-t_L$ and the preparation at $t=-t_L-t_R$. For example, the time $t0-t_L$ is negative, but the interval time $t_L$ is positive.  
 
\begin{eqnarray}
\text{FDLHD} &=&\frac{ \BW\big(\opO,0,t_R\big||\evf\rangle,|\p\rangle \big)-\BW\big(\opO,-t_L,t_R\big||\evf\rangle,|\p\rangle \big)}{t_L}.\nonumber
 \label{ap4.1b} 
\end{eqnarray} 

\subsection{What is the gauge condition of the FDLHD? }

To further elaborate this expression, we assume that the experimental weak value (not its time-derivative) can be approximated by its theoretical weak value, so that:
\begin{eqnarray}
\text{FDLHD} &\approx &\frac{1}{t_L} \left( \Re{  \frac{\langle \evf | \opO  \opUt{t_R}|\p\rangle }{\langle \evf | \opUt{t_R}|\p\rangle }}-\Re{  \frac{\langle \evf| \opUt{t_L} \opO(-t_L)  \opUt{t_R}|\p\rangle }{\langle \evf | \opUt{t_L} \opUt{t_R}|\p\rangle }}\right).
 \label{ap4.2b} 
\end{eqnarray} 
as far as \textbf{C1}, \textbf{C2}, and \textbf{C3} are satisfied, it is quite simple to realize that $\Re{  \frac{\langle \evf| \opUt{t_L} \opO(-t_L)  \opUt{t_R}|\p\rangle }{\langle \evf | \opUt{t_L} \opUt{t_R}|\p\rangle }}$ in \eqref{ap4.2b} is gauge invariant. The denominator is gauge invariant by construction so that only the gauge invariance of the numerator needs to be checked. Using \eqref{utg}, we can rewrite the inner product $\langle \evf | \opUt{t_L } \opO(-t_L )  \opUt{t_R}|\p\rangle$ in another gauge as:
\begin{eqnarray}
&&\langle \evf^g | \opUtg{t_L } \opO^g(-t_L )  \opUtg{t_R}|\pg\rangle=\nonumber\\
&&\langle \evf |\cancel{\opG^{\dagger}(0)} \cancel{\opG(0) }\opUt{t_L }\cancel{\opG^{\dagger}(-t_L ) }\cancel{\opG(-t_L )} \opO(-t_L ) \cancel{\opG^{\dagger}(-t_L )} \cancel{\opG(-t_L )} \opUt{t_R}\cancel{\opG^{\dagger}(-t_L -t_R)}\cancel{\opG(-t_L -t_R)}|\p\rangle\nonumber\\
&&=\langle \evf | \opUt{t_L } \opO(-t_L )  \opUt{t_R}|\p\rangle.
 \label{ap4.3b} 
\end{eqnarray}
Of course, this is exactly the same argumentation that we have already elaborated to show that the weak values in \eqref{ewv} is gauge invariant. This result states that the subtraction of two gauge-invariant numbers is also a gauge-invariant number.

\subsection{What is the general expression of the FDLHD? }

We want to evaluate the term \eqref{ap4.3b}, and similar ones present in \eqref{ap4.2b}, for small $t_L$. Since FDLHD is gauge invariant (the outputs are independent of any gauge), one can use the approximation of the time evolution operator $\opUt{t_L } = e^{-\frac{i}{\hbar}\int^0\_{-t_L } \opH , dt}$ in the Coulomb gauge for small $t_L$, giving:
\begin{eqnarray}
\opUt{t_L } = e^{-\frac{i}{\hbar}\int^0_{-t_L } \opH , dt} \approx \unit - t_L \frac{i}{\hbar} \opH.
\label{utgfd}
\end{eqnarray}

However, to fully appreciate the differences between LHD and FDLHD, it is important to realize that the gauge dependence in the left-hand term in \eqref{utgfd}, i.e., $\opUt{t_L }$, given by Eq. (6), is different from the gauge dependence of the right-hand side term in \eqref{utgfd} because of the presence of $\opH$, whose gauge dependence is given by Eq. (4). This result is not surprising, since \eqref{utgfd} is merely an approximation, not an identity. Let us clarify this point by rewriting the exact expression of the time-evolution operator for an arbitrary gauge:
\begin{eqnarray}
e^{i\frac{q}{\hbar}\opg(t_L)} \opUt{t_L }e^{-i\frac{q}{\hbar}\opg(0)}&=& \lim_{M,N\to\infty} e^{i\frac{q}{\hbar}\left( \sum_{m=0}^{M} \tau_M \left.\frac{d^{m}\opg(t)}{dt^{m}}\right|_{t=0} \right)}\left(\Pi_{n=0}^{N-1}e^{-i\tau_N \frac{\opH^g(n \tau_N)}{\hbar}}\right) e^{-i\frac{q}{\hbar}\opg(0)}\nonumber\\
\label{utgfd2}
\end{eqnarray}
with $\tau_M=t_L/M$ and $\tau_N=t_L/N$. The gauge dependence of the left- and right-hand sides of \eqref{utgfd2} is the same, but this is no longer true when we approximate the right-hand side, for $N = 1$ and $M = 1$, as:
\begin{eqnarray}
e^{i\frac{q}{\hbar}\opg(t_L)} \opUt{t_L }e^{-i\frac{q}{\hbar}\opg(0)}&\approx& e^{i\frac{q}{\hbar}\opg(0)}\left(\unit+i\frac{q}{\hbar} t_L \left.\frac{d\opg(t)}{dt}\right|_{t=0} \right) \left(\unit-t_L \frac{i}{\hbar} \opH \right) e^{-i\frac{q}{\hbar}\opg(0)}
\end{eqnarray}
Since \eqref{ap4.2b} is gauge invariant, we choose the simplest Coulomb gauge approximation, given by \eqref{utgfd}, to evaluate it. The only caveat in making this approximation is that it introduces an artificial gauge dependence in the subsequent terms. However, this modification is not significant in the present context, as we are not concerned with the gauge invariance of \eqref{ap4.2b} itself, but rather with its (approximate) structural form. Then,  we can rewrite:
\begin{equation}
\langle \evf | \opUt{t_L } \opO(-t_L )  \opUt{t_R}|\p\rangle\approx\langle \evf |  \opO(-t_L ) \opUt{t_R}|\p\rangle -\frac{i}{\hbar} t_L  \langle \evf |  \opH \opO(-t_L )  \opUt{t_R} |\p\rangle.
 \label{ap4.4b} 
\end{equation}
We also assume that $\opO({-t_L })=\opO-\frac{\partial \opO}{\partial t} t_L $ . Finally, we can rewrite:
\begin{equation}
\langle \evf | \opUt{t_L } \opO(-t_L ) \opUt{t_R}|\p\rangle\approx\langle \evf |  \opO  \opUt{t_R}|\p\rangle-t_L  \langle \evf | \frac{\partial \opO}{\partial t} \opUt{t_R}|\p\rangle -\frac{i}{\hbar} t_L  \langle \evf |  \opH \opO  \opUt{t_R}|\p\rangle,
 \label{ap4.5b} 
\end{equation}
where we disregard terms of the order of $(t_L )^2$. Identically, we get:
\begin{eqnarray}
\frac{1}{\langle \evf | \opUt{t_L } \opUt{t_R}|\p\rangle}&&\approx\frac{1}{\langle \evf | \opUt{t_R}|\p\rangle -\frac{i}{\hbar} t_L  \langle \evf |  \opH  \opUt{t_R}|\p\rangle}\approx \frac{1}{\langle \evf | \opUt{t_R}|\p\rangle \left(1 - \frac{i}{\hbar} t_L  \frac{\langle \evf |  \opH  \opUt{t_R}|\p\rangle}{\langle \evf | \opUt{t_R}|\p\rangle}\right)}\nonumber\\
&&\approx\frac{1}{\langle \evf | \opUt{t_R}|\p\rangle}  \left(1 + \frac{i}{\hbar} t_L  \frac{\langle \evf |  \opH  \opUt{t_R}|\p\rangle}{\langle \evf | \opUt{t_R}|\p\rangle}\right),
\label{ap4.6b}
\end{eqnarray}
where we have used the approximation $\frac{1}{1-x} \approx 1+x$ when $x\ll1$. Putting \eqref{ap4.5b}  and \eqref{ap4.6b}  into \eqref{ap4.4b}, we get:
\begin{eqnarray}
&&\Re{  \frac{\langle \evf | \opUt{t_L } \opO(-t_L )  \opUt{t_R}|\p\rangle }{\langle \evf |\opUt{t_L } \opUt{t_R}|\p\rangle }}\nonumber\\
&&\approx \Re{ \left(   \frac{\langle \evf |  \opO  \opUt{t_R}|\p\rangle}{\langle \evf | \opUt{t_R}|\p\rangle }-t_L  \frac{\langle \evf |  \frac{\partial \opO}{\partial t} \opUt{t_R}|\p\rangle}{\langle \evf | \opUt{t_R}|\p\rangle } - \frac{i}{\hbar} t_L  \frac{ \langle \evf |  \opH \opO  \opUt{t_R}|\p\rangle}{\langle \evf | \opUt{t_R}|\p\rangle }\right) \left(1 + \frac{i}{\hbar} t_L  \frac{\langle \evf |  \opH  \opUt{t_R}|\p\rangle}{\langle \evf | \opUt{t_R}|\p\rangle}\right)} 
\nonumber\\
&&\approx \Re{   \frac{\langle \evf |  \opO  \opUt{t_R}|\p\rangle}{\langle \evf | \opUt{t_R}|\p\rangle }-t_L  \frac{\langle \evf |  \frac{\partial \opO}{\partial t} \opUt{t_R}|\p\rangle}{\langle \evf | \opUt{t_R}|\p\rangle } - \frac{i}{\hbar} t_L  \frac{ \langle \evf |  \opH \opO  \opUt{t_R}|\p\rangle}{\langle \evf | \opUt{t_R}|\p\rangle }+ \frac{i}{\hbar} t_L  \frac{\langle \evf |  \opO  \opUt{t_R}|\p\rangle}{\langle \evf | \opUt{t_R}|\p\rangle }  \frac{\langle \evf |  \opH  \opUt{t_R}|\p\rangle}{\langle \evf | \opUt{t_R}|\p\rangle}},\nonumber 
 \label{ap4.7b} 
\end{eqnarray} 
where we have disregarded the terms of the order of $(t_L )^2$.  By subtracting the two weak values, \eqref{ap4.2b}, and dividing by $t_L$ we get:
\begin{eqnarray}
\text{FDLHD} &&\approx \frac{1}{t_L} \left( \Re{  \frac{\langle \evf | \opO  \opUt{t_R}|\p\rangle }{\langle \evf | \opUt{t_R}|\p\rangle }}-\Re{  \frac{\langle \evf| \opUt{t_L} \opO(-t_L)  \opUt{t_R}|\p\rangle }{\langle \evf | \opUt{t_L} \opUt{t_R}|\p\rangle }}\right)\nonumber\\
 &&\approx \cancel{\frac{1}{t_L}}  \Re{  \cancel{t_L} \frac{\langle \evf |  \frac{\partial \opO}{\partial t} \opUt{t_R}|\p\rangle}{\langle \evf | \opUt{t_R}|\p\rangle } + \frac{i}{\hbar} \cancel{t_L} \frac{ \langle \evf |  \opH \opO  \opUt{t_R}|\p\rangle}{\langle \evf | \opUt{t_R}|\p\rangle }- \frac{i}{\hbar} \cancel{t_L}\frac{\langle \evf |  \opO  \opUt{t_R}|\p\rangle}{\langle \evf | \opUt{t_R}|\p\rangle }  \frac{\langle \evf |  \opH  \opUt{t_R}|\p\rangle}{\langle \evf | \opUt{t_R}|\p\rangle}}. \nonumber
  \label{ap4.9b} 
\end{eqnarray} 
By using  $\opH\opO=[\opH,\opO]+\opO\opH$ and $\opC:=\frac{i}{\hbar}[\opH,\opO]+\frac{\partial \opO}{\partial t}$, we can rewrite:
\begin{eqnarray}
\text{FDLHD}\approx \Re{ \frac{\langle \evf | \opC \opUt{t_R}|\p\rangle }{\langle \evf | \opUt{t_R}|\p\rangle }}+ \Re{   \frac{i}{\hbar}  \frac{ \langle \evf | \opO  \opH   \opUt{t_R}|\p\rangle}{\langle \evf | \opUt{t_R}|\p\rangle }- \frac{i}{\hbar} \frac{\langle \evf |  \opH  \opUt{t_R}|\p\rangle}{\langle \evf | \opUt{t_R}|\p\rangle}\frac{\langle \evf |  \opO  \opUt{t_R}|\p\rangle}{\langle \evf | \opUt{t_R}|\p\rangle } }. \nonumber\\
 \label{ap4.10b} 
\end{eqnarray}
The final result \eqref{ap4.10b} is mathematically equivalent to \eqref{ap4.3}, as both have been derived in the Coulomb gauge. However, there is an important distinction. If we wish to evaluate \eqref{ap4.10b} in a different gauge, we must apply the new gauge to  \eqref{ap4.2b}, which, as shown in \eqref{ap4.3b}, is gauge invariant. In other words, as explained in \eqref{utgfd} and \eqref{utgfd2}, the gauge-dependent term $\langle \evf |\opH  \opUt{t_R}|\p\rangle$ in \eqref{ap4.10b} is merely a numerical approximation to the exact gauge-independent term $\langle \evf | \opUt{t_L} \opO(-t_L)  \opUt{t_R}|\p\rangle$ in \eqref{ap4.2b}. In contrast, the gauge-dependent term $\langle \evf |\opH  \opUt{t_R}|\p\rangle$ in \eqref{ap4.1} is an exact result of LHD and inherently carries its gauge dependence.

\subsection{Necessary and sufficient condition for FDLHD being a weak value}

In this context, the final result \eqref{ap4.10b} is mathematically equivalent to \eqref{ap4.3}. Generally, however, the approximate expression of FDLHD does not naturally take the form of a weak value. Nevertheless, by extending the reasoning applied to LHD, we can conclude that FDLHD can also be compactly expressed as a weak value when condition \textbf{C4} is satisfied. A summary of these findings regarding the gauge condition and the structural form of FDLHD is provided in Table \ref{summary}.

\section{Finite-difference right-hand derivative (FDRHD) of weak values in \eqref{FDRHD}}
\label{ap5b}

We compute the finite-difference right-hand derivative of weak values in~\eqref{FDRHD}. We consider two weak values whose difference is the time interval between the preparation (or pre-selection) and the weak perturbation (the time $t_L$ is fixed in both). In the first weak value, such time interval is $t_R$ and in the second is $0$. Notice that we define $t=0$ as the initial time (pre-selection), as seen in Fig.~\ref{f0}(e).   
\begin{eqnarray}
\text{FDRHD} &=&\frac{ \BW\big(\opO,t_L,t_R\big||\evf\rangle,|\p\rangle \big)-\BW\big(\opO,t_L,0\big||\evf\rangle,|\p\rangle \big)}{t_R}.\nonumber
 \label{ap5.1b} 
\end{eqnarray} 

\subsection{What is the gauge condition of the FDRHD? }

To further elaborate this expression, we assume that the weak value can be written as in \eqref{ewv}, so that:
\begin{eqnarray}
\text{FDRHD} &\approx &\frac{1}{t_R} \left( \Re{  \frac{\langle \evf | \opUt{t_L} \opO(t_R)  \opUt{t_R}|\p\rangle }{\langle \evf |\opUt{t_L} \opUt{t_R}|\p\rangle }}-\Re{  \frac{\langle \evf |\opUt{t_L} \opO  |\p\rangle }{\langle \evf | \opUt{t_L} |\p\rangle }}\right),
 \label{ap5.2b} 
\end{eqnarray} 
as far as \textbf{C1}, \textbf{C2}, and \textbf{C3} are satisfied, it is quite simple to realize that \eqref{ap5.2b} is gauge invariant. We have to follow the same argument done in \eqref{ap4.2b}. For example, using \eqref{utg}, we can rewrite the inner product $\langle \evf |\opUt{t_L} \opO(t_R)  \opUt{t_R}|\p\rangle $ in another gauge as:
\begin{eqnarray}
&&\langle \evf^g | \opUtg{t_L} \opO^g(t_R)  \opUtg{t_R}|\pg\rangle=\nonumber\\
&&\langle \evf |\cancel{\opG^{\dagger}(t_L+t_R)} \cancel{\opG(t_L+t_R) }\opUt{t_L}\cancel{\opG^{\dagger}(t_R) }\cancel{\opG(t_R)} \opO(t_R) \cancel{\opG^{\dagger}(t_R)} \cancel{\opG(t_R)} \opUt{t_R}\cancel{\opG^{\dagger}(0)}\cancel{\opG(0)}|\p\rangle\nonumber\\
&&=\langle \evf | \opUt{t_L} \opO(t_R)  \opUt{t_R}|\p\rangle.\nonumber
 \label{ap5.3b} 
\end{eqnarray}
Again the demonstration why FDRHD is gauge invariant follows exactly the same demonstration that we have already elaborated to show that the weak values in \eqref{ewv} is gauge invariant. 																																											  

\subsection{What is the general shape of the FDRHD? }

Taking into account the reflection done in \eqref{utgfd2} and \eqref{utgfd2},  since \eqref{ap5.2b} is gauge invariant, we select the simplest Coulomb gauge to evaluate this expression further.  We assume that $t_R$ is small enough so that, using \eqref{utg}, can be evaluated as $\opUt{t_R}= e^{-\frac{i}{\hbar}\int^{t_R}_{0} \opH dt}\approx e^{-\frac{i}{\hbar} \opH {t_R}}\approx\unit-\frac{i}{\hbar} \opH t_R$. Then,  we can rewrite:
\begin{equation}
\langle \evf | \opUt{t_L} \opO(t_R)  \opUt{t_R}|\p\rangle\approx\langle \evf |  \opUt{t_L} \opO(t_R) |\p\rangle -\frac{i}{\hbar} t_R \langle \evf |  \opUt{t_L} \opO(t_R) \opH |\p\rangle.
 \label{ap5.4b} 
\end{equation}
We also assume that $\opO({t_R})=\opO+\frac{\partial \opO}{\partial t} t_R$. Finally, we can rewrite:
\begin{equation}
\langle \evf | \opUt{t_L} \opO(t_R) \opUt{t_R} |\p\rangle\approx \langle \evf |  \opUt{t_L} \opO  |\p\rangle+t_R \langle \evf | \opUt{t_L} \frac{\partial \opO}{\partial t}|\p\rangle -\frac{i}{\hbar} t_R \langle \evf | \opUt{t_L} \opO  \opH |\p\rangle
 \label{ap5.5b} 
\end{equation}
where we disregard the terms of the order of $(t_R)^2$. Identically, we get:
\begin{eqnarray}
\frac{1}{\langle \evf | \opUt{t_L} \opUt{t_R} |\p\rangle}&&\approx\frac{1}{\langle \evf | \opUt{t_L}|\p\rangle -\frac{i}{\hbar} t_R \langle \evf |  \opUt{t_L} \opH |\p\rangle}\approx \frac{1}{\langle \evf | \opUt{t_L}|\p\rangle \left(1 - \frac{i}{\hbar} t_R \frac{\langle \evf | \opUt{t_L}  \opH |\p\rangle}{\langle \evf | \opUt{t_L}|\p\rangle}\right)}\nonumber\\
&&\approx\frac{1}{\langle \evf |\opUt{t_L}|\p\rangle}  \left(1 + \frac{i}{\hbar} t_R \frac{\langle \evf | \opUt{t_L} \opH |\p\rangle}{\langle \evf | \opUt{t_L}|\p\rangle}\right),
\label{ap5.6b} 
\end{eqnarray}
where we have used the approximation $\frac{1}{1-x} \approx 1+x$ when $x\ll1$. Putting \eqref{ap5.5b}  and \eqref{ap5.6b}  into \eqref{ap5.4b}, we get:
\begin{eqnarray}
&&\Re{  \frac{\langle \evf | \opUt{t_L} \opUt{t_R} \opO(-t_R)  \opUt{t_R}|\p\rangle }{\langle \evf |  \opUt{t_L}\opUt{t_R}|\p\rangle }}\nonumber\\
&&\approx \Re{ \left(   \frac{\langle \evf |   \opUt{t_L} \opO  |\p\rangle}{\langle \evf |  \opUt{t_L}|\p\rangle }+t_R \frac{\langle \evf |  \opUt{t_L} \frac{\partial \opO}{\partial t} |\p\rangle}{\langle \evf | \opUt{t_L}|\p\rangle } - \frac{i}{\hbar} t_R \frac{ \langle \evf |  \opUt{t_L} \opO  \opH |\p\rangle}{\langle \evf |  \opUt{t_L}|\p\rangle }\right) \left(1 + \frac{i}{\hbar} t_R \frac{\langle \evf |  \opUt{t_L} \opH |\p\rangle}{\langle \evf |  \opUt{t_L}|\p\rangle}\right)} 
\nonumber\\
&&\approx \Re{   \frac{\langle \evf |  \opUt{t_L} \opO |\p\rangle}{\langle \evf |  \opUt{t_L}|\p\rangle }+t_R \frac{\langle \evf |  \opUt{t_L} \frac{\partial \opO}{\partial t}|\p\rangle}{\langle \evf | \opUt{t_L} |\p\rangle } - \frac{i}{\hbar} t_R \frac{ \langle \evf |   \opUt{t_L} \opH \opH |\p\rangle}{\langle \evf | \opUt{t_L}|\p\rangle }+ \frac{i}{\hbar} t_R \frac{\langle \evf |  \opUt{t_L} \opO  |\p\rangle}{\langle \evf |  \opUt{t_L}|\p\rangle }  \frac{\langle \evf |   \opUt{t_L}\opH  |\p\rangle}{\langle \evf |  \opUt{t_L}|\p\rangle}},\nonumber 
 \label{ap5.7b} 
\end{eqnarray} 
where we have disregarded the terms of the order of $(t_R)^2$.  By subtracting the two weak values in \eqref{ap5.2b}, and dividing by $t_R$, we get:
\begin{eqnarray}
\text{FDRHD} &\approx &\frac{1}{t_R} \left( \Re{  \frac{\langle \evf | \opUt{t_L} \opO(t_R)  \opUt{t_R}|\p\rangle }{\langle \evf |\opUt{t_L} \opUt{t_R}|\p\rangle }}-\Re{  \frac{\langle \evf |\opUt{t_L} \opO|\p\rangle }{\langle \evf | \opUt{t_L} |\p\rangle }}\right)\nonumber\\
 && \approx \cancel{\frac{1}{t_R}} \Re{  \cancel{t_R} \frac{\langle \evf | \opUt{t_L} \frac{\partial \opO}{\partial t} |\p\rangle}{\langle \evf | \opUt{t_L}|\p\rangle } - \frac{i}{\hbar} \cancel{t_R} \frac{ \langle \evf |  \opUt{t_L} \opO \opH |\p\rangle}{\langle \evf | \opUt{t_L}|\p\rangle }+ \frac{i}{\hbar} \cancel{t_R}\frac{\langle \evf | \opUt{t_L} \opO |\p\rangle}{\langle \evf | \opUt{t_L}|\p\rangle }  \frac{\langle \evf |  \opUt{t_L} \opH  |\p\rangle}{\langle \evf | \opUt{t_L}|\p\rangle}}. \nonumber
  \label{ap5.9b} 
\end{eqnarray} 
By using  $-\opO\opH=[\opH,\opO]-\opH\opO$ and $\opC:=\frac{i}{\hbar}[\opH,\opO]+\frac{\partial \opO}{\partial t}$, we can rewrite:
\begin{eqnarray}
\text{FDRHD}\approx \Re{ \frac{\langle \evf | \opUt{t_L} \opC |\p\rangle }{\langle \evf | \opUt{t_L}|\p\rangle }}+ \Re{   -\frac{i}{\hbar}  \frac{ \langle \evf | \opUt{t_L} \opH \opO |\p\rangle}{\langle \evf | \opUt{t_L}|\p\rangle }+ \frac{i}{\hbar} \frac{\langle \evf |  \opUt{t_L}\opH |\p\rangle}{\langle \evf | \opUt{t_L}|\p\rangle}\frac{\langle \evf | \opUt{t_L} \opO  |\p\rangle}{\langle \evf | \opUt{t_L}|\p\rangle } }.\nonumber 
 \label{ap5.10b} 
\end{eqnarray} 
he final result is mathematically equivalent to \eqref{ap5.3}, as both expressions have been derived in the Coulomb gauge. The gauge independence of FDRHD follows the same reasoning outlined in Appendix~\ref{ap5b} for FDLHD,

\subsection{Necessary and sufficient condition for FDLHD being a weak value}

In this case, the final result \eqref{ap5.10b} is mathematically equivalent to \eqref{ap5.3}. In general, however, the approximate expression of FDRHD does not naturally take the form of a weak value. Nevertheless, by applying the same reasoning established for RHD, we can conclude that FDRHD can also be compactly expressed as a weak value when condition \textbf{C5} is satisfied. These findings regarding the gauge condition and the structural form of FDRHD are summarized in Table \ref{summary}.

\subsection{On the sum of FDLHD plus FDRHD}

Another important result appears from the sum of FDLHD plus FDRHD evaluated at $t_R=0$ and $t_L=0$ respectively, giving:
\begin{eqnarray}
\left.\text{FDLHD}\right|_{ t_R =0}+\left.\text{FDRHD}\right|_{t_L =0}=\BW\big(\opC,0,0\big||\evf\rangle,|\p\rangle \big)+\frac{\langle \evf | \frac{\partial  \opO}{\partial t}| \p\rangle }{\langle \evf | \p\rangle }.
\label{aptwovalues}
\end{eqnarray}
Notice that \eqref{aptwovalues} introduces the same deep relationship between FDLHD and FDRHD that we have presviously seen for LHD and RHD.


\section{Gauge invariance of Bohmian velocities}
\label{ap6}

The gauge invariance of the Bohmian velocity is already demonstrated in the paper as a by-product of the gauge invariance of the weak value in \eqref{weakvvelo}. Hereafter, we show an alternative demonstration dealing directly with the general gauge Schrödinger equation, which is the left-hand side of \eqref{ap1.1},   
\begin{equation}
i\hbar \frac{\partial \pg}{\partial t}=\left(\frac{1}{2m^*}\left( \opp -q\A^g\right)^2+ q\Vg\right)\pg.
\label{ap6bis}
\end{equation}
We use  $\pg=R^ge^{iS^g/\hbar}$ with $R^g$ the modulus and $S^g$ the phase of the wave function, both being real functions. Then, 
\begin{equation}
i\hbar \frac{\partial \pg}{\partial t}=e^{iS^g/\hbar} i\hbar \frac{\partial R^g}{\partial t}-e^{iS^g/\hbar}R^g \frac{\partial S^g}{\partial t}\nonumber.
\end{equation}
We evaluate:
\begin{eqnarray}
&&\frac{1}{2m^*}\left( \opp -q\A^g\right)^2\pg =-\frac{\hbar^2}{2m^*}\nabla^2\pg+i\frac{q}{2m^*}\hbar (\grad \A^g)\pg+i2\frac{q}{2m^*}\hbar \A^g \grad \pg+\frac{q^2}{2m^*}(\A^g)^2\pg\nonumber.
\end{eqnarray}
By acknowledging $\grad \pg=e^{iS^g/\hbar} \grad R^g+e^{iS^g/\hbar}R^g \frac{i}{\hbar}\grad S^g$ and $\nabla^2 \pg=e^{iS^g/\hbar} \nabla^2 R^g+e^{iS^g/\hbar}2\frac{i}{\hbar}\grad R^g \grad S^g -e^{iS^g/\hbar} R^g \frac{1}{\hbar^2}(\grad S^g)^2+e^{iS^g/\hbar}\frac{i}{\hbar} R^g \nabla^2 S^g$, eliminating $e^{iS^g/\hbar}$ and dividing by $R^g$, the real part of the Schrödinger Equation in (\ref{ap6bis}) is:
\begin{eqnarray}
-\frac{\partial S^g}{\partial t}=-\frac{\hbar^2}{2m^*}\frac{\nabla^2 R^g}{ R^g}+\frac{1}{2m^*}(\grad S^g)^2+\frac{q^2}{2m^*}(\A^g)^2-2\frac{q}{2m^*}\A^g  \grad S^g+q\Vg\nonumber, 
\end{eqnarray}
which can be rewritten as a quantum Hamilton-Jacobi equation:
\begin{eqnarray}
0=\frac{\partial S^g}{\partial t}-\frac{\hbar^2}{2m^*}\frac{\nabla^2 R^g}{ R^g}+\frac{\left(\grad S^g-q\A^g\right)^2}{2m^*}+q\Vg.
\label{ap6.4}
\end{eqnarray}
The imaginary part of the Schrödinger Equation in (\ref{ap6bis}), on the other hand, is $\frac{\partial R^g}{\partial t}=\frac{q}{2m^*} (\grad \A^g)R^g+\frac{q}{m^*} \A^g \grad R^g
-\frac{1}{^*m}\grad R^g \grad S^g-\frac{1}{2m^*} R^g \nabla^2 S^g$, which can be rewritten as a continuity equation: 
\begin{eqnarray}
\frac{\partial (R^g)^2}{\partial t}=-\grad \left((R^g)^2 \left(\frac{\grad S^g-q\A^g}{m^*}\right) \right).
\label{ap6.5}
\end{eqnarray}
In both cases, \eqref{ap6.4} and \eqref{ap6.5}, the Bohmian velocity is defined as:
\begin{eqnarray}
\vB{|\p\rangle}=\frac{\grad S^g-q\A^g}{m^*}=\frac{\grad S+\cancel{q\nabla\g} -q\A -\cancel{q \grad \g} }{m^*}
     =\frac{\grad S -q\A }{m^*}
\label{ap6.6}
\end{eqnarray}
which is gauge invariant when $\A^g$ is defined as in \eqref{vg} and $S^g$ following \eqref{wg} in the paper. Such a gauge invariance of Bohmian trajectories is not usually mentioned in the literature, but it is also elaborated, in a different way, in Appendix~A of Ref. \cite{deotto1998}. 

As a byproduct, the above development shows that the continuity equation in (\ref{ap6.5}) is gauge invariant because $R=R^g$ and $\vB{|\p\rangle}=\vB{|\p\rangle,g}$ as seen in \eqref{ap6.6}. Identically, the quantum Hamilton-Jacobi equation (\ref{ap6.4}) is also gauge invariant. The kinetic energy is gauge invariant because \eqref{ap6.6} and the quantum potential is also gauge invariant because $R=R^g$, while the gauge dependence of $\frac{\partial S^g}{ \partial t}=\frac{\partial S}{ \partial t}+q\frac{\partial \g}{ \partial t}$ and $q\Vg=q\V-q\frac{\partial \g}{ \partial t}$ compensate each other.

\section{ state of $\opv=\frac{i}{\hbar}[\opH,\opx]$}
\label{ap7}

From the Hamiltonian in \eqref{hami} in the paper,  we can evaluate:
\begin{eqnarray}
\frac{i}{\hbar}[\opH,\opx]&&=\frac{i}{\hbar}[{1}/({2m^*})\left( \opp -q\A\right)^2+q\V,\opx]
=\frac{i}{2m^*\hbar}[ \opp^2-\opp q\A -q\A \opp +(q\A)^2,\opx]\nonumber\\
&&=\frac{i}{2m^*\hbar}\left([ \opp^2,\opx]-q[\opp \A,\opx]-q[\A \opp,\opx]\right)\nonumber\\
&&=\frac{i}{2m^*\hbar}\left(\opp[ \opp,\opx]+[ \opp,\opx]\opp-q[\opp \A,\opx]-q[\A \opp,\opx]\right)\nonumber\\
&&=\frac{i}{2m^*\hbar}\left(-2 i \hbar \opp-q\opp \A\opx+q\opx\opp\A-q\A \opp\opx+q\opx\A\opp \right)\nonumber\\
&&=\frac{i}{2m^*\hbar}\left(-2 i \hbar \opp-q[\opp,\opx] \A-q\A [\opp,\opx] \right)\nonumber\\
&&=\frac{i}{2m^*\hbar}\left(-2 i \hbar \opp+2i\hbar q \A\right)
=\frac{1}{m^*}\left(\opp- q \A\right)=\opv,\nonumber
\end{eqnarray}
where we have used $[\opp,\opx]=-i\hbar\unit$. Notice that $\opx$ is time-independent (in the Coulomb Gauge for example) and gauge invariant so that $\frac{\partial \opx}{\partial t}=0$. Because of \eqref{ap6.0}, the same result will be found in any gauge.   A similar development can be found in Ref. \cite{Ballentine2014}.

\section{Derivation of $\frac{q}{2}\left(\opE \cdot \opv+\opv\cdot  \opE \right)=\frac{i}{\hbar}[\opH,\frac{1}{2}m^*\opv^2]+\frac{\partial \frac{1}{2}m^*\opv^2}{\partial t}$}
\label{ap8}

From the Hamiltonian in \eqref{hami} in the paper, we can evaluate:
\begin{eqnarray}
\frac{i}{\hbar}\left[\opH,\frac{1}{2}m^*\opv^2\right]&=&i\frac{m^*}{2\hbar}\left[\frac{m^*}{2} \opv^2+q\V,\opv^2\right]=i\frac{qm^*}{2\hbar}[\V,\opv^2]+i\frac{ {m^*}^2}{2^2\hbar}[\opv^2,\opv^2]\nonumber\\
&=&i\frac{qm^*}{2\hbar}\left(\opv[\V,\opv]+[\V,\opv]\opv\right).\nonumber
\label{ap8.1}
\end{eqnarray}
We also evaluate:
\begin{eqnarray}
i\frac{qm^*}{2\hbar}[\V,\opv^2]&=&i\frac{qm^*}{2\hbar}[\V,\opvn{1}^2+\opvn{2}^2+\opvn{3}^2]=i\frac{qm^*}{2\hbar}\sum_{\alpha} \left(\opvn{\alpha}[\V,\opvn{\alpha}]+[\V,\opvn{\alpha}]\opvn{\alpha}\right)\nonumber\\
&=&-\frac{q}{2}\sum_{\alpha} \left(\opvn{\alpha}\frac{\partial \V}{\partial x_{\alpha}}+\frac{\partial \V}{\partial x_{\alpha}}\opvn{\alpha}\right),
\label{ap8.2}
\end{eqnarray}
with $i\frac{qm^*}{2\hbar}[\V,\opvn{\alpha}]=-\frac{q}{2}\frac{\partial \V}{\partial x_{\alpha}}$.  Then, we can develop  $\frac{1}{2}m^* \frac{\partial \opv^2}{\partial t}$ as:
\begin{eqnarray}
\frac{m^*}{2} \frac{\partial \opv^2}{\partial t}&=&\frac{m^*}{2} \frac{\partial \opv\cdot \opv}{\partial t}=\frac{m^*}{2} \left( \frac{\partial \opv}{\partial t}\cdot \opv+ \opv\cdot  \frac{\partial \opv}{\partial t}\right)\nonumber\\
&=&-\frac{q}{2}\frac{\partial \opA}{\partial t}\cdot \opv-\frac{q}{2} \opv\cdot  \frac{\partial \opA}{\partial t}.
\label{ap8.3}
\end{eqnarray}

Thus, the sum $\frac{i}{\hbar}[\opH,\frac{1}{2}m^*\opv^2]+\frac{\partial \frac{1}{2}m^*\opv^2}{\partial t}$, using \eqref{ap8.2} and \eqref{ap8.3}, gives the (gauge invariant) definition of electromagnetic work as:

\begin{equation}
\frac{i}{\hbar}\frac{1}{2}m^* [\opH,\opv^2]+\frac{1}{2}m^* \frac{\partial \opv^2}{\partial t}=\frac{q}{2}\left(\opE \cdot \opv+\opv\cdot  \opE \right).\nonumber
\end{equation}

Because of \eqref{ap6.0}, the same result will be found in any gauge.  A similar development can be found in Ref. \cite{Ballentine2014}.

\section{Local work-energy theorem from weak values in \eqref{power}}
\label{ap9}

Using the weak value in (\ref{ewv}) for $t_R=0$ and $t_L=0$, we want to evaluate:
\begin{equation}
\frac{\langle \egx | \frac{i}{2\hbar}{m^*} [\opH,\opv^2]+\frac{1}{2}{m^*} \frac{\partial \opv^2}{\partial t}|\egv\rangle}{\langle \egx | \egv\rangle}.\nonumber
\end{equation}
In particular, we evaluate $\frac{\langle \egx |\En{\alpha}\opvn{\alpha}|\egv\rangle}{\langle \egx | \egv\rangle}=\En{\alpha}(\egx)\frac{\langle \egx |\opvn{\alpha}|\egv\rangle}{\langle \egx | \egv\rangle}=\En{\alpha}(\egx)\vBn{\alpha}{|\egv\rangle} (\egx)+i \En{\alpha}(\egx)\vOn{\alpha}{|\egv\rangle} (\egx)$ with the Bohmian velocity defined in (\ref{weakvvelo}) as \eqref{ap9.5} and the osmotic velocity as \eqref{ap9.6}. Using $\opvn{\alpha}=\frac{\oppn{\alpha}-q\An{\alpha}}{m^*}$, we also evaluate, 
\begin{eqnarray}
\frac{\langle \egx |\opvn{\alpha} \En{\alpha}|\egv\rangle}{\langle \egx | \egv\rangle}&&=\frac{1}{m^*}\frac{\langle \egx |\oppn{\alpha} \En{\alpha}|\egv\rangle}{\langle \egx | \egv\rangle}-\frac{q}{m^*} \An{\alpha}(\egx)\En{\alpha}(\egx)=-\frac{i\hbar}{m^*}\frac{\frac{\partial }{\partial \egxn{\alpha}} \En{\alpha}(\egx) \langle \egx|\egv\rangle}{\langle \egx | \egv\rangle}-\frac{q}{m^*} \An{\alpha}(\egx)\En{\alpha}(\egx)\nonumber\\
&&=-\frac{i\hbar}{m^*}\En{\alpha}(\egx) \frac{\frac{\partial }{\partial \egxn{\alpha}} \langle \egx|\egv\rangle}{\langle \egx | \egv\rangle}-\frac{i\hbar}{m^*}\frac{\partial \En{\alpha}(\egx)}{\partial \egxn{\alpha}}-\frac{q}{m^*} \An{\alpha}(\egx)\En{\alpha}(\egx) \nonumber\\
&&=\vBn{\alpha}{|\egv\rangle} (\egx)\En{\alpha}(\egx)+i\vOn{\alpha}{|\egv\rangle} (\egx)\En{\alpha}(\egx)-\frac{i\hbar}{m^*}\frac{\partial \En{\alpha}(\egx)}{\partial \egxn{\alpha}}, \nonumber
\end{eqnarray}
where we have use the definition of the Bohmian velocity  \eqref{ap9.5} and the osmotic velocity as \eqref{ap9.6}  for the state $|\p\rangle=|\egv\rangle$. Finally, 
\begin{equation}
\frac{q}{2}\frac{\langle \egx |\left(\opE \cdot \opv+ \opv\cdot  \opE \right) |\egv\rangle}{\langle \egx | \egv\rangle}=q\vB{|\egv\rangle}(\egx) \cdot \textbf{E}(\egx)+iq\vO{|\egv\rangle}(\egx) \cdot \textbf{E}(\egx)-\frac{i\hbar q}{2m^*} \grad \cdot \textbf{E}(\egx)\nonumber, 
\end{equation}
such that, 
\begin{equation}
\Re{\frac{\langle \egx | \frac{i}{\hbar}[\opH,\frac{1}{2}{m^*}\opv^2]+\frac{\partial \frac{1}{2}{m^*}\opv^2}{\partial t} |\egv\rangle}{\langle \egx | \egv\rangle}}=q\;\vB{|\egv\rangle}(\egx) \cdot \textbf{E}(\egx)\nonumber,
\end{equation}
which can also be written as:
\begin{equation}
\frac{\partial}{\partial t_R }\BW \big(\egw,0,t_R\big||\egx\rangle,|\egv\rangle\big)=q\;\vB{|\egv\rangle}(\egx) \cdot \textbf{E}(\egx)\nonumber.
\end{equation}
where $\egw$ is related to the operator $\opW:=\frac{1}{2}{m^*}\opv^2$ which is gauge invariant and satisfies \textbf{C5} because $|\egv\rangle$ is an eigenstate of $\opO=\opW$.

\section{Derivation of the operator $\opa:=\frac{i}{\hbar}[\opH,\opv]+\frac{\partial \opv}{\partial t}=q\left(\frac{\opE}{m^*}+\frac{1}{2m^*}\left(\opv \times \opB-\opB \times \opv\right)\right)$}
\label{ap10}

From the Hamiltonian in (\ref{hami}) in the paper, using the velocity operator $\opv=\frac{1}{m^*}\left(-i\hbar \grad -q\A \right)$, we can evaluate:
\begin{eqnarray}
\frac{i}{\hbar}[\opH,\opv]=\frac{i}{\hbar}\left[\frac{m^*}{2} \opv^2+q\V,\opv\right]=i\frac{m^*}{2\hbar}\left[\opv^2,\opv\right]+\frac{iq}{\hbar}[\V,\opv].\nonumber
\end{eqnarray}
We then evaluate  $\frac{iq}{\hbar}[\V,\opv]=-\frac{q}{m^*}\grad \V$ and $\frac{\partial \opv}{\partial t}=-\frac{q}{m^*}\frac{\partial \opA}{\partial t}$. Thus, the sum $\frac{iq}{\hbar}[\V,\opv]+\frac{\partial \opv}{\partial t}$ gives the (gauge invariant) definition of electric field operator $\opE$ as:
\begin{equation}
\frac{q \opE}{m^*}:=-\frac{q}{m^*}\grad \opV-\frac{q}{m^*}\frac{\partial \opA}{\partial t}.
\label{ap7.2}
\end{equation}
 
On the other hand, we consider the components of the velocity $\opv=\{\opvn{1},\opvn{2},\opvn{3}\}$ so that
\begin{eqnarray}
i\frac{m^*}{2\hbar}[\opv^2,\opvn{\beta}]=i\frac{m^*}{2\hbar}[\opvn{1}^2+\opvn{2}^2+\opvn{3}^2,\opvn{\beta}]
=i\frac{m^*}{2\hbar}\sum_{\alpha\ne \beta} \left( \opvn{\alpha} [\opvn{\alpha},\opvn{\beta}]+[\opvn{\alpha},\opvn{\beta}]\opvn{\alpha}\right)\nonumber.
\end{eqnarray}
for $\beta=1,2,3$. Next, we evaluate:
\begin{eqnarray}
[\opvn{\alpha},\opvn{\beta}]&=&\frac{1}{{m^*}^2} [\oppn{\alpha},\oppn{\beta}]-\frac{q}{{m^*}^2}[\oppn{\alpha},\An{\beta}]
-\frac{q}{{m^*}^2} [\An{\alpha},\oppn{\beta}]+\frac{q^2}{{m^*}^2}[\An{\alpha},\An{\beta}]\nonumber\\
&=&\frac{i\hbar q}{{m^*}^2}\left( \frac{\partial \An{\beta}}{\partial \egxn{\alpha}}-  \frac{\partial \An{\alpha}}{\partial \egxn{\beta}}\right)
=\sum_{\gamma}\frac{i\hbar q}{m^2}\epsilon_{\alpha,\beta,\gamma}\Bn{\gamma}\nonumber,
\end{eqnarray}
where we have used $[\An{\alpha},\An{\beta}]=0$, $[\oppn{\alpha},\oppn{\beta}]=0$ and $[\oppn{\alpha},\An{\beta}]=-i\hbar \partial \An{\beta}/\partial \egxn{\alpha}$ and the Levi-Civita symbol $\epsilon_{\alpha,\beta,\gamma}$ is $1$ if $\{\alpha,\beta,\gamma\}$ is an even permutation of $\{1,2,3\}$, $-1$  if it is an odd permutation and $0$ if any index is repeated.

Hence, we have:
\begin{eqnarray}
&&i\frac{m^*}{2\hbar}[\opv^2,\opvn{\beta}]=-\frac{q}{2m^*}\sum_{\alpha,\gamma} \left(  \epsilon_{\alpha,\beta,\gamma}\opvn{\alpha} \Bn{\gamma}+\epsilon_{\alpha,\beta,\gamma}\Bn{\gamma}\opvn{\alpha}\right).\nonumber
\end{eqnarray}
By noticing that the cross product can be written as $\textbf{a} \times \textbf{b}|_{\beta}=-\sum_{\alpha,\gamma} \epsilon_{\alpha,\beta,\gamma} a_{\alpha} b_{\gamma}$  and $\textbf{b} \times \textbf{a}|_{\beta}=\sum_{\alpha,\gamma} \epsilon_{\alpha,\beta,\gamma} b_{\gamma} a_{\alpha}$ (without the negative sign in the last expression because changing the order implies an additional permutation in the Levi-Civita coefficients), we write:
\begin{equation}
i\frac{m^*}{2\hbar}[\opv^2,\opv] =\frac{q}{2m^*}\left(\opv \times \opB-\opB \times \opv\right). 
\label{ap7.8}
\end{equation}

Finally, adding \eqref{ap7.2} and \eqref{ap7.8}, we get:
\begin{equation}
\frac{i}{\hbar}[\opH,\opv]+\frac{\partial \opv}{\partial t} =\frac{q \opE}{m^*}+\frac{q}{2m^*}\left(\opv \times \opB-\opB \times \opv\right).\nonumber 
\end{equation}
Because of \eqref{ap6.0}, the same result will be found in any gauge. A similar development can be found in Ref. \cite{Ballentine2014}.

\section{Local Lorentz force from weak values in \eqref{lorentz}}
\label{ap11}

Using the weak value in \eqref{ewv} with $t_R=0$ and $t_L=0$, and the velocity operator $\opv=\frac{1}{m^*}\left(-i\hbar \grad -q\A \right)$, we want to evaluate:

\begin{equation}
\frac{\langle \egx | \frac{i}{\hbar} [\opH,\opv]+ \frac{\partial \opv}{\partial t}|\egv\rangle}{\langle \egx | \egv\rangle}.
\label{ap9.0}
\end{equation}

First of all, for all the terms, except the kinetic energy operator, in \eqref{ap9.0}, we can write as $\frac{\langle \egx | \frac{iq}{\hbar}[\V,\opv]+\frac{\partial \opv}{\partial t}|\egv\rangle}{\langle \egx | \egv\rangle}= \frac{1}{m^*}  \frac{\langle \egx |q \opE |\egv\rangle}{\langle \egx | \egv\rangle}=\frac{q}{m^*} \E(\egx)$ where we have used that the electrical field operator is position dependent so that the weak value post-selected in position is a pure real number.

Second, we evaluate \eqref{ap9.0} for the kinetic energy operator:
\begin{equation}
\frac{\langle \egx |i\frac{m^*}{2\hbar}[\opv^2,\opvn{\beta}]|\egv\rangle}{\langle \egx | \egv\rangle}=-\frac{q}{2m^*}\sum_{\alpha,\gamma} \left(  \epsilon_{\alpha,\beta,\gamma}\frac{\langle \egx |\opvn{\alpha} \Bn{\gamma}|\egv\rangle}{\langle \egx | \egv\rangle}+\epsilon_{\alpha,\beta,\gamma}\frac{\langle \egx |\Bn{\gamma}\opvn{\alpha}|\egv\rangle}{\langle \egx | \egv\rangle}\right)\nonumber,
\end{equation}
where we have used $\frac{\langle \egx |\Bn{\gamma}\opvn{\alpha}|\egv\rangle}{\langle \egx | \egv\rangle}=\Bn{\gamma}(\egx)\frac{\langle \egx |\opvn{\alpha}|\egv\rangle}{\langle \egx | \egv\rangle}=\Bn{\gamma}(\egx)\vBn{\alpha}{|\egv\rangle} (\egx)-i \Bn{\gamma}(\egx)\vOn{\alpha}{|\egv\rangle} (\egx)$ with the Bohmian velocity \cite{wiseman2007,durr2009} defined in  Eq. (25) in the paper, for $|\egv\rangle=|\pg \rangle$, as:
\begin{eqnarray}
\Re{\frac{\langle \egx | \opv^g |\pg \rangle}{\langle \egx|\pg \rangle}}=\frac{1}{m^*}\Re{-i\hbar \frac{\grad \pg }{\pg } -q \textbf{A}^g}= \frac{\grad S-\cancel{q \grad G}-q\textbf{A} +\cancel{q\grad G}}{m^*}:=\vB{|\egv\rangle}\nonumber,\\
\label{ap9.5}
\end{eqnarray}
and the osmotic velocity \cite{destefani2023,hiley2012,nelson1966,bohm1989} $\vO{|\egv\rangle}$ as:
\begin{eqnarray}
&&\Im{\frac{\langle \egx | \opv^g |\pg \rangle}{\langle \egx|\pg \rangle}}=\frac{1}{m^*}\Im{-i\hbar \frac{\grad \pg }{\pg } -q \textbf{A}^g}=-\frac{\hbar}{m^*} \frac{\grad R}{R}=-\vO{|\egv\rangle},
\label{ap9.6}
\end{eqnarray}
with the position dependence of the magnetic field $\Bn{\gamma}(\egx)$. Using $\opvn{\alpha}=\frac{\oppn{\alpha}-q\An{\alpha}}{m^*}$, we also evaluate, 
\begin{eqnarray}
\frac{\langle \egx |\opvn{\alpha} \Bn{\gamma}|\egv\rangle}{\langle \egx | \egv\rangle}&&=\frac{1}{m^*}\frac{\langle \egx |\oppn{\alpha} \Bn{\gamma}|\egv\rangle}{\langle \egx | \egv\rangle}-\frac{q}{m^*} \An{\alpha}(\egx)\Bn{\gamma}(\egx)=-\frac{i\hbar}{m^*}\frac{\frac{\partial }{\partial \egxn{\alpha}} \Bn{\gamma}(\egx) \langle \egx|\egv\rangle}{\langle \egx | \egv\rangle}-\frac{q}{m^*} \An{\alpha}(\egx)\Bn{\gamma}(\egx)\nonumber\\
&&=-\frac{i\hbar}{m^*}\Bn{\gamma}(\egx) \frac{\frac{\partial }{\partial \egxn{\alpha}} \langle \egx|\egv\rangle}{\langle \egx | \egv\rangle}-\frac{i\hbar}{m^*}\frac{\partial \Bn{\gamma}(\egx)}{\partial \egxn{\alpha}}-\frac{q}{m^*} \An{\alpha}(\egx)\Bn{\gamma}(\egx) \nonumber\\
&&=\vBn{\alpha}{|\egv\rangle} (\egx)\Bn{\gamma}(\egx)-i\vOn{\alpha}{|\egv\rangle} (\egx)\Bn{\gamma}(\egx)-\frac{i\hbar}{m^*}\frac{\partial \Bn{\gamma}(\egx)}{\partial \egxn{\alpha}}, \nonumber
\end{eqnarray}
where we have used the definition of the Bohmian velocity given in the weak value of \eqref{weakvvelo} in the paper for $|\p\rangle=|\egv\rangle$ and the cross product written as $\textbf{a} \times \textbf{b}|_{\beta}=-\sum_{\alpha,\gamma} \epsilon_{\alpha,\beta,\gamma} a_{\alpha} b_{\gamma}$  and $\textbf{b} \times \textbf{a}|_{\beta}=\sum_{\alpha,\gamma} \epsilon_{\alpha,\beta,\gamma} b_{\gamma} a_{\alpha}$ (without the negative sign in the last expression because changing the order implies an additional permutation in the Levi-Civita coefficients).  Finally:
\begin{equation}
\frac{\langle \egx | \frac{i m^*}{2\hbar}[\opv^2,\opv] |\egv\rangle}{\langle \egx | \egv\rangle}=\frac{q}{m^*}\vB{|\egv\rangle}(\egx) \times \textbf{B}(\egx)-i\frac{q}{m^*}\vO{|\egv\rangle}(\egx) \times \textbf{B}(\egx)-\frac{i\hbar q}{{ 2 \;m^*}^2} \grad \times \textbf{B}(\egx), \nonumber
\end{equation}
such that, 
\begin{equation}
m^* \Re{\frac{\langle \egx | \frac{i}{\hbar}[\opH,\opv] |\egv\rangle}{\langle \egx | \egv\rangle}}=q \E(\egx) +q \vB{|\egv\rangle}(\egx) \times \textbf{B}(\egx), \nonumber
\end{equation}
which can also be written as:
\begin{equation}
m^* \left.\frac{\partial^2}{\partial t_R \partial t_L}\BW \big(\egx,t_L,t_R\big||\egx\rangle,|\egv\rangle\big)\right|{\begin{matrix}_{ t_R=0}\\ _{t_L=0} \end{matrix}}=q \E(\egx) +q \vB{|\egv\rangle}(\egx) \times \textbf{B}(\egx). 
\label{final}
\end{equation}


\section{Details of the numerical simulations}
\label{apnumerical}

In all the numerical simulations in this work, we consider an electron with a charge of $q=-1.6\cdot10^{-19}$ C and an effective mass of $m^*=0.067 m_0$, with $m_0$ being the free electron mass. Such an effective mass corresponds, for example, to that of a transport electron inside a GaAs semiconductor. The reason for selecting a massive particle (instead of a massless photon, where typical experiments on weak values are conducted) is that the connection between a weak value of the momentum post-selected in position and the Bohmian velocity, is well justified within the Bohmian theory of massive particles, which are solutions of the non-relativistic Schrödinger equation, as we have studied in this paper. On the contrary, such a connection remains under debate for massless photons. Some possible protocols to measure weak values for electrons inside solid-state devices are mentioned in Ref.~\cite{marian2016}.  

In this appendix, we describe the numerical details of how the states are defined at the preparation time and how their time evolution is numerically computed.  We consider a mesh in the time variable with a constant step, $\Delta t$, that is, $t_j = j \; \Delta t $ for $j = 1,\ldots,N_t$. Additionally, we define a mesh on the 1D spatial degree of freedom $x$ with a constant step, $\Delta x$, that is, $x_k = k \; \Delta x $ for $k = 1,\ldots,N_x$ (in the evolution in a 2D space a similar discretization is done for the degree of freedom $y$).  For all numerical simulations, we use a spatial step of $\Delta x=0.2$ nm and a temporal step of $\Delta t=0.01$ fs. Once the grid is defined, any function of time and space is associated with an $N_t \times N_x$ matrix. For example, we define $\psi(x,t)|_{x = x_k;t = t_j} = \psi(t_{j},x_k)$. 

\subsection{Time evolution of the state with the Coulomb gauge without magnetic field}

In order to define the wave function at its preparation time $t_p$, we have to define two consecutive times $t =t_p$ and $t=t_p+\Delta t$. We assume that during these initial times, the system evolves in a flat potential region and that the quantum state can  be defined by an initial Gaussian wave packet \cite{cohen1986} given by:
\begin{eqnarray}
\psi_{E_c,x_c,\sigma_x}(x,t_p) &=& {{\left( \frac{1}{\pi {\sigma_x^2}} \right)}^{1/4}}{{e}^{i\left( {{k}_{c}}(x-{{x}_c}) \right)}} \exp \left( -\frac{{{(x-{{x}_c})}^{2}}}{{2\sigma_x^2}} \right),\\
\psi_{E_c,x_c,\sigma_x}(x,t_p+\Delta t) &=& {{\left( \frac{4 \sigma_x^2}{\pi } \right)}^{1/4}}\frac{{{e}^{i\zeta }}{e}^{i(k_c(x-x_c))}}{{{\left( 4 \sigma_x^4 + \frac{4{{\hbar }^{2}}{{(\Delta t)}^{2}}}{{{m^*}^{2}}} \right)}^{1/4}}}\exp \left(-\frac{{{\left[x - {{x}_c} - \frac{\hbar {{k}_{c}}}{m^*}(\Delta t) \right]}^{2}}}{{2 \sigma_x^2} + \frac{2i\hbar (\Delta t)}{m^*}} \right),\nonumber
\label{apnum5}
\end{eqnarray}
where $x_c$ is the central position of the wave packet at the preparation time $t_p$, $E_c$ is the central (kinetic) energy related to the central wave vector ${k}_{c} = \sqrt{\frac{2m^*E_c}{{\hbar }^{2}}}$, and ${{\sigma }_{x}}$ is the wave packet spatial dispersion that can be related to the wave packet width in the reciprocal space as ${{\sigma }_{k}} = 1/{{\sigma }_{x}}$.  We also define  $\zeta = -\kappa - {\hbar {k_c}^{2} \Delta t}/{(2m^*)}$ with $\tan (2 \kappa ) = {\hbar \Delta t}/{(m^* \sigma_x^2)}$. 

The time-evolution of the Gaussian wave packet is computed from the time-dependent Schr\"odinger equation: 
\begin{equation}
i\hbar \frac{\partial \p}{\partial t}=\left(\frac{1}{2m^*}\left( -i\hbar \frac{\partial}{\partial x} \right)^2+ q\V(x,t)\right)\p=\left(-\frac{\hbar^2}{2m^*}\frac{\partial^2}{\partial x^2}+ q\V(x,t)\right)\p.
\label{apnum1}
\end{equation}
The spatial and time derivatives in our temporal and spatial meshes can be approximated by:
\begin{eqnarray}
\left.\frac{\partial \psi \left(x,t \right)}{\partial t}\right|_{x = x_k;t = t_j} &\approx& \frac{\psi(t_{j + 1},x_k) - \psi(t_{j - 1},x_k)} {2\Delta t}, \label{apnum2} \\
\left.\frac{{{\partial }^{2}}\psi \left( x,t \right)}{\partial {{x}^{2}}}\right|_{x = x_k;t = t_j} &\approx& \frac{\psi(t_{j},x_{k + 1}) - 2\psi(t_{j},x_{k}) + \psi(t_{j},x_{k - 1})}{{\Delta x}^{2}}.\nonumber
\end{eqnarray}
Inserting \eqref{apnum2}  into \eqref{apnum1}, we obtain the following simple recursive expression to evaluate the evolution of the quantum state:
\begin{eqnarray}
\psi(t_{j + 1},x_{k}) &=& \psi(t_{j - 1},x_{k}) + i\frac{\hbar \Delta t}{{{\Delta x}^{2}}m^*}\left(\psi(t_{j},x_{k + 1}) - 2\psi(t_{j},x_{k}) + \psi(t_{j},x_{k - 1})\right)\nonumber\\
&-&\,i\frac{2 q \Delta t}{\hbar }A(t_{j},x_k) \psi(t_{j},x_{k}).
\label{apnum4}
\end{eqnarray}
Once we know the wave function at the particular times $t_j$ and $t_{j - 1}$ for all spatial positions in the mesh, we can compute
the wave function at next time $t_{j + 1}$, using \eqref{apnum4}. This algorithm solves an initial-value problem. To avoid discussions on the boundary conditions, we use a very large
spatial simulating box so that the entire wave packet is contained in it at any time, i.e., the wave function at the boundaries is always negligible.

The time-dependent Schr\"odinger equation in \eqref{apnum1} gives the following continuity equation: 
\begin{eqnarray}
\frac{\partial |\psi(x,t)|^2}{\partial t}+\frac{\partial }{\partial x}\left( \frac {\hbar}{m^*} \Im{ \psi^*(x,t) \frac{\partial \psi(x,t)}{\partial x}} \right)=0,
\label{apnum5bis}
\end{eqnarray}
which gives the Bohmian velocity:
\begin{eqnarray}
v_B^{\psi}(x,t)= \frac {\hbar}{m^*} \Im{\frac{\frac{\partial \psi(x,t)}{\partial x}}{\psi(x,t)}}.
\label{apnum6}
\end{eqnarray}
Thus, the Bohmian trajectory $x_B(t_j)$ is computed from a numerical time-integration of the velocity \eqref{apnum6} given by:
\begin{eqnarray}
x_B(t_{j+1})=x_B(t_j)+v_B^{\psi}(t_j,x_{B,j}) \Delta t,
\label{apnum7}
\end{eqnarray}
where $x_{B,j}$ is the spatial grid point corresponding to $x_B(t_j)$. Technically, to minimize errors, if $v_B^{\psi}(t_j,x_j) \Delta t$ is greater than $\Delta x$ a smaller $\Delta t'$ (that just ensures that no grid is changed) is used and a new velocity is computed when reaching the new spatial cell.    

\subsection{Time evolution of the state in an arbitrary gauge}

In this paper, we use the following gauge function: 
\begin{equation}
g(x,t)=g_o\cos(k_g x + w_g t +\theta),
\label{apnum7gauge}
\end{equation}
with $g_o=1\cdot 10^{-14}$ V/s  , $k_g=8\cdot 10^{6}$ $m^{-1}$ and $w_g=10$ Trad/s. The angle $\theta$ is changed to define a set of different gauge functions.  Then, the states at the preparation time in \eqref{apnum5} are re-defined as:
\begin{eqnarray}
\label{apnum8}
\Psi^g_{E_c,x_c,\sigma_x}(x,t_p) &=& e^{i\frac{q}{\hbar}g(x,t)}\Psi_{E_c,x_c,\sigma_x}(x,t_p), \\
\Psi^g_{E_c,x_c,\sigma_x}(x,t_p+\Delta t)&=&e^{i\frac{q}{\hbar}g(x,t)}\Psi_{E_c,x_c,\sigma_x}(x,t_p+\Delta t).\nonumber
\end{eqnarray}
The Schrödinger equation in an arbitrary gauge for a particle in free space is obtained by changing $\V\to\V^g$, $\A\to\A^g$ and $\p\to\p^g$ so that: 
\begin{equation}
i\hbar \frac{\partial \pg}{\partial t}=\left(\frac{1}{2m^*}\left( -i\hbar \frac{\partial}{\partial x} -q\frac{\partial  g(x,t)}{\partial x}\right)^2+ q\V(x,t)-q\frac{\partial g(x,t)}{\partial t}\right)\pg,\nonumber
\label{apnum10}
\end{equation}
that can be rewritten as:
\begin{eqnarray}
i\hbar \frac{\partial \pg}{\partial t}=\left( -\frac{\hbar}{2 m^*} \frac{\partial^2}{\partial x^2} +i\frac{\hbar q}{2 m^*}\frac{\partial^2  g(x,t)}{\partial x^2}+i\frac{\hbar q}{ m^*}\frac{\partial  g(x,t)}{\partial x}\frac{\partial}{\partial x}\right.\nonumber\\
\left.
+\frac{q^2}{2 m^*}\left(\frac{\partial  g(x,t)}{\partial x}\right)^2+  q\V(x,t)-q\frac{\partial g(x,t)}{\partial t}\right)\pg.
\label{apnum11}
\end{eqnarray}
From \eqref{apnum7gauge}, the following discrete functions are defined:
\begin{eqnarray}
\left.\frac{\partial g \left(x,t \right)}{\partial t}\right|_{x = x_k;t = t_j} &\approx& \frac{g(t_{j + 1},x_k) - g(t_{j - 1},x_k)} {2\Delta t} \label{apnum12}, \\
\left.\frac{\partial g \left(x,t \right)}{\partial x}\right|_{x = x_k;t = t_j} &\approx& \frac{g(t_{j },x_{k+1}) - g(t_{j},x_{k-1})} {2\Delta x} \label{apnum13}, \\
\left.\frac{{{\partial }^{2}}g\left( x,t \right)}{\partial {{x}^{2}}}\right|_{x = x_k;t = t_j} &\approx& \frac{g(t_{j},x_{k + 1}) - 2g(t_{j},x_{k}) + g(t_{j},x_{k - 1})}{{\Delta x}^{2}}. \label{apnum14}
\end{eqnarray}
Inserting \eqref{apnum12}, \eqref{apnum13}  and \eqref{apnum14} into \eqref{apnum11}, we obtain the following simple recursive expression:
\begin{eqnarray}
&&\psi^g(t_{j + 1},x_{k}) = \psi^g(t_{j - 1},x_{k}) + i\frac{\hbar \Delta t}{{{\Delta x}^{2}}m^*}\left(\psi^g(t_{j},x_{k + 1}) - 2\psi^g(t_{j},x_{k}) + \psi^g(t_{j},x_{k - 1})\right)-\nonumber\\
&&i\frac{2 q \Delta t}{\hbar }A(t_{j},x_k) \psi^g(t_{j},x_{k})
+\frac{q \Delta t}{m^*}\frac{g(t_{j},x_{k + 1}) - 2g(t_{j},x_{k}) + g(t_{j},x_{k - 1})}{{\Delta x}^{2}} \psi^g(t_{j},x_{k})\nonumber\\
&&+\frac{q \Delta t}{ m^* \Delta x} \frac{g(t_{j },x_{k+1}) 
- g(t_{j},x_{k-1})} {2\Delta x}(\psi^g(t_{j },x_{k+1}) - \psi^g(t_{j},x_{k-1}))\\
&&-i\frac{q^2 \Delta t}{m^* \hbar}\left(\frac{g(t_{j },x_{k+1}) - g(t_{j},x_{k-1})} {2\Delta x}\right)^2 \psi^g(t_{j},x_{k}) + i\frac{2 q \Delta t}{\hbar}\frac{g(t_{j + 1},x_k) - g(t_{j - 1},x_k)} {2\Delta t}\psi^g(t_{j},x_{k}).\nonumber
\label{apnum15}
\end{eqnarray}

\subsection{Time evolution of the state in the Coulomb gauge with magnetic field}

The Schrödinger equation in \eqref{hamiB} can be re-written as:
\begin{eqnarray}
\frac{1}{2m^*}\left( -{\hbar^2} \frac{\partial^2}{\partial x^2} -{\hbar^2} \frac{\partial^2}{\partial y^2}+i 2 qBx {\hbar} \frac{\partial}{\partial y}+\left(qBx  \right)^2\right) \Psi\rt=E \Psi(x,y,t).
\label{apn2}
\end{eqnarray}
Then, the continuity equation of the presence probability density of \eqref{apn2} is given by: 
\begin{eqnarray}
\frac{\partial |\Psi(x,y,t)|^2}{\partial t}&+&\frac{\partial }{\partial x}\left( \frac {\hbar}{m^*} \Im{ \Psi^*(x,y,t) \frac{\partial \Psi(x,y,t)}{\partial x}} \right)\nonumber\\
&+&\frac{\partial }{\partial x}\left( \frac {\hbar}{m^*} \Im{ \Psi^*(x,y,t)\frac{\partial \Psi(x,y,t)}{\partial y}} -\frac{q\; B\; x}{m^*} |\Psi(x,y,t)|^2\right)=0, \nonumber
\end{eqnarray}
where the Bohmian velocity in the x-component is:
\begin{eqnarray}
v_x(x,y,t)= \frac {\hbar}{m^*} \Im{\frac{\frac{\partial \Psi(x,y,t)}{\partial x}}{\Psi(x,y,t)}},\nonumber
\end{eqnarray}
and the Bohmian velocity in the y-component is:
\begin{eqnarray}
v_y(x,y,t)= \frac {\hbar}{m^*} \Im{\frac{\frac{\partial \Psi(x,y,t)}{\partial y}}{\Psi(x,y,t)}} -\frac{q\; B\; x}{m^*}.\nonumber
\end{eqnarray}
The Hamiltonian of \eqref{hamiB} has translational invariance in the $y$ directions because $[ -i{\hbar} \frac{\partial}{\partial y}, H]=0$, so that one can look for energy eigenstates that are also eigenstates of the canonical momentum in the $y$ direction $\Psi\rt=e^{ik_y y} \psi(x)$. Then, \eqref{apn2} becomes:
\begin{eqnarray}
e^{ik_y y} \frac{1}{2m^*}\left( -{\hbar^2} \frac{\partial^2}{\partial x^2} +{\hbar^2} k_y^2-2 qBx {\hbar} k_y+q^2B^2x^2  \right)  \psi(x)=e^{ik_y y} E  \psi(x),\nonumber\\
e^{ik_y y} \left( -\frac{\hbar^2}{2m^*} \frac{\partial^2}{\partial x^2} +\frac{q^2B^2}{2m^*}\left( \frac{{\hbar} k_y}{qB}-x \right)^2 \right) \psi(x)=e^{ik_y y} E  \psi(x).\nonumber
\end{eqnarray}
We define $\omega_B=\frac{|q|B}{m^*}$ and $l_B=\sqrt{\frac{\hbar}{|q|B}}$ so that:
\begin{eqnarray}
\left(-\frac{\hbar^2}{2m^*} \frac{\partial^2}{\partial x^2} +\frac{1}{2}m^* \omega_B^2 \left(x + k_y l_B^2 \right)^2 \right) \psi(x)= E  \psi(x).\nonumber
\end{eqnarray}
We define $x'=x + k_y l_B^2$ to get the equation of a displaced harmonic oscillator: 
\begin{eqnarray}
\left(-\frac{\hbar^2}{2m^*} \frac{\partial^2}{\partial x'^2} +\frac{1}{2}m^* \omega_B^2 (x')^2  \right) \psi(x'-k_y l_B^2)= E  \psi(x'-k_y l_B^2).\nonumber
\end{eqnarray}
The global energy eigenstates of \eqref{apn2} is by two quantum numbers $n$ and $k_y$, as:
\begin{equation}
\psi_{n,k_y}(x,y)=\frac{1}{\BN} H_n\left((x- x_y)/l_B\right) e^{-\frac{m^* \omega}{2\hbar}(x- x_y)^2}e^{ik_y y},
\label{egvh}
\end{equation}
with $\BN$ a normalization constant,  $\omega_B=\frac{|q|B}{m^*}=0.49$ Trad/s the cyclotron frequency of the harmonic oscillator,  $x_y=k_y l_B^2=40.8$ nm the displacement of the harmonic oscillator,  $l_B=\sqrt{\frac{\hbar}{|q|B}}=58$ nm the length and $H_n(x)$ the usual Hermite polynomial wave functions  of order $n$ of the harmonic oscillator \cite{landau1,landau2}. The energy of such a Hamiltonian eigenstate is given by $E_{n,k_y}=\hbar \omega\left(\frac{1}{2}+n \right)$. Then, the time evolution of each Hamilton eigenstate in \eqref{egvh}  can be easily computed analytically as:
\begin{equation}
c_{n,k_y}(t)=c_{n,k_y}(0)e^{-i\frac{E_{n,k_y}t}{\hbar}}.
\label{chb}
\end{equation}
We assume $c_{n,k_y}(0)=\BN'$ for all $n=0,...,9$ components where $\BN'$ is another normalization constant.

\section{Weak values post-selected in position satisfy the \textit{time-dependent consistency} in \eqref{which4}}
\label{apenfinal}

We provide here a quite simple alternative demonstration that the selection $|\egf\rangle = |x\rangle$ satisfies \eqref{which4} (and \eqref{which1}) for a one-dimensional system. It is straightforward to show that: $\Re\left\{\frac{\langle x | \opx  |\p(t)\rangle}{\langle x |\p(t)\rangle}\right\} = x$ in \eqref{which1}. Notice that the final result, $x$, is the eigenvalue of the post-selected state $|x\rangle$, with no dependency on any property of the pre-selected state $|\p(t)\rangle$.  Then, $\frac{d}{dt} \langle \p(t)|\opx|\p(t)\rangle = \int dx \; x \; \frac{d}{dt} |\langle \p(t)|x \rangle|^2.$ We can use the continuity equation to express the following time derivative as $\frac{d}{dt} |\langle \p(t)|x \rangle|^2 = -\frac{d}{dx} J(x,t),$ where $J(x,t)$ is the probability current density. Therefore, $\int dx \; x \; \frac{d}{dt} |\langle \p(t)|x \rangle|^2 = -\int dx \; x \; \frac{d}{dx} J(x,t) = \int dx \; J(x,t),$ where we have performed an integration by parts, assuming that $x J(x,t) \to 0$ as $x \to \pm \infty$. Finally, using the definition of the Bohmian velocity $v_B^{|\p\rangle}(x,t) = \frac{J(x,t)}{|\p(x,t)|^2}$, we obtain the expected result:
\begin{equation}
\frac{d}{dt} \langle \p(t)|\opx|\p(t)\rangle = \int dx \; |\langle \p(t)|x \rangle|^2 v_B^{|\p\rangle}(x,t),
\label{which4bis}
\end{equation}
which is expression \eqref{which4} written for $|\egf\rangle = |x\rangle$. The fundamental requirement in this simple demonstration has been the result  $\Re\left\{\frac{\langle x | \opx  |\p(t)\rangle}{\langle x |\p(t)\rangle}\right\} = x$, which specifies that if you want to use weak values to extract local information (i.e. $x-$density) on $\langle \p(t)|\opx|\p(t)\rangle$ and simultaneously use the time-derivative of such weak values to extract also local information (i.e. $x-$density) on $\frac{d}{dt} \langle \p(t)|\opx|\p(t)\rangle$ you must use weak values of the position post-selection in position. This is exactly condition \textbf{C4}.

\end{appendix}

\nocite{*}


\begin{thebibliography}{99}

\bibitem{aharonov1988} Aharonov Y, Albert DZ, Vaidman L. How the result of a measurement of a component of the spin of a spin-1/2 particle can turn out to be 100. \textit{Phys Rev Lett}. 1988;60:1351. \href{https://doi.org/10.1103/PhysRevLett.60.1351}{https://doi.org/10.1103/PhysRevLett.60.1351}

\bibitem{aharonov1990} Aharonov Y, Vaidman L. Properties of a quantum system during the time interval between two measurements. \textit{Phys Rev A}. 1990;41:11. \href{https://doi.org/10.1103/PhysRevA.41.11}{https://doi.org/10.1103/PhysRevA.41.11}

\bibitem{vaidman1996} Vaidman L. Weak-measurement elements of reality. \textit{Found Phys}. 1996;26:895. \href{https://doi.org/10.1007/BF02148832}{https://doi.org/10.1007/BF02148832}

\bibitem{aharonov1964} Aharonov Y, Bergmann PG, Lebowitz JL. Time symmetry in the quantum process of measurement. \textit{Phys Rev}. 1964;134:B1410. \href{https://doi.org/10.1103/PhysRev.134.B1410}{https://doi.org/10.1103/PhysRev.134.B1410}

\bibitem{wiseman2007} Wiseman HM. Grounding Bohmian mechanics in weak values and Bayesianism. \textit{New J Phys}. 2007;9:165. \href{https://doi.org/10.1088/1367-2630/9/6/165}{https://doi.org/10.1088/1367-2630/9/6/165}

																																														   
 
\bibitem{durr2009} Dürr D, Goldstein S, Zanghì N. On the weak measurement of velocity in Bohmian mechanics. \textit{J Stat Phys}. 2009;134:1023. \href{https://doi.org/10.1007/s10955-008-9674-0}{https://doi.org/10.1007/s10955-008-9674-0}

\bibitem{foo2022} Foo J, Asmodelle E, Lund AP, et al. Relativistic Bohmian trajectories of photons via weak measurements. \textit{Nat Commun}. 2022;13:4002. \href{https://doi.org/10.1038/s41467-022-31608-6}{https://doi.org/10.1038/s41467-022-31608-6}

\bibitem{destefani2023b} Destefani CF, Oriols X. Assessing quantum thermalization in physical and configuration spaces via many-body weak values. \textit{Phys Rev A}. 2023;107:012213. \href{https://doi.org/10.1103/PhysRevA.107.012213}{https://doi.org/10.1103/PhysRevA.107.012213}

\bibitem{destefani2023} Destefani CF, Oriols X. Kinetic energy equipartition: A tool to characterize quantum thermalization. \textit{Phys Rev Research}. 2023;5:033168. \href{https://doi.org/10.1103/PhysRevResearch.5.033168}{https://doi.org/10.1103/PhysRevResearch.5.033168}

\bibitem{oriols1996} Oriols X, Martín F, Sune J. Implications of the noncrossing property of Bohm trajectories in one-dimensional tunneling configurations. \textit{Phys Rev A}. 1996;54:2594. \href{https://doi.org/10.1103/PhysRevA.54.2594}{https://doi.org/10.1103/PhysRevA.54.2594}

\bibitem{ruseckas2002} Ruseckas J, Kaulakys B. Weak measurement of arrival time. \textit{Phys Rev A}. 2002;66:052106. \href{https://doi.org/10.1103/PhysRevA.66.052106}{https://doi.org/10.1103/PhysRevA.66.052106}


\bibitem{ahnert2004} Ahnert SE, Payne MC. Weak measurement of the arrival times of single photons and pairs of entangled photons. \textit{Phys Rev A}. 2004;69:042103. \href{https://doi.org/10.1103/PhysRevA.69.042103}{https://doi.org/10.1103/PhysRevA.69.042103}

\bibitem{steinberg1995} Steinberg AM. How much time does a tunneling particle spend in the barrier region? \textit{Phys Rev Lett}. 1995;74:2405. \href{https://doi.org/10.1103/PhysRevLett.74.2405}{https://doi.org/10.1103/PhysRevLett.74.2405}

\bibitem{campo2004} Campo D, Parentani R. Space-time correlations in inflationary spectra: A wave-packet analysis. \textit{Phys Rev D}. 2004;70:105020. \href{https://doi.org/10.1103/PhysRevD.70.105020}{https://doi.org/10.1103/PhysRevD.70.105020}

\bibitem{review2012} Kofman AG, Ashhab S, Nori F. Nonperturbative theory of weak pre- and post-selected measurements. \textit{Phys Rep}. 2012;520(2):43. \href{https://doi.org/10.1016/j.physrep.2012.07.001}{https://doi.org/10.1016/j.physrep.2012.07.001}

\bibitem{diosi2023} Adam P, Diosi L. Sequential unsharp measurement of photon polarization. \textit{Phys Rev A}. 2023;107(6):063706. \href{https://doi.org/10.1103/PhysRevA.107.063706}{https://doi.org/10.1103/PhysRevA.107.063706}
 
																																																		 

\bibitem{brout1995} Brout R, Massar S, Parentani R, Spindel P. A primer for black hole quantum physics. \textit{Phys Rep}. 1995;260:329. \href{https://doi.org/10.1016/0370-1573(95)00008-5}{https://doi.org/10.1016/0370-1573(95)00008-5}

\bibitem{brun2008} Brun T, Diosi L, Strunz WT. Test of weak measurement on a two- or three-level system. \textit{Phys Rev A}. 2008;77:032101. \href{https://doi.org/10.1103/PhysRevA.77.032101}{https://doi.org/10.1103/PhysRevA.77.032101}

\bibitem{botero2000} Botero A, Reznik B. Quantum-communication protocol with correlated weak measurements. \textit{Phys Rev A}. 2000;61:050301(R). \href{https://doi.org/10.1103/PhysRevA.61.050301}{https://doi.org/10.1103/PhysRevA.61.050301}

\bibitem{hofmann2010} Hofmann HF. Complete characterization of post-selected quantum statistics using weak measurement tomography. \textit{Phys Rev A}. 2010;81:012103. \href{https://doi.org/10.1103/PhysRevA.81.012103}{https://doi.org/10.1103/PhysRevA.81.012103}

\bibitem{kastner2004} Kastner RE. Weak values and consistent histories in quantum theory. \textit{Stud Hist Philos M P}. 2004;35:57. \href{https://doi.org/10.1016/j.shpsb.2003.02.001}{https://doi.org/10.1016/j.shpsb.2003.02.001}

\bibitem{leavens2005} Leavens CR. Weak measurements from the point of view of Bohmian mechanics. \textit{Found Phys}. 2005;35:469. \href{https://doi.org/10.1007/s10701-004-1984-8}{https://doi.org/10.1007/s10701-004-1984-8}

\bibitem{devashish2021} Pandey D, Sampaio R, Ala-Nissila T, Albareda G, Oriols X. Identifying weak values with intrinsic dynamical properties in modal theories. \textit{Phys Rev A}. 2021;103:052219. \href{https://doi.org/10.1103/PhysRevA.103.052219}{https://doi.org/10.1103/PhysRevA.103.052219}

\bibitem{agarwal2007} Agarwal GS, Pathak PK. Realization of quantum-mechanical weak values of observables using entangled photons. \textit{Phys Rev A}. 2007;75:032108. \href{https://doi.org/10.1103/PhysRevA.75.032108}{https://doi.org/10.1103/PhysRevA.75.032108}

\bibitem{marian2016} Marian D, Zanghi N, Oriols X. Weak values from displacement currents in multiterminal electron devices. \textit{Phys Rev Lett}. 2016;116:110404. \href{https://doi.org/10.1103/PhysRevLett.116.110404}{https://doi.org/10.1103/PhysRevLett.116.110404}

\bibitem{sbramanian2023} Subramanian M, Mathew A, Muralidharan B. Resonant weak-value enhancement for solid-state quantum metrology. \textit{Phys Rev Appl}. 2023;20:044065. \href{https://doi.org/10.1103/PhysRevApplied.20.044065}{https://doi.org/10.1103/PhysRevApplied.20.044065}

\bibitem{pryde2005} Pryde GJ, O’Brien JL, White AG, Ralph TC, Wiseman HM. Measurement of quantum weak values of photon polarization. \textit{Phys Rev Lett}. 2005;94:220405. \href{https://doi.org/10.1103/PhysRevLett.94.220405}{https://doi.org/10.1103/PhysRevLett.94.220405}

\bibitem{jordan2010} Jordan AN, Korotkov AN. Uncollapsing the wavefunction by undoing quantum measurements. \textit{Contemp Phys}. 2010;51:125. \href{https://doi.org/10.1080/00107510903385292}{https://doi.org/10.1080/00107510903385292}

\bibitem{jordan2007} Jordan AN, Trauzettel B, Burkard G. Weak-measurement theory of quantum-dot spin qubits. \textit{Phys Rev B}. 2007;76:155324. \href{https://doi.org/10.1103/PhysRevB.76.155324}{https://doi.org/10.1103/PhysRevB.76.155324}

\bibitem{korotkov2006} Korotkov AN, Jordan AN. Undoing a weak quantum measurement of a solid-state qubit. \textit{Phys Rev Lett}. 2006;97:166805. \href{https://doi.org/10.1103/PhysRevLett.97.166805}{https://doi.org/10.1103/PhysRevLett.97.166805}

\bibitem{romito2008} Romito A, Gefen Y, Blanter YM. Weak values of electron spin in a double quantum dot. \textit{Phys Rev Lett}. 2008;100:056801. \href{https://doi.org/10.1103/PhysRevLett.100.056801}{https://doi.org/10.1103/PhysRevLett.100.056801}

\bibitem{zilberberg2011} Zilberberg O, Romito A, Gefen Y. Null values and quantum state discrimination. \textit{Phys Rev Lett}. 2011;106:080405. \href{https://doi.org/10.1103/PhysRevLett.106.080405}{https://doi.org/10.1103/PhysRevLett.106.080405}
 
																																																														 

\bibitem{hariri2019experimental} Hariri A, Curic D, Giner L, Lundeen JS. Experimental realization of weak value amplification using a single photon source. \textit{Phys Rev A}. 2019;100:032119. \href{https://doi.org/10.1103/PhysRevA.100.032119}{https://doi.org/10.1103/PhysRevA.100.032119}

\bibitem{ramos2020measurement} Ramos R, Spierings D, Racicot I, Steinberg AM. Measurement of the time spent by a tunneling atom within the barrier region. \textit{Nature}. 2020;583:529. \href{https://doi.org/10.1038/s41586-020-2490-7}{https://doi.org/10.1038/s41586-020-2490-7}

\bibitem{review2014} Dressel J, Malik M, Miatto FM, Jordan AN, Boyd RW. Colloquium: Understanding quantum weak values: Basics and applications. \textit{Rev Mod Phys}. 2014;86:307. \href{https://doi.org/10.1103/RevModPhys.86.307}{https://doi.org/10.1103/RevModPhys.86.307}

\bibitem{lundeen2011} Lundeen JS, Sutherland B, Patel A, Stewart C, Bamber C. Direct measurement of the quantum wavefunction. \textit{Nature}. 2011;474(7350):188. \href{https://doi.org/10.1038/nature10120}{https://doi.org/10.1038/nature10120}

\bibitem{zhu2021} Zhu J, Wang A, Liu X, Liu Y, Zhang Z, Gao F. Reconstructing the wave function through the momentum weak value. \textit{Phys Rev A}. 2021;104(3):032221. \href{https://doi.org/10.1103/PhysRevA.104.032221}{https://doi.org/10.1103/PhysRevA.104.032221}

\bibitem{ricthie1991} Ritchie NWM, Story JG, Hulet RG. Realization of a measurement of a ‘weak value’. \textit{Phys Rev Lett}. 1991;66(9):1107. \href{https://doi.org/10.1103/PhysRevLett.66.1107}{https://doi.org/10.1103/PhysRevLett.66.1107}

\bibitem{resch2004} Resch KJ, Lundeen JS, Steinberg AM. Experimental realization of the quantum box problem. \textit{Phys Lett A}. 2004;324:125. \href{https://doi.org/10.1016/j.physleta.2004.02.042}{https://doi.org/10.1016/j.physleta.2004.02.042}

\bibitem{williams2008} Williams NS, Jordan AN. Weak values and the Leggett-Garg inequality in solid-state qubits. \textit{Phys Rev Lett}. 2008;100:026804. \href{https://doi.org/10.1103/PhysRevLett.100.026804}{https://doi.org/10.1103/PhysRevLett.100.026804}

\bibitem{brunner2004} Brunner N, Scarani V, Wegmuller M, Legré M, Gisin N. Direct measurement of superluminal group velocity and signal velocity in an optical fiber. \textit{Phys Rev Lett}. 2004;93:203902. \href{https://doi.org/10.1103/PhysRevLett.93.203902}{https://doi.org/10.1103/PhysRevLett.93.203902}

\bibitem{kocsis2011observing} Kocsis S, Braverman B, Ravets S, Stevens MJ, Mirin RP, Shalm LK, Steinberg AM. Observing the average trajectories of single photons in a two-slit interferometer. \textit{Science}. 2011;332:1170. \href{https://doi.org/10.1126/science.1202218}{https://doi.org/10.1126/science.1202218}

\bibitem{hosten2008} Hosten O, Kwiat P. Observation of the spin Hall effect of light via weak measurements. \textit{Science}. 2008;319:787. \href{https://doi.org/10.1126/science.1152697}{https://doi.org/10.1126/science.1152697}

\bibitem{starling2009} Starling DJ, Dixon PB, Jordan AN, Howell JC. Optimizing the signal-to-noise ratio of a beam-deflection measurement with interferometric weak values. \textit{Phys Rev A}. 2009;80:041803(R). \href{https://doi.org/10.1103/PhysRevA.80.041803}{https://doi.org/10.1103/PhysRevA.80.041803}

\bibitem{starling2010} Starling DJ, Dixon PB, Jordan AN, Howell JC. Precision frequency measurements with interferometric weak values. \textit{Phys Rev A}. 2010;82:063822. \href{https://doi.org/10.1103/PhysRevA.82.063822}{https://doi.org/10.1103/PhysRevA.82.063822}

\bibitem{braverman2013} Braverman B, Simon C. Proposal to observe the nonlocality of Bohmian trajectories with entangled photons. \textit{Phys Rev Lett}. 2013;110:060406. \href{https://doi.org/10.1103/PhysRevLett.110.060406}{https://doi.org/10.1103/PhysRevLett.110.060406}

\bibitem{healey2007} Healey R. \textit{Gauging what's real: The conceptual foundations of contemporary gauge theories}. Oxford: Oxford University Press; 2007.

\bibitem{kobe1982} Kobe DH, Wen ET. Gauge invariance in quantum mechanics: charged harmonic oscillator in an electromagnetic field. \textit{J Phys A: Math Gen}. 1982;15(3):787. \href{https://doi.org/10.1088/0305-4470/15/3/018}{https://doi.org/10.1088/0305-4470/15/3/018}

\bibitem{stokes2021} Stokes A, Nazir A. Identification of Poincaré-gauge and multipolar nonrelativistic theories of QED. \textit{Phys Rev A}. 2021;104(3):032227. \href{https://doi.org/10.1103/PhysRevA.104.032227}{https://doi.org/10.1103/PhysRevA.104.032227}

\bibitem{jackson2001} Jackson JD, Okun LB. Historical roots of gauge invariance. \textit{Rev Mod Phys}. 2001;73(3):663. \href{https://doi.org/10.1103/RevModPhys.73.663}{https://doi.org/10.1103/RevModPhys.73.663}

\bibitem{aharonov1959} Aharonov Y, Bohm D. Significance of electromagnetic potentials in the quantum theory. \textit{Phys Rev}. 1959;115(3):485. \href{https://doi.org/10.1103/PhysRev.115.485}{https://doi.org/10.1103/PhysRev.115.485}

\bibitem{cohen1986} Cohen-Tannoudji C, Diu B, Laloe F. \textit{Quantum mechanics, volume 1}. New York: John Wiley and Sons; 1986.

\bibitem{optics} Grynberg G, Aspect A, Fabre C. \textit{Introduction to quantum optics: from the semi-classical approach to quantized light}. New York: Cambridge University Press; 2010.

\bibitem{Ballentine2014} Ballentine LE. \textit{Quantum mechanics: a modern development}. Singapore: World Scientific Publishing Company; 2000.

\bibitem{sokolovski2016} Sokolovski D. Weak measurements measure probability amplitudes (and very little else). \textit{Phys Lett A}. 2016;380(18-19):1593. \href{https://doi.org/10.1016/j.physleta.2016.02.051}{https://doi.org/10.1016/j.physleta.2016.02.051}
																																																																																																																														  

\bibitem{matzkin2019} Matzkin A. Weak values and quantum properties. \textit{Found Phys}. 2019;49(3):298. \href{https://doi.org/10.1007/s10701-019-00245-3}{https://doi.org/10.1007/s10701-019-00245-3}

\bibitem{reznik1995} Reznik B, Aharonov Y. Time-symmetric formulation of quantum mechanics. \textit{Phys Rev A}. 1995;52:2538. \href{https://doi.org/10.1103/PhysRevA.52.2538}{https://doi.org/10.1103/PhysRevA.52.2538}

\bibitem{hiley2012} Hiley BJ. Weak values: Approach through the Clifford and Moyal algebras. \textit{J Phys: Conf Ser}. 2012;361:012014. \href{https://doi.org/10.1088/1742-6596/361/1/012014}{https://doi.org/10.1088/1742-6596/361/1/012014}

\bibitem{kastner2017} Kastner RE. Demystifying weak measurements. \textit{Found Phys}. 2017;47:697. \href{https://doi.org/10.1007/s10701-017-0085-4}{https://doi.org/10.1007/s10701-017-0085-4}

\bibitem{cohen2017} Cohen E. What weak measurements and weak values really mean: Reply to Kastner. \textit{Found Phys}. 2017;47(10):1261. \href{https://doi.org/10.1007/s10701-017-0107-2}{https://doi.org/10.1007/s10701-017-0107-2}

\bibitem{sinclair2019} Sinclair J, Spierings D, Brodutch A, Steinberg AM. Interpreting weak value amplification with a toy realist model. \textit{Phys Lett A}. 2019;383(24):2839. \href{https://doi.org/10.1016/j.physleta.2019.06.010}{https://doi.org/10.1016/j.physleta.2019.06.010}

\bibitem{garretson2004} Garretson JL, Wiseman HM, Pope DT, Pegg DT. The uncertainty relation in 'which-way' experiments: how to observe directly the momentum transfer using weak values. \textit{J Opt B: Quantum Semiclass Opt}. 2004;6:S506. \href{https://doi.org/10.1088/1464-4266/6/6/008}{https://doi.org/10.1088/1464-4266/6/6/008}

\bibitem{rebufello2021} Rebufello E, Piacentini F, Avella A, Souza MA, de Gramegna M, Dziewior J, Cohen E, Vaidman L, Degiovanni IP, Genovese M. Anomalous weak values via a single photon detection. \textit{Light Sci Appl}. 2021;10:106. \href{https://doi.org/10.1038/s41377-021-00539-0}{https://doi.org/10.1038/s41377-021-00539-0}

\bibitem{cohen2018} Cohen E, Pollak E. Determination of weak values of quantum operators using only strong measurements. \textit{Phys Rev A}. 2018;98:042112. \href{https://doi.org/10.1103/PhysRevA.98.042112}{https://doi.org/10.1103/PhysRevA.98.042112}

\bibitem{beche2016} Berche B, Malterre D, Medina E. Gauge transformations and conserved quantities in classical and quantum mechanics. \textit{Am J Phys}. 2016;84(8):616. \href{https://doi.org/10.1119/1.4955153}{https://doi.org/10.1119/1.4955153}

\bibitem{tumulka2022} Tumulka R. Limitations to genuine measurements in ontological models of quantum mechanics. \textit{Found Phys}. 2022;52(5):110. \href{https://doi.org/10.1007/s10701-022-00633-2}{https://doi.org/10.1007/s10701-022-00633-2}

\bibitem{kobe1985} Kobe DH, Yang KH. Gauge transformation of the time-evolution operator. \textit{Phys Rev A}. 1985;32(2):952. \href{https://doi.org/10.1103/PhysRevA.32.952}{https://doi.org/10.1103/PhysRevA.32.952}

\bibitem{svensson2013pedagogical} Svensson BEY. Pedagogical review of quantum measurement theory with an emphasis on weak measurements. \textit{Quanta}. 2013;2:18. \href{https://doi.org/10.12743/quanta.v2i1.12}{https://doi.org/10.12743/quanta.v2i1.12}

\bibitem{wiseman2009} Wiseman H, Milburn GJ. \textit{Quantum measurement and control}. New York: Cambridge University Press; 2009.

\bibitem{ehrenfest1927} Ehrenfest P. Bemerkung über die angenäherte Gültigkeit der klassischen Mechanik innerhalb der Quantenmechanik. \textit{Zeitschrift fur Physik}. 1927;45:455. \href{https://doi.org/10.1007/BF01329203}{https://doi.org/10.1007/BF01329203}

\bibitem{dressel2012} Dressel J, Jordan AN. Significance of the imaginary part of the weak value. \textit{Phys Rev A}. 2012;85:012107. \href{https://doi.org/10.1103/PhysRevA.85.012107}{https://doi.org/10.1103/PhysRevA.85.012107}

\bibitem{jozsa2007} Jozsa R. Complex weak values in quantum measurement. \textit{Phys Rev A}. 2007;76(4):044103. \href{https://doi.org/10.1103/PhysRevA.76.044103}{https://doi.org/10.1103/PhysRevA.76.044103}

\bibitem{freericks2023} Freericks JK. How to measure the momentum of single quanta. \textit{arXiv}. 2023. \href{https://doi.org/10.48550/arXiv.2302.12303}{https://doi.org/10.48550/arXiv.2302.12303}

\bibitem{deotto1998} Deotto E, Ghirardi GC. Bohmian mechanics revisited. \textit{Found Phys}. 1998;28:1. \href{https://doi.org/10.1023/A:1018752202576}{https://doi.org/10.1023/A:1018752202576}

\bibitem{nelson1966} Nelson E. Derivation of the Schrödinger equation from Newtonian mechanics. \textit{Phys Rev}. 1966;150:1079. \href{https://doi.org/10.1103/PhysRev.150.1079}{https://doi.org/10.1103/PhysRev.150.1079}

\bibitem{bohm1989} Bohm D, Hiley BJ. Non-locality and locality in the stochastic interpretation of quantum mechanics. \textit{Phys Rep}. 1989;172:93. \href{https://doi.org/10.1016/0370-1573(89)90160-9}{https://doi.org/10.1016/0370-1573(89)90160-9}

\bibitem{pang2020} Pang AO, et al. Experimental comparison of Bohm-like theories with different guidance equations. \textit{Quantum}. 2020;4:365. \href{https://doi.org/10.22331/q-2020-11-26-365}{https://doi.org/10.22331/q-2020-11-26-365}

\bibitem{destefani2022} Destefani CF, Villani M, Cartoixa X, Feiginov M, Oriols X. Resonant tunneling diodes in semiconductor microcavities: modeling polaritonic features in the THz displacement current. \textit{Phys Rev B}. 2022;106:205306. \href{https://doi.org/10.1103/PhysRevB.106.205306}{https://doi.org/10.1103/PhysRevB.106.205306}

\bibitem{landau1} Landau LD. Diamagnetismus der metalle. \textit{Zeitschrift für Physik}. 1930;64:629. \href{https://doi.org/10.1007/BF01397213}{https://doi.org/10.1007/BF01397213}

\bibitem{landau2} Landau LD, Lifshitz EM. \textit{Quantum mechanics: non-relativistic theory (Vol. 3)}. Oxford: Elsevier; 2013.

\end{thebibliography}
\end{document}